\documentclass[%
 reprint,
 superscriptaddress,
 amsmath,amssymb,
 aps,
 prb,
 twocolumn
]{revtex4-2}

\usepackage{graphicx}
\usepackage{dcolumn}
\usepackage{bm}
\usepackage{colortbl}
\usepackage{float}
\usepackage[table,dvipsnames]{xcolor}
\usepackage{makecell}
\usepackage[mathscr]{euscript}
\usepackage{multirow}
\usepackage{braket}
\usepackage[pagewise]{lineno}
\usepackage{enumitem}
\usepackage{amssymb}
\usepackage{makecell}
\usepackage{tikz}
\usepackage{bbding}
\usepackage{pifont}
\usepackage{dsserif}
\usepackage{adjustbox}
\usepackage[colorlinks=true, urlcolor=black, linkcolor=black, citecolor=black]{hyperref}

\begin{document}

\title{Zeeman-type spin splittings in strained $d$-wave altermagnets}
\author{Yahui Zhai}
\thanks{These authors contributed equally to this work.}
\affiliation{Key Laboratory of Material Simulation Methods and Software of Ministry of Education, College of Physics, Jilin University, Changchun 130012, China}

\author{Longju Yu}
\thanks{These authors contributed equally to this work.}
\affiliation{Key Laboratory of Material Simulation Methods and Software of Ministry of Education, College of Physics, Jilin University, Changchun 130012, China}

\author{Jian Lv}
\affiliation{Key Laboratory of Material Simulation Methods and Software of Ministry of Education, College of Physics, Jilin University, Changchun 130012, China}

\author{Wei Zhang}
\affiliation{Key Laboratory of Material Simulation Methods and Software of Ministry of Education, College of Physics, Jilin University, Changchun 130012, China}

\author{Hong Jian Zhao}
\affiliation{Key Laboratory of Material Simulation Methods and Software of Ministry of Education, College of Physics, Jilin University, Changchun 130012, China}
\affiliation{Key Laboratory of Physics and Technology for Advanced Batteries (Ministry of Education), College of Physics, Jilin University, Changchun 130012, China}
\affiliation{International Center of Future Science, Jilin University, Changchun 130012, China}

\begin{abstract}
Recently, altermagnetic materials have become rather attractive because such materials showcase combined advantages of ferromagnets (e.g., spin current) and antiferromagnets (e.g., low stray field and ultrafast spin dynamics). Symmetry arguments imply that $d$-wave altermagnets may host strain-induced nonrelativistic Zeeman-type spin splittings (ZSSs), and a theoretical, numerical, and experimental justification of such phenomena are of high necessity. In the present work, we work with collinear spin point groups (SPGs) and use symmetry analysis to identify 15 SPGs that host strain-induced nonrelativistic ZSSs. These 15 SPGs coincide with the cases associated with $d$-wave alternating spin splittings reported in literature. We further corroborate our analysis by first-principles numerical simulations, which indicate that a shear strain of $2\%$ creates sizable nonrelativistic ZSSs of up to 177, 100, and 102 meV in CoF$_2$, LiFe$_2$F$_6$ and $\mathrm{La_{2}O_{3}Mn_{2}Se_{2}}$ $d$-wave altermagnetic semiconductors, respectively. Our work suggests an alternative route toward creating spin current in altermagnets, which may be used to design altermagnetic-based spintronic devices.

\end{abstract}

\maketitle
\noindent

\section{Introduction}
The creation of spin current is of particular importance in spintronics~\cite{wolf2001spintronics, vzutic2004spintronics, sharma2005create, maekawa2017spin}.
Conventional ferromagnets or canted antiferromagnets (with weak magnetization) host Zeeman-type spin splittings (ZSSs)~\cite{liu2023zeeman,milivojevic2024interplay,liu2024unconventional,zhao2022zeeman,tao2024layer,lou2020tunable}, and therefore accommodate spin-polarized carriers and enable spin current~\cite{wolf2001spintronics, maekawa2013spin, schapers2021semiconductor,maekawa2017spin}.  Unlike conventional ferromagnets or antiferromagnets, the recently discovered altermagnets exhibit various ferromagnetic-like phenomena such as broken Kramers degeneracy~\cite{hayami2019momentum, krempasky2024altermagnetic,lee2024broken, reimers2024direct,
hajlaoui2024temperature, zeng2024observation,yang2025three,ding2024large,fedchenko2024observation,zhang2025crystal,vsmejkal2022emerging,mazin2022altermagnetism,tamang2024newly}, the anomalous Hall/Nernst effect~\cite{vsmejkal2020crystal,reichlova2024observation,feng2022anomalous,gonzalez2023spontaneous,han2025nonvolatile,reichlova2024observation,zhou2024crystal,zhou2025manipulation,han2024electrical,song2025altermagnets,takahashi2025elasto}, and giant/tunneling magnetoresistance effect~\cite{shao2021spin,shao2024antiferromagnetic,jiang2023prediction,samanta2025spin,vsmejkal2022giant,liu2024giant}. Moreover, altermagnets also retain a variety of advantages of antiferromagnets (i.e., low stray fields and ultrafast spin dynamics~\cite{baltz2018antiferromagnetic,jungwirth2016antiferromagnetic,jungwirth2018multiple}). Because of the combined advantages of ferromagnets and antiferromagnets, altermagnetic materials offer an innovative and unique platform for the development of future spintronic devices with high performance. Usually, altermagnetic materials are associated with $d$-, $g$-, or $i$-wave nonrelativistic alternating spin splittings (ASSs)~\cite{vsmejkal2022beyond}, where $d$-wave (respectively, $g$-, or $i$-wave) spin splitting can be utilized to generate linear (respectively, nonlinear) spin current~\cite{gonzalez2021efficient,bose2022tilted,bai2022observation,karube2022observation,naka2019spin,naka2021perovskite,ezawa2025third,belashchenko2025giant}. As shown in Ref.~\cite{vsmejkal2022beyond}, the collinear altermagnetic materials with $d$-wave nonrelativistic ASSs can be attributed to 15 collinear spin point groups (SPGs). Generally, the $d$-wave nonrelativistic ASSs between two momentum directions are described by $\lambda_{\alpha\beta}k_\alpha k_\beta \sigma_\chi$~\cite{vsmejkal2022beyond, roig2024minimal}, where $k_\alpha$ and $k_\beta$ are wave vector components along $\alpha$ and $\beta$ directions, and $\sigma_\chi$ is the Pauli matrix component along the $\chi$ direction. From symmetry point of view, $k_\alpha k_\beta$ transforms identically as the $\eta_{\alpha\beta}$ strain~\cite{voon2009kp}. This implies that altermagnets with alternating $\lambda_{\alpha\beta}k_\alpha k_\beta \sigma_\chi$ spin splittings naturally host strain-induced nonrelativistic ZSSs described by $\tilde{\lambda}_{\alpha\beta}\eta_{\alpha\beta} \sigma_\chi$. 
Recent studies indicate that external strain can induce piezomagnetic effects in $d$-wave altermagnets~\cite{ma2021multifunctional,zhu2023multipiezo,wu2024valley,yu2024typical,guo2023piezoelectric,xie2025piezovalley,li2024strain,chen2024strain,xun2025stacking,hu2025valley,guo2023quantum,tian2025spin,fernandes2024topological,hu2025catalog,wu2024valley, chang2025valley}. A sequence of candidate materials (e.g., monolayer V$_2$Se$_2$O~\cite{ma2021multifunctional}, CoF$_2$~\cite{ma2021multifunctional}, V$_2$SeTeO~\cite{zhu2023multipiezo}, and Cr$_2$S$_2$~\cite{chen2024strain}) with such intriguing phenomena has been proposed as well. The physical underpinnings for piezomagnetic effects and strain-induced ZSSs are similar, which highlights the possibility to achieve strain-induced nonrelativistic ZSSs in $d$-wave altermagnets.

Focusing on $d$-wave altermagnetic materials, the present work aims at exploring the nonrelativistic ZSSs that are rooted in mechanical strain. We carry out a systematic symmetry analysis with respect to 58 collinear antiferromagnetic SPGs, and identify 15 SPGs that host strain-induced nonrelativistic ZSSs [coinciding with the aforementioned 15 cases associated with $d$-wave nonrelativistic ASSs (see Ref.~\cite{vsmejkal2022beyond})]. We also derive two-band effective Hamiltonians for these SPGs by considering the terms regarding effective masses, nonrelativistic ASSs, and strain-induced nonrelativistic ZSSs. We further numerically check our analysis regarding strain-induced nonrelativistic ZSSs by working with three typical $d$-wave altermagnetic semiconductors (i.e., $\mathrm{CoF_{2}}$, $\mathrm{LiFe_{2}F_{6}}$ and $\mathrm{La_{2}O_{3}Mn_{2}Se_{2}}$). By first-principles simulations, we find that shear strain $\eta_{xy}$ creates nonrelativistic ZSSs in these three materials. Strikingly, a $\eta_{xy}$ strain of $2\%$ yields nonrelativistic ZSSs as giant as 177 meV in $\mathrm{CoF_{2}}$, and creates sizable nonrelativistic ZSSs of $\sim$ 100 meV in  $\mathrm{LiFe_{2}F_{6}}$ and $\mathrm{La_{2}O_{3}Mn_{2}Se_{2}}$. Furthermore, the band dispersions in $\mathrm{CoF_{2}}$, $\mathrm{LiFe_{2}F_{6}}$ and $\mathrm{La_{2}O_{3}Mn_{2}Se_{2}}$ are well described by our two-band  effective Hamiltonians.

\section{Methods}
We performed first-principles simulations by using the Vienna Ab initio Simulation Package (VASP)~\cite{kresse1996efficient, kresse1999ultrasoft}. We first relaxed the crystal structures of $\mathrm{CoF_{2}}$, $\mathrm{LiFe_{2}F_{6}}$, and $\mathrm{La_{2}O_{3}Mn_{2}Se_{2}}$  under various strains with the force convergence criterion of 5 meV/$\text{Å}$, and then computed the band structures for these systems. During our calculations, we adopted the PBEsol~\cite{perdew1996generalized} exchange-correlation functional and Projector Augmented Wave (PAW) potential~\cite{blochl1994projector}, without involving spin-orbit interaction. We used kinetic cutoff energies of 550 eV for $\mathrm{CoF_{2}}$, 650 eV for $\mathrm{LiFe_{2}F_{6}}$, and 550 eV for $\mathrm{La_{2}O_{3}Mn_{2}Se_{2}}$, and solve the following electronic configurations: 
$3s^{2}3p^{6}3d^{8}4s^{1}$ for $\mathrm{Co}$, $2s^{2}2p^{5}$ for $\mathrm{F}$, $1s^{2}2s^{1}$ for $\mathrm{Li}$, $3s^{2}3p^{6}3d^{7}4s^{1}$ for $\mathrm{Fe}$, $5s^{2}5p^{6}5d^{1}6s^{2}$ for $\mathrm{La}$,  $3s^{2}3p^{6}3d^{6}4s^{1}$ for $\mathrm{Mn}$, $2s^{2}2p^{4}$ for $\mathrm{O}$, and $4s^{2}4p^{4}$ for $\mathrm{Se}$. We considered the correlation effects by adding effective Hubbard $U$ corrections (by the Dudarev approach\cite{dudarev1998electron}) of
3 eV, 4 eV and 4 eV to the $3d$ orbitals of $\mathrm{Co}$, $\mathrm{Fe}$ and $\mathrm{Mn}$, respectively, following Refs.~\cite{wei20252,yu2024typical,lin2017ferroelectric}. We selected the $k$-point meshes of $8\times8\times11$ for $\mathrm{CoF_{2}}$, $8\times8\times4$ for $\mathrm{LiFe_{2}F_{6}}$ and $6\times6\times2$ for $\mathrm{La_{2}O_{3}Mn_{2}Se_{2}}$. We employed Vaspkit~\cite{wang2021vaspkit, vaspkit}, Vesta~\cite{momma2011vesta}, Mathematica~\cite{mathematica}, 
Bilbao Crystallographic Server~\cite{aroyo2011crystallography, aroyo2006bilbao1, aroyo2006bilbao2}, Magndata~\cite{gallego2016magndata1, gallego2016magndata2, magndata},  Pyprocar~\cite{herath2020pyprocar,pyprocar}, The Material Project~\cite{jain2013commentary}, Find Spin Group~\cite{chen2024enumeration}
and Matplotlib~\cite{hunter2007matplotlib} in the present work as well.

\section{Results and discussion}

Our results and discussion contain two subsetions. In Section A, we analyze the symmetries of SPGs and derive the effective Hamiltonians that describe the band dispersions in $d$-wave altermagnets. In Section B, we use first-principles simulations to numerically verify our analysis, by working with $\mathrm{CoF_{2}}$, $\mathrm{LiFe_{2}F_{6}}$ and $\mathrm{La_{2}O_{3}Mn_{2}Se_{2}}$ altermagnetic semiconductors.

\subsection{Effective Hamiltonians}  
 
\begin{figure}[!ht]
    \centering
    \begin{adjustbox}{center}
        \includegraphics[width=10cm]{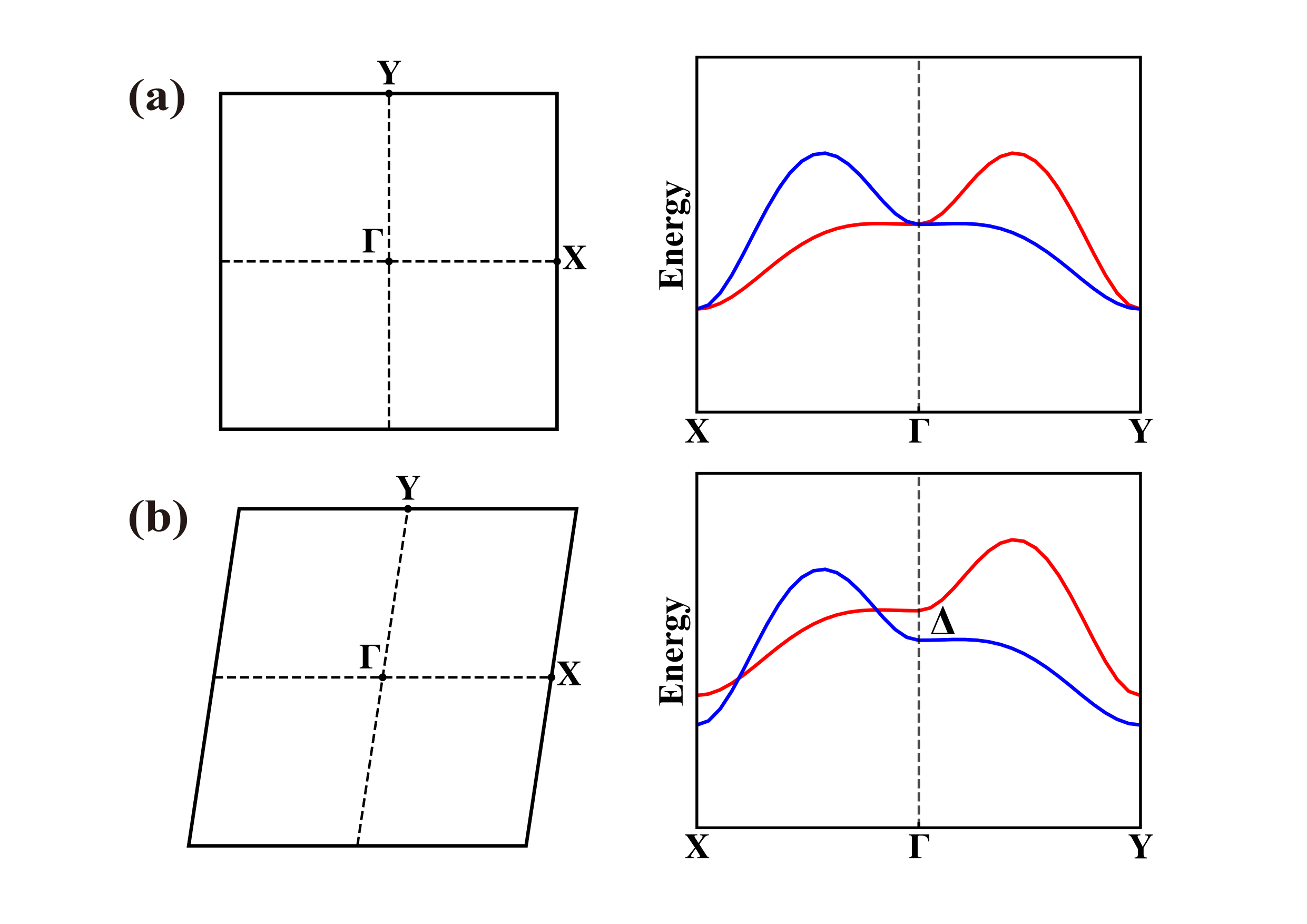}
    \end{adjustbox}
    \vspace{-1cm}
    \caption{Schematic illustrations of the Brillouin zones and band structures for a $d$-wave altermagnet with nonrelativistic ASSs (a) and strain-induced nonrelativistic ZSSs (b). The spin-up and spin-down electronic states are colored in red and blue, respectively.  In panel (b), $\mathbf{\Delta}$ indicates strain-induced nonrelativistic ZSSs at the zone center (i.e., $\Gamma$ point).}
    \label{fig_am_zeeman}
\end{figure}

In $d$-wave altermagnets, the two-band effective Hamiltonian takes the form~\cite{vsmejkal2022beyond, roig2024minimal} 
\begin{equation}\label{eq_hami_unstrain}
\mathcal{H}_{0}(\bm{k})=\sum_{\gamma}\rho_{\gamma}k_{\gamma}^{2}+\sum_{\alpha\beta}\lambda_{\alpha\beta}k_\alpha k_\beta \sigma_\chi,
\end{equation}
where $\bm{k}$ is the wave vector, $k_i$ ($i=\alpha,\beta,\gamma$) its Cartesian components, and $\sigma_\chi$ the Pauli matrix component along the $\chi$ direction. The magnetic moments in the $d$-wave altermagnets are aligned along the $\pm\chi$ directions ($\rho_{\gamma}$ and $\lambda_{\alpha\beta}$ being cofficients). In Eq.~(\ref{eq_hami_unstrain}), the first and second terms represent the effective mass terms and the momentum-dependent nonrelativistic ASS terms [see Fig.~\ref{fig_am_zeeman}(a)], respectively. There are 15 SPGs that host such $d$-wave nonrelativistic ASSs, and this is demonstrated in Table SI of the Supplementary Material in Ref.~\cite{vsmejkal2022beyond}.
Since $k_\alpha k_\beta$ transforms identically as the mechanical strain $\eta_{\alpha\beta}$ under spatial and temporal symmetry operations~\cite{voon2009kp},
Eq.~(\ref{eq_hami_unstrain}) implies another effective Hamiltonian given as
\begin{equation}\label{eq_hami_tot}
\mathcal{H}(\bm{k}, \eta)\!=\!\sum_{\gamma}\rho_{\gamma}k_{\gamma}^{2}\!+\!\sum_{\alpha\beta}\lambda_{\alpha\beta}k_\alpha k_\beta \sigma_\chi\!+\!\sum_{\alpha\beta}\tilde{\lambda}_{\alpha\beta}\eta_{\alpha \beta} \sigma_\chi,
\end{equation} 
where the second-order symmetric strain tensor $\eta_{\alpha\beta}$ contains six independent components (i.e., $\eta_{xx}$, $\eta_{yy}$, $\eta_{zz}$, $\eta_{xy}$, $\eta_{yz}$, and $\eta_{xz}$), and $\tilde{\lambda}_{\alpha\beta}$ denotes constant coefficient. The third term in Eq.~(\ref{eq_hami_tot}) is associated with strain-induced nonrelativistic ZSSs in $d$-wave altermagnets [see Fig.~\ref{fig_am_zeeman}(b)]. The SPGs listed in the aforementioned Table SI~\cite{vsmejkal2022beyond} should host such a phenomenon as well. Notably, we simplify our discussion by omitting the higher-order terms such as $\eta_{\alpha\beta}k_\mu k_\nu$ and $\eta_{\alpha\beta}k_\mu k_\nu \sigma_\chi$ that characterize the strain-dependent effective masses and nonrelativistic ASSs. Such a simplification enables us to capture the essential physics of the strain-induced band dispersions in $d$-wave altermagnets. 
In the following, we extract the SPGs associated with strain-induced nonrelativistic ZSSs, and compare our results with Table SI of Ref.~\cite{vsmejkal2022beyond}. The SPGs contain 90 collinear cases and 58 of them describe the symmetries of collinear antiferromagnets~\cite{litvin1974spin,litvin1977spin,liu2022spin,chen2024enumeration,schiff2025crystallographic}. We focus on the 58 collinear antiferromagnetic SPGs and derive the symmetry-adapted effective Hamiltonians for them.

\begin{table}[b]
    \centering
\caption{\label{tab:trans}Transformation rules for $\eta_{\alpha\beta}$ strain and $\sigma_{\chi}$ Pauli matrix with respect to the nontrivial SPG ($G_{ns}$) for  $^{\bar{1}}4^{\infty m}1$ SPG. Here, the symmetry operations in the spin space (left side of `$||$') include identity operation ($\mathfrak{1}$) and time reversal operation ($\mathcal{T}$), while the symmetry operations in the spatial part (right side of `$||$') contain identity operation ($\mathfrak{1}$), two-fold rotation along $z$-axis ($\mathfrak{2_{001}}$), and rotation of $\pm\frac{\pi}{4}$ along the $z$ axis ($\mathfrak{4^{\pm}_{001}}$).
    }
    \begin{ruledtabular}
    \begin{tabular}{l r r r r r r r}
         & $\eta_{xx}$ & $\eta_{yy}$ & $\eta_{zz}$ & $\eta_{xy}$ & $\eta_{xz}$ & $\eta_{yz}$ & $\sigma_{\chi}$\\
         \colrule
$P_{[\mathfrak{1}||\mathfrak{1}]}$ & $\eta_{xx}$ & $\eta_{yy}$ & $\eta_{zz}$ & $\eta_{xy}$ & $\eta_{xz}$ & $\eta_{yz}$ & $\sigma_{\chi}$\\
$P_{[\mathfrak{1}||\mathfrak{2_{001}}]}$ & $\eta_{xx}$ & $\eta_{yy}$ & $\eta_{zz}$ & $\eta_{xy}$ & $-\eta_{xz}$ & $-\eta_{yz}$ & $\sigma_{\chi}$\\
$P_{[\mathcal{T}||\mathfrak{4^{+}_{001}}]}$ & $\eta_{yy}$ & $\eta_{xx}$ & $\eta_{zz}$ & $-\eta_{xy}$ & $\eta_{yz}$ & $-\eta_{xz}$ & $-\sigma_{\chi}$\\ 
$P_{[\mathcal{T}||\mathfrak{4^{-}_{001}}]}$ 
& $\eta_{yy}$ & $\eta_{xx}$ & $\eta_{zz}$ & $-\eta_{xy}$ & $-\eta_{yz}$ & $\eta_{xz}$ & $-\sigma_{\chi}$\\
    \end{tabular}
    \end{ruledtabular}
\end{table}

\begin{table*}[!ht]
\centering
\renewcommand{\arraystretch}{1.4}
\caption{\label{tab:AM} {
The SPGs that are associated with the $d$-wave altermagnetism and host the strain-induced nonrelativistic ZSSs. The effective Hamiltonians describing the strain-induced nonrelativistic ZSSs, the nonrelativistic ASSs, and the effective mass are shown in the second, third and fourth columns, respectively. The last column lists the candidate $d$-wave altermagnetic materials associated with the corresponding SPGs. The effective Hamiltonians that has been derived in literature are endowed with a reference number after the corresponding formula.}
}
\begin{tabular}{
  >{\centering\arraybackslash}m{2.6cm} | 
  >{\centering\arraybackslash}m{3.6cm} | 
  >{\centering\arraybackslash}m{4.6cm} 
  |>{\centering\arraybackslash}m{2.6cm}|>{\centering\arraybackslash}m{4cm}
}
\toprule
\textbf{SPGs} & \textbf{Nonrelativistic ZSSs} & \textbf{Nonrelativistic ASSs} & \textbf{Effective mass} & \textbf{Candidate Materials}\\
\colrule
$^{\bar{1}}m^{\infty m}1$ &   $(\tilde{\lambda}_{xy}\eta_{xy}\!+\!\tilde{\lambda}_{yz}\eta_{yz})\sigma_{\chi}$ & $(\lambda_{xy}k_{x}k_{y}\!+\!\lambda_{yz}k_{y}k_{z})\sigma_{\chi}$ & $\rho_{x}k_{x}^{2}\!+\!\rho_{y}k_{y}^{2}\!+\!\rho_{z}k_{z}^{2}$ & V$_2$ClBrI$_2$O$_2$\cite{sodequist2024two} \\

$^{\bar{1}}2^{\infty m}1$ &   $(\tilde{\lambda}_{xy}\eta_{xy}\!+\!\tilde{\lambda}_{yz}\eta_{yz})\sigma_{\chi}$ & $(\lambda_{xy}k_{x}k_{y}\!+\!\lambda_{yz}k_{y}k_{z})\sigma_{\chi}$ & $\rho_{x}k_{x}^{2}\!+\!\rho_{y}k_{y}^{2}\!+\!\rho_{z}k_{z}^{2}$ & OsNNaSCl$_5$\cite{sodequist2024two} \\

$^{\bar{1}}2/^{\bar{1}}m^{\infty m}1$ &   $(\tilde{\lambda}_{xy}\eta_{xy}\!+\!\tilde{\lambda}_{yz}\eta_{yz})\sigma_{\chi}$ & $(\lambda_{xy}k_{x}k_{y}\!+\!\lambda_{yz}k_{y}k_{z})\sigma_{\chi}$\cite{vsmejkal2022beyond, roig2024minimal} & $\rho_{x}k_{x}^{2}\!+\!\rho_{y}k_{y}^{2}\!+\!\rho_{z}k_{z}^{2}$ & CuF$_2$\cite{vsmejkal2022beyond}
\\

$^{\bar{1}}2^{\bar{1}}2^{1}2^{\infty m}1$ & $\tilde{\lambda}_{xy}\eta_{xy}\sigma_{\chi}$
&  $\lambda_{xy} k_{x}k_{y}\sigma_{\chi}$ & $\rho_{x}k_{x}^{2}\!+\!\rho_{y}k_{y}^{2}\!+\!\rho_{z}k_{z}^{2}$ & BaCrF$_5$\cite{guo2023spin}
\\

$^{\bar{1}}m^{\bar{1}}m^{1}2^{\infty m}1$  & $\tilde{\lambda}_{xy}\eta_{xy}\sigma_{\chi}$
&  $\lambda_{xy} k_{x}k_{y}\sigma_{\chi}$ & $\rho_{x}k_{x}^{2}\!+\!\rho_{y}k_{y}^{2}\!+\!\rho_{z}k_{z}^{2}$ & Ca$_3$Mn$_2$O$_7$\cite{gu2025ferroelectric}
\\

$^{\bar{1}}m^{1}m^{\bar{1}}2^{\infty m}1$  & $\tilde{\lambda}_{xz}\eta_{xz}\sigma_{\chi}$
&  $\lambda_{xz} k_{x}k_{z}\sigma_{\chi}$ & $\rho_{x}k_{x}^{2}\!+\!\rho_{y}k_{y}^{2}\!+\!\rho_{z}k_{z}^{2}$ & Y$_2$Cu$_2$O$_5$\cite{guo2023spin} 
\\

$^{\bar{1}}m^{\bar{1}}m^{1}m^{\infty m}1$ & $\tilde{\lambda}_{xy}\eta_{xy}\sigma_{\chi}$
& $\lambda_{xy} k_{x}k_{y}\sigma_{\chi}$~\cite{roig2024minimal} & $\rho_{x}k_{x}^{2}\!+\!\rho_{y}k_{y}^{2}\!+\!\rho_{z}k_{z}^{2}$ & LaMnO$_3$\cite{bai2024altermagnetism}
\\

$^{\bar{1}}4^{\infty m}1$ 
& $[\tilde{\lambda}_{xy}\eta_{xy}\!+\!\tilde{\lambda}_{xx}(\eta_{xx}\!-\!\eta_{yy})]\sigma_{\chi}$
&  $[\lambda_{xy} k_{x}k_{y}\!+\!\lambda_{xx}(k_{x}k_{x}\!-\!k_{y}k_{y})]\sigma_{\chi}$ & $\rho_{x}(k_{x}^{2}\!+\!k_{y}^{2})\!+\!\rho_{z}k_{z}^{2}$ & \textbf{---}
\\

$^{\bar{1}}\bar{4}^{\infty m}1$ 
& $[\tilde{\lambda}_{xy}\eta_{xy}\!+\!\tilde{\lambda}_{xx}(\eta_{xx}\!-\!\eta_{yy})]\sigma_{\chi}$
&  $[\lambda_{xy} k_{x}k_{y}\!+\!\lambda_{xx}(k_{x}k_{x}\!-\!k_{y}k_{y})]\sigma_{\chi}$ & $\rho_{x}(k_{x}^{2}\!+\!k_{y}^{2})\!+\!\rho_{z}k_{z}^{2}$ & CsCoF$_4$\cite{bai2024altermagnetism}\\

$^{\bar{1}}4/^{1}m^{\infty m}1$ 
& $[\tilde{\lambda}_{xy}\eta_{xy}\!+\!\tilde{\lambda}_{xx}(\eta_{xx}\!-\!\eta_{yy})]\sigma_{\chi}$
& $[\lambda_{xy} k_{x}k_{y}\!+\!\lambda_{xx}(k_{x}k_{x}\!-\!k_{y}k_{y})]\sigma_{\chi}$\cite{roig2024minimal}
& $\rho_{x}(k_{x}^{2}\!+\!k_{y}^{2})\!+\!\rho_{z}k_{z}^{2}$ & $\alpha$-MnO$_2$\cite{bai2024altermagnetism}
\\

$^{\bar{1}}4^{\bar{1}}2^{1}2^{\infty m}1$ 
& $\tilde{\lambda}_{xy}\eta_{xy}\sigma_{\chi}$ 
&  $\lambda_{xy} k_{x}k_{y}\sigma_{\chi}$ & $\rho_{x}(k_{x}^{2}\!+\!k_{y}^{2})\!+\!\rho_{z}k_{z}^{2}$ & NaFeO$_2$\cite{gao2025ai}
\\

$^{\bar{1}}4^{\bar{1}}m^{1}m^{\infty m}1$ 
& $\tilde{\lambda}_{xy}\eta_{xy}\sigma_{\chi}$ 
&  $\lambda_{xy} k_{x}k_{y}\sigma_{\chi}$
& $\rho_{x}(k_{x}^{2}\!+\!k_{y}^{2})\!+\!\rho_{z}k_{z}^{2}$ & LiFe$_2$F$_6$\cite{guo2023altermagnetic}
\\

$^{\bar{1}}\bar{4}^{\bar{1}}2^{1}m^{\infty m}1$ 
& $\tilde{\lambda}_{xy}\eta_{xy}\sigma_{\chi}$
& $\lambda_{xy} k_{x}k_{y}\sigma_{\chi}$ 
& $\rho_{x}(k_{x}^{2}\!+\!k_{y}^{2})\!+\!\rho_{z}k_{z}^{2}$ & Ca(CoN)$_2$\cite{zhang2024predictable}
\\

$^{\bar{1}}\bar{4}^{\bar{1}}m^{1}2^{\infty m}1$ 
& $\tilde{\lambda}_{xy}\eta_{xy}\sigma_{\chi}$ 
&  $\lambda_{xy} k_{x}k_{y}\sigma_{\chi}$ & $\rho_{x}(k_{x}^{2}\!+\!k_{y}^{2})\!+\!\rho_{z}k_{z}^{2}$ & \textbf{---}\\

$^{\bar{1}}4/^{1}m^{\bar{1}}m^{1}m^{\infty m}1$ &
$\tilde{\lambda}_{xy}\eta_{xy}\sigma_{\chi}$ &
$\lambda_{xy} k_{x}k_{y}\sigma_{\chi}$ \cite{vsmejkal2022beyond,roig2024minimal} &
$\rho_{x}(k_{x}^{2}+k_{y}^{2})+\rho_{z}k_{z}^{2}$ &
\makecell[c]{CoF$_2$\cite{guo2023spin}, MnF$_2$\cite{guo2023spin},\\ La$_2$O$_3$Mn$_2$Se$_2$\cite{wei20252}} \\
\hline
\hline
\end{tabular}
\end{table*}

\begin{figure*}[!ht]
    \centering
    \begin{adjustbox}{center}
        \includegraphics[width=17cm]{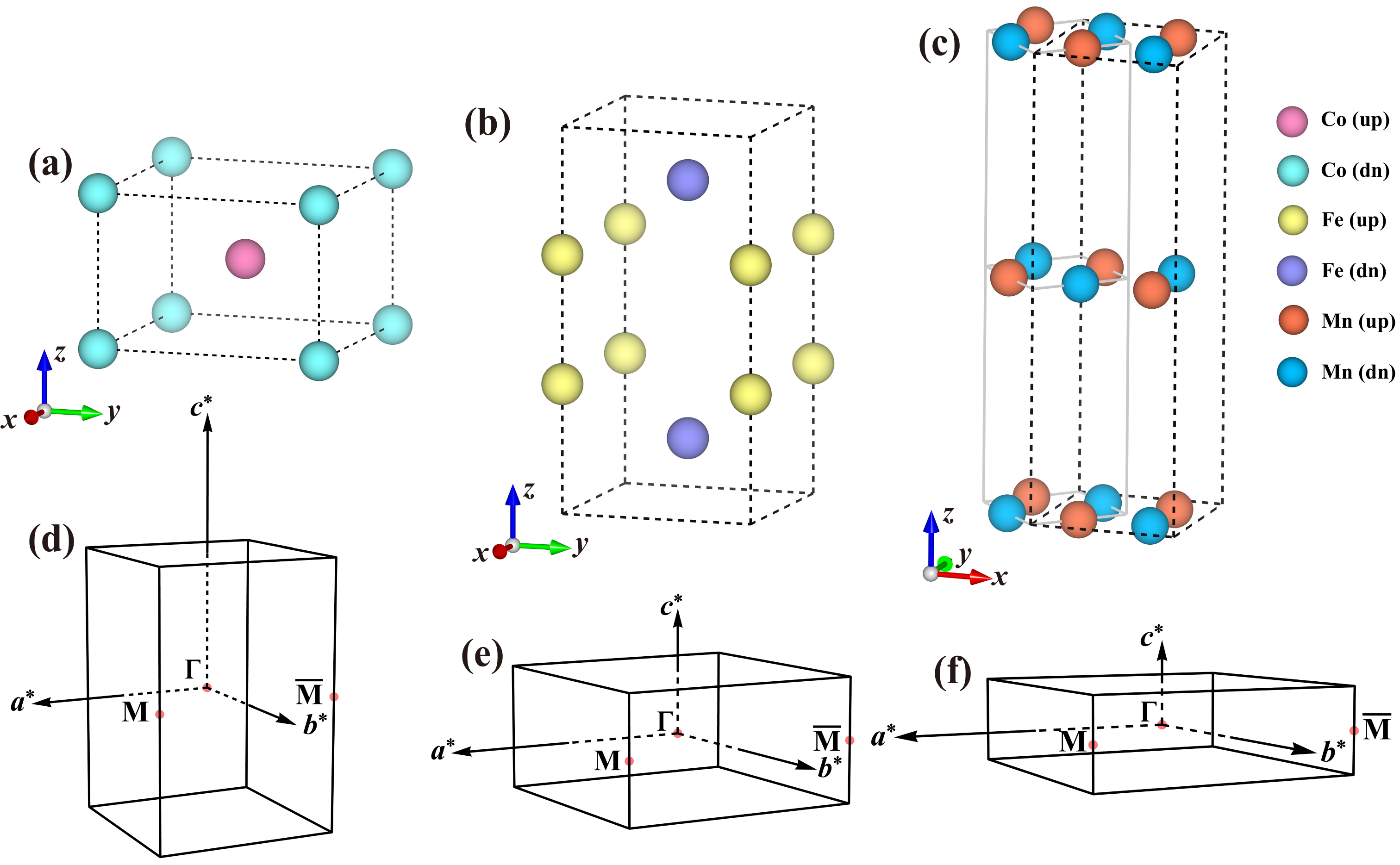}
    \end{adjustbox}
    \caption{Panels (a), (b), and (c) sketch the collinear magnetic structures for $\mathrm{CoF_{2}}$, $\mathrm{LiFe_{2}F_{6}}$, and $\mathrm{La_{2}O_{3}Mn_{2}Se_{2}}$, respectively. The non-magnetic F, Li, La, Se and O ions are not displayed. Panels (d), (e), and (f) sketch the Brillouin zones for  $\mathrm{CoF_{2}}$, $\mathrm{LiFe_{2}F_{6}}$, and $\mathrm{La_{2}O_{3}Mn_{2}Se_{2}}$, respectively. The black dashed boxes in panels (a), (b), and (c) demonstrate the cells that are used in our simulations, while the grey solid box in panel (c) represents the conventional unit cell that is employed in literature (see e.g., Ref.~\cite{wei20252}).}
    \label{fig_crystal}
\end{figure*}

\begin{figure*}[!ht]
    \centering  
    \begin{adjustbox}{center}
\includegraphics[width=18cm]{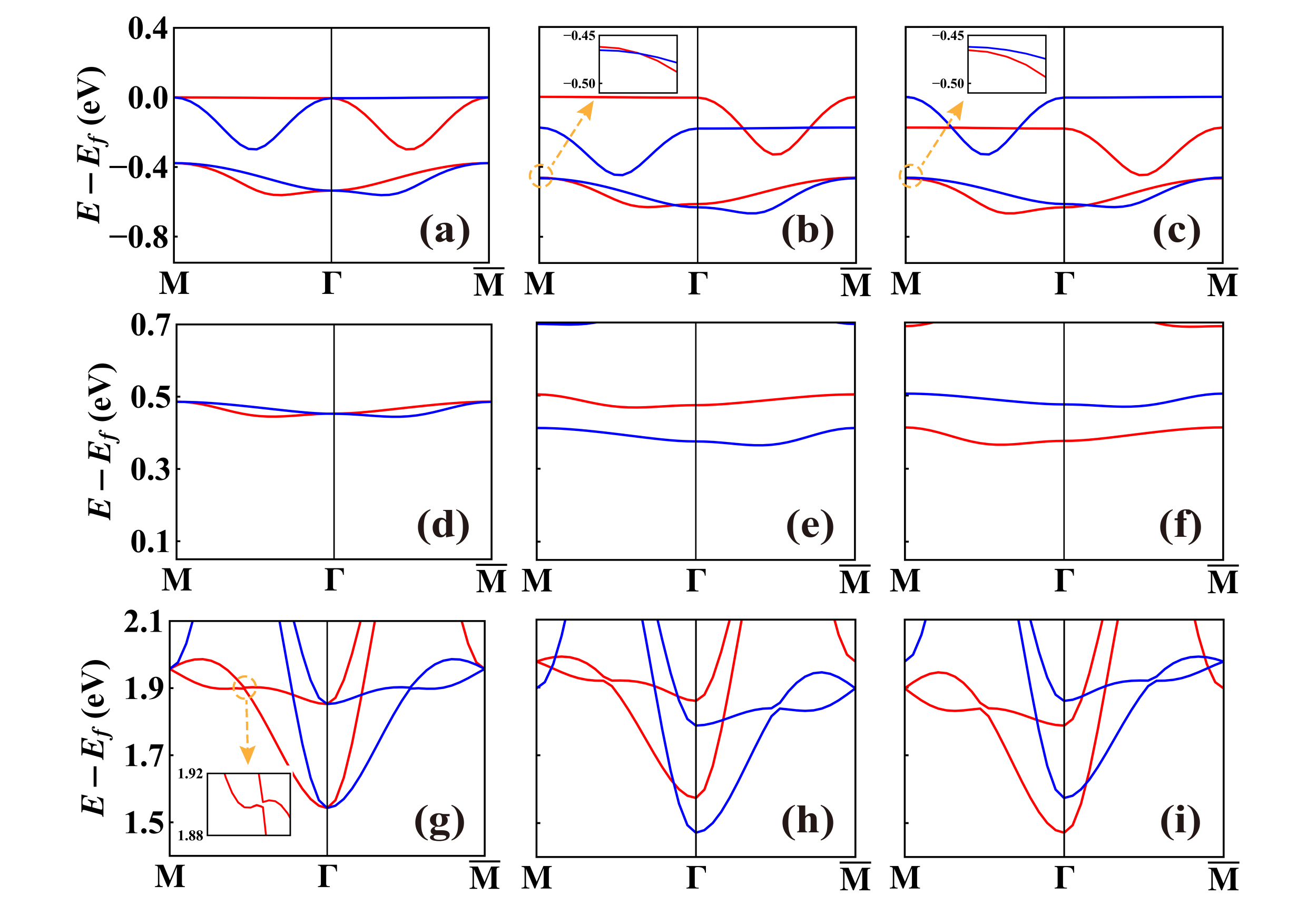} 
    \end{adjustbox}
    \vspace{-0.8cm} 
   \caption{The band structures along $\mathrm{M}-\Gamma-\overline{\mathrm{M}}$ high-symmetry lines for $\mathrm{CoF_2}$ [(a), (b), and (c)], $\mathrm{LiFe_{2}F_{6}}$ [(d), (e), and (f)], and $\mathrm{La_{2}O_{3}Mn_{2}Se_{2}}$ [(g), (h), and (i)]. Panels (a), (d), and (g) are the band structures for the bulk materials without strain; Panels (b), (e), and (h) [respectively, (c), (f), and (i)] are the band structures for the strained materials with shear strain $\eta_{xy}$ of $+2\%$ [respectively, $-2\%$]. Insets in panels (b), (c) and (g) show the zoom-in band structures within the corresponding orange dashed circles. The spin-up and spin-down electronic states are colored in red and blue, respectively. The $E_{f}$ level is set as the top of the valence band.}\label{fig_band}
\end{figure*}

To begin with, we briefly introduce the collinear SPGs following the conventions shown in Refs.~\cite{litvin1974spin,litvin1977spin,liu2022spin,chen2024enumeration,schiff2025crystallographic}. 
According to Refs.~\cite{litvin1974spin,litvin1977spin,liu2022spin,chen2024enumeration,schiff2025crystallographic}, the symmetry operations in SPGs can be written as $[g_{s}||g_{l}]$, with $g_{s}$ and $g_{l}$ being symmetry operations in spin and spatial spaces, respectively. In particular, the $[g_{s}||g_{l}]$ operation transforms $\eta_{\alpha\beta}$ and $\sigma_{\chi}$ as $[g_{s}||g_{l}]:\eta_{\alpha\beta}\rightarrow g_{l}\eta_{\alpha\beta}$ and $[g_{s}||g_{l}]:\sigma_{\chi}\rightarrow g_{s}\sigma_{\chi}$. We recall that the SPGs ($G_{sp}$) are formulated as~\cite{litvin1974spin,litvin1977spin,liu2022spin,chen2024enumeration,schiff2025crystallographic} 
\begin{equation}
    G_{sp}=G_{so}\otimes G_{ns}.
\end{equation}
Here, $G_{so}$ is the so-called spin-only group and $G_{ns}$ is known as the nontrivial spin point group~\cite{litvin1974spin,litvin1977spin,liu2022spin,chen2024enumeration,schiff2025crystallographic}. For collinear magnets, the spin-only group can be expressed as 
\begin{equation}
    G_{so}=\left\{U_{\parallel}(\phi)||\mathfrak{1}\right\}\oplus\left\{\mathcal{T}U_{\parallel}(\phi)U_{\perp}(\pi)||\mathfrak{1}\right\},
\end{equation}
where $U_{\parallel}(\phi)$ ($U_{\perp}(\pi)$) is spin rotation operation along the axis parallel (perpendicular) to $\chi$, $\phi$ and $\pi$ characterize the rotation angle, $\mathcal{T}$ denotes the time-reversal operation, and $\mathfrak{1}$ represents the identity spatial operation. As a matter of fact, the symmetry operations in $G_{so}$ preserve both $\eta_{\alpha\beta}$ and $\sigma_{\chi}$. Therefore, the $G_{so}$ are not considered in our following symmetry analysis.

For each SPG, we shall work with its nontrivial spin point group $G_{ns}$ and generate the symmetry-adapted  
effective Hamiltonian [i.e., Eq.~(\ref{eq_hami_tot})] by the projection operator technique (see e.g., Refs.~\cite{dresselhaus2007group,geilhufe2018gtpack}).
 The effective Hamiltonian must be invariant with respect to the symmetry operations in $G_{ns}$. Therefore, the projection operator is defined according to the identity representation of the $G_{ns}$ group, given by
\begin{equation}\label{eq_projection}
\hat{\Pi} = \frac{1}{n(G_{ns})} \sum_{g\in G_{ns}}\hat{P}_g, 
\end{equation}
with $n(G_{ns})$ being the order of the $G_{ns}$ group and $\hat{P}_g$ being the symmetry operation in $G_{ns}$. To demonstrate our derivations, we take the $^{\bar{1}}4^{\infty m}1$ SPG as an example, and derive the symmetry-allowed $\tilde{\lambda}_{xy}\eta_{\alpha\beta}\sigma_{\chi}$ terms for this group. The $G_{ns}$ group for the $^{\bar{1}}4^{\infty m}1$ SPG contains four symmetry operations, and the effects of these operations on the $\eta_{\alpha\beta}$ strain and the $\sigma_\chi$ are shown in Table~\ref{tab:trans}. Accordingly, we construct the projection operator as
\begin{align*}
\hat{\Pi}=\frac{1}{4}(P_{[\mathfrak{1}||\mathfrak{1}]}+P_{[\mathfrak{1}||\mathfrak{2_{001}}]}
+P_{[\mathcal{T}||\mathfrak{4^{+}_{001}}]}+P_{[\mathcal{T}||\mathfrak{4^{-}_{001}}]}),
\end{align*}
and work with $\eta_{xx}\sigma_{\chi}$, $\eta_{yy}\sigma_{\chi}$, $\eta_{zz}\sigma_{\chi}$, $\eta_{xy}\sigma_{\chi}$, $\eta_{xz}\sigma_{\chi}$, and $\eta_{yz}\sigma_{\chi}$ as follows
\begin{itemize}
\item Case (i):
\begin{align*}
\hat{\Pi}\eta_{xx}\sigma_{\chi}
&=\frac{1}{4}(\eta_{xx}\sigma_{\chi}+\eta_{xx}\sigma_{\chi}-\eta_{yy}\sigma_{\chi}-\eta_{yy}\sigma_{\chi})\\
&=\frac{1}{2}(\eta_{xx}-\eta_{yy})\sigma_{\chi},    
\end{align*}
\item Case (ii): \begin{align*}
\hat{\Pi}\eta_{yy}\sigma_{\chi}
&=\frac{1}{4}(\eta_{yy}\sigma_{\chi}+\eta_{yy}\sigma_{\chi}-\eta_{xx}\sigma_{\chi}-\eta_{xx}\sigma_{\chi})\\
&=\frac{1}{2}(\eta_{yy}-\eta_{xx})\sigma_{\chi},    
\end{align*}
\item Case (iii): 
\begin{align*}
\hat{\Pi}\eta_{zz}\sigma_{\chi}
&=\frac{1}{4}(\eta_{zz}\sigma_{\chi}+\eta_{zz}\sigma_{\chi}-\eta_{zz}\sigma_{\chi}-\eta_{zz}\sigma_{\chi})\\
&=0,    
\end{align*}
\item Case (iv): 
\begin{align*}
\hat{\Pi}\eta_{xy}\sigma_{\chi}
&=\frac{1}{4}(\eta_{xy}\sigma_{\chi}+\eta_{xy}\sigma_{\chi}+\eta_{xy}\sigma_{\chi}+\eta_{xy}\sigma_{\chi})\\
&=\eta_{xy}\sigma_{\chi},  
\end{align*}
\item Case (v): 
\begin{align*}
\hat{\Pi}\eta_{xz}\sigma_{\chi}
&=\frac{1}{4}(\eta_{xz}\sigma_{\chi}-\eta_{xz}\sigma_{\chi}-\eta_{yz}\sigma_{\chi}+\eta_{yz}\sigma_{\chi})\\
&=0,   
\end{align*}
\item Case (vi): 
\begin{align*}
\hat{\Pi}\eta_{yz}\sigma_{\chi}
&=\frac{1}{4}(\eta_{yz}\sigma_{\chi}-\eta_{yz}\sigma_{\chi}+\eta_{xz}\sigma_{\chi}-\eta_{xz}\sigma_{\chi})\\
&=0.    
\end{align*}
\end{itemize}
The aforementioned results indicate that $\eta_{zz}\sigma_{\chi}$, $\eta_{xz}\sigma_{\chi}$, and $\eta_{yz}\sigma_{\chi}$ couplings are not symmetrically allowed in the $^{\bar{1}}4^{\infty m}1$ SPG. In contrast, $\eta_{xx}\sigma_{\chi}$, $\eta_{yy}\sigma_{\chi}$, and $\eta_{xy}\sigma_{\chi}$ are allowed by symmetry. Here, cases (i) and (ii) suggest the $\tilde{\lambda}_{xx}(\eta_{xx}-\eta_{yy})\sigma_{\chi}$ coupling, while case (iv) implies the $\tilde{\lambda}_{xy}\eta_{xy}\sigma_{\chi}$ coupling. Consequently, the terms regarding strain-induced nonrelativistic ZSSs are derived as $\tilde{\lambda}_{xx}(\eta_{xx}-\eta_{yy})\sigma_{\chi}+\tilde{\lambda}_{xy}\eta_{xy}\sigma_{\chi}$ for the $^{\bar{1}}4^{\infty m}1$ SPG.
As shown in Table~\ref{tab:trans}, $\eta_{xx}$, $\eta_{yy}$, and $\eta_{xy}$ strains are not compatible with $[\mathcal{T}||\mathfrak{4^{+}_{001}}]$ and $[\mathcal{T}||\mathfrak{4^{-}_{001}}]$ symmetry operations. Therefore, the application of $\eta_{xx}$, $\eta_{yy}$, or $\eta_{xy}$ strain will break the symmetries of the $^{\bar{1}}4^{\infty m}1$ SPG, driving a transition from $d$-wave altermagnets ($^{\bar{1}}{4}^{\infty m}1$ SPG) to fully compensated ferrimagnets ($^{1}2^{\infty m}1$ SPG). This interprets the strain-induced nonrelativistic ZSSs in materials with $^{\bar{1}}4^{\infty m}1$ SPG.

Similarly, we work with the 58 SPGs for collinear antiferromagnets and identify 15 SPGs that are compatible with strain-induced nonrelativistic ZSSs (see Table~\ref{tab:AM}). As shown in Ref.~\cite{vsmejkal2022beyond}, these 15 SPGs belong to the category of $d$-wave altermagnetism. We also derive the two-band effective Hamiltonians [containing all terms in Eq.~(\ref{eq_hami_tot})] for these SPGs. In Table~\ref{tab:AM}, we summarize the effective Hamiltonians and the corresponding material candidates (extracted from Refs.~\cite{vsmejkal2022beyond, wei20252, gao2025ai,bai2024altermagnetism,gu2025ferroelectric,guo2023altermagnetic, sodequist2024two, guo2023spin,zhang2024predictable}) for these SPGs. Our derived $\lambda_{\alpha\beta}k_{\alpha}k_{\beta}\sigma_{\chi}$, which describes the nonrelativistic ASSs, coincide with those shown in Refs.~\cite{vsmejkal2022beyond,roig2024minimal}. Note that the symbols marking the SPGs in our work are different from those in Ref.~\cite{vsmejkal2022beyond}. 
The symbols in Table~\ref{tab:AM} follow the conventions demonstrated in Refs.~\cite{litvin1974spin,litvin1977spin,liu2022spin,chen2024enumeration,schiff2025crystallographic}.

\begin{figure}[!ht]
    \centering
\centerline{\includegraphics[width=0.95\linewidth]{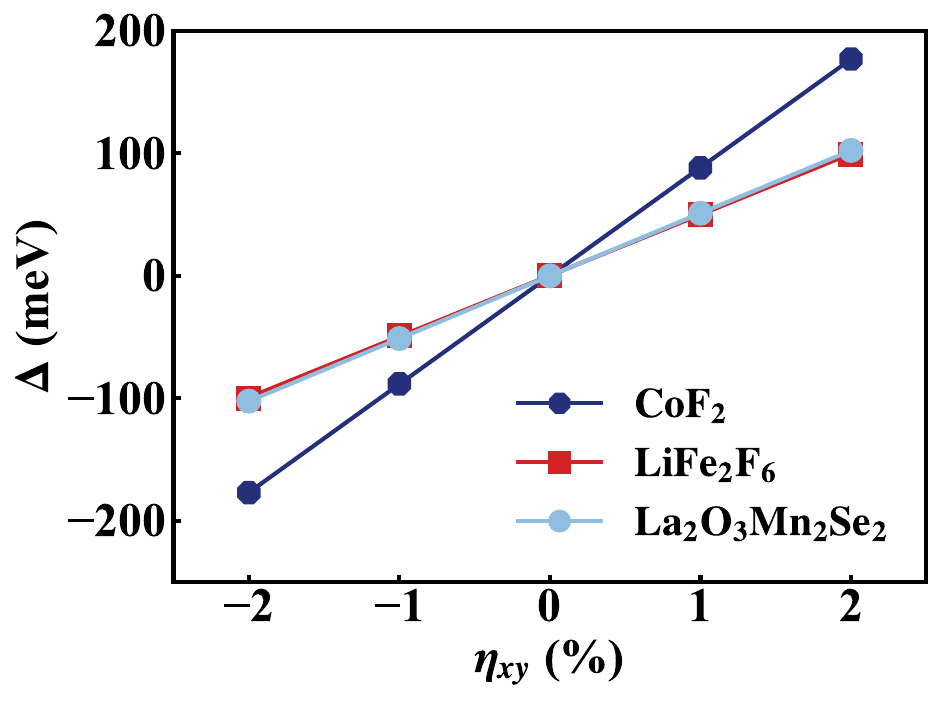}}
    \vspace{-0.2cm}
    \caption{The nonrelativistic ZSSs for $\mathrm{CoF_{2}}$, $\mathrm{LiFe_{2}F_{6}}$, and $\mathrm{La_{2}O_{3}Mn_{2}Se_{2}}$ induced by shear strain $\eta_{xy}$. The nonrelativistic ZSSs are extracted at the $\Gamma$ point for $\mathrm{CoF_{2}}$ (the two highest occupied valence bands), and at the $\Gamma$ point for  $\mathrm{LiFe_{2}F_{6}}$ and $\mathrm{La_{2}O_{3}Mn_{2}Se_{2}}$ (the two lowest unoccupied conduction bands).}
    \label{fig_strain}
\end{figure}

\subsection{$d$-wave altermagnets with strain-induced nonrelativistic ZSSs}

We now verify our aforementioned theory by working with several typical $d$-wave altermagnets. According to Refs.~\cite{guo2023altermagnetic, guo2023spin, wei20252, ma2021multifunctional}, $\mathrm{CoF_{2}}$, $\mathrm{LiFe_{2}F_{6}}$, and $\mathrm{La_{2}O_{3}Mn_{2}Se_{2}}$ are $d$-wave altermagnetic semiconductors that might showcase strain-induced nonrelativistic ZSSs. Experimentally, $\mathrm{CoF_{2}}$ is an antiferromagnet with $P4_{2}/mnm$ crystallographic space group and $P4^{\prime}_{2}/mnm^{\prime}$ magnetic space group~\cite{strempfer2004magnetic}. Figs.~\ref{fig_crystal}(a) and (d) show the collinear magnetic structure and the Brillouin zone for bulk CoF$_2$. By the Find Spin Group code~\cite{chen2024enumeration}, we identify the spin point group of bulk CoF$_2$ as $^{\bar{1}}4/^{1}m^{\bar{1}}m^{1}m^{\infty m}1$. Figs.~\ref{fig_crystal}(c) and (f) sketch the magnetic structure and Brillouin zone for bulk $\mathrm{La_{2}O_{3}Mn_{2}Se_{2}}$. The crystallographic and magnetic space groups for $\mathrm{La_{2}O_{3}Mn_{2}Se_{2}}$ are $I4/mmm$ and $I4^{\prime}/mmm^{\prime}$, respectively~\cite{wei20252}. Bulk $\mathrm{La_{2}O_{3}Mn_{2}Se_{2}}$ has a $^{\bar{1}}4/^{1}m^{\bar{1}}m^{1}m^{\infty m}1$ SPG as well. As for bulk $\mathrm{LiFe_{2}F_{6}}$, early experiment shows that $\mathrm{LiFe_{2}F_{6}}$ has a space group of $P4_{2}/mnm$~\cite{shachar1972neutron}. Recent calculations and experiments suggest that $\mathrm{LiFe_{2}F_{6}}$ is possibly multiferroic with a polar crystallographic space group of $P4_{2}nm$~\cite{fourquet1988trirutile} and a polar magnetic space group of $P4^{\prime}_{2}nm^{\prime}$~\cite{lin2017ferroelectric}. The spin point group of the polar $\mathrm{LiFe_{2}F_{6}}$ is identified (by the Find Spin Group code~\cite{chen2024enumeration}) as $^{\bar{1}}4^{\bar{1}}m^{1}m^{\infty m}1$. The magnetic structure and Brillouin zone for bulk $\mathrm{LiFe_{2}F_{6}}$ are shown in Figs.~\ref{fig_crystal}(b) and (e).

\begin{figure*}[!ht]
    \centering  
    \begin{adjustbox}{center}
\includegraphics[width=16cm]{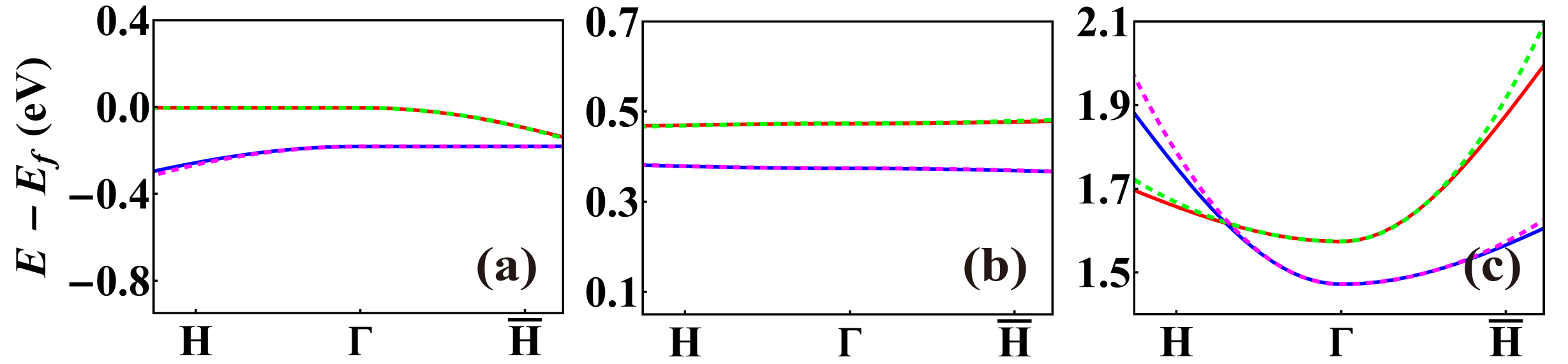} 
    \end{adjustbox}
   \caption{Panels (a), (b) and (c) are the band structures for $\mathrm{CoF_2}$, $\mathrm{LiFe_2F_6}$, and $\mathrm{La_2O_3Mn_2Se_2}$ under +2\% shear strain. The solid and dashed lines indicate the results calculated by first-principles and fitted by Eq.~(\ref{eq_alter_cof}), respectively. The $E_{f}$ level is set as the top of the valence band. The coordinates for the $\mathrm{H}$ and $\overline{\mathrm{H}}$ points, with respect to the corresponding reciprocal lattice vectors, are $(0.1, 0.1, 0)$ and $(-0.1, 0.1, 0)$, respectively. 
  }\label{fig_data_fitting_str}
\end{figure*}

As shown in Table~\ref{tab:AM}, the $^{\bar{1}}4/^{1}m^{\bar{1}}m^{1}m^{\infty m}1$ and $^{\bar{1}}4^{\bar{1}}m^{1}m^{\infty m}1$ SPGs share the same effective Hamiltonian, that is,
\begin{equation}\label{eq_alter_cof}
 \mathcal{H}(\bm{k},\eta)\!=\!\varepsilon_0\!+\!\rho_{x}(k_{x}^{2}\!+\!k_{y}^{2})\!+\!\rho_{z}k_{z}^{2}\!+\!\lambda_{xy}k_{x}k_{y}\sigma_{\chi}\!+\!\tilde{\lambda}_{xy}\eta_{xy}\sigma_{\chi}.  
\end{equation}
Note that $\varepsilon_0$ is introduced in Eq.~(\ref{eq_alter_cof}) to characterize the energy correction relative to the Fermi level of the studied system.
The $\tilde{\lambda}_{xy}\eta_{xy}\sigma_{\chi}$ term in Eq.~(\ref{eq_alter_cof}) implies the $\eta_{xy}$-induced nonrelativistic ZSSs in $\mathrm{CoF_{2}}$, $\mathrm{LiFe_{2}F_{6}}$, and $\mathrm{La_{2}O_{3}Mn_{2}Se_{2}}$. In Figs.~\ref{fig_band}(a), (b), and (c), we show the band structures (valence bands) for bulk CoF$_2$ and $\eta_{xy}$-strained CoF$_2$. Without shear strain $\eta_{xy}$, the spin-up and spin-down energy levels at the $\Gamma$ point are degenerate. The shear strain of $\pm2\%$ creates nonrelativistic ZSSs and modifies the band structures near the $\Gamma$ point, where switching $\eta_{xy}$ between $+2\%$ and $-2\%$ switches the spin feature of the electronic bands in CoF$_2$. Figs.~\ref{fig_band}(d)--(i) show the band structures for $\mathrm{LiFe_{2}F_{6}}$ and $\mathrm{La_{2}O_{3}Mn_{2}Se_{2}}$. Their energy bands around the $\Gamma$ point show similar features as those for CoF$_2$. In Fig.~\ref{fig_strain}, we show the nonrelativistic ZSSs at the $\Gamma$ point (denoted by $\Delta$) for $\mathrm{CoF_{2}}$, $\mathrm{LiFe_{2}F_{6}}$ and $\mathrm{La_{2}O_{3}Mn_{2}Se_{2}}$ as functions of shear strain $\eta_{xy}$. As predicted by our effective Hamiltonian [see Eq.~(\ref{eq_alter_cof})], the $\Delta$ splitting magnitude depends on $\eta_{xy}$ linearly. More striking is that a shear strain of $\pm 2\%$ can induce giant nonrelativistic ZSSs as large as 177 meV, 100 meV, 102meV for $\mathrm{CoF_{2}}$, $\mathrm{LiFe_{2}F_{6}}$ and $\mathrm{La_{2}O_{3}Mn_{2}Se_{2}}$. To conclude this paragraph, our first-principles band structure calculations confirm our aforementioned analyses regarding strain-induced nonrelativistic ZSSs.

\begin{table}[h]   
    \centering
    \renewcommand{\arraystretch}{1.3}
    \caption{\label{tab:paras}
    {The fitted $\rho_{x}$, $\lambda_{xy}$, and $\tilde{\lambda}_{xy}$ parameters [see Eq.~(\ref{eq_alter_cof})] for $\mathrm{CoF_{2}}$, $\mathrm{LiFe_{2}F_{6}}$ and $\mathrm{La_{2}O_{3}Mn_{2}Se_{2}}$ under a shear strain of $2\%$. The energy bands for $\mathrm{CoF_{2}}$ (respectively, $\mathrm{LiFe_{2}F_{6}}$ and $\mathrm{La_{2}O_{3}Mn_{2}Se_{2}}$) locate in the vicinity of the two highest occupied valence bands (respectively, the two lowest unoccupied conduction bands).} }
    \begin{ruledtabular}
    \begin{tabular}{l c c c c c}  
\textbf{Materials} & $\varepsilon_0$ (eV) & $\rho_{x}$ (eV \AA$^{2}$) & $\lambda_{xy}$ (eV \AA$^{2}$) & $\tilde{\lambda}_{xy}$ (eV) \\
\hline 
CoF$_2$ & -0.093 & -1.205  & 2.403 & 4.424 \\
LiFe$_2$F$_6$ & 0.424 & 0.005 & -0.257 & 2.485  \\
La$_2$O$_3$Mn$_2$Se$_2$  & 1.523 & 8.975  & -9.742  & 2.559
    \end{tabular}
    \end{ruledtabular}
\end{table}

To further validate our effective Hamiltonian, we fit the first-principles band structures for $\mathrm{CoF_{2}}$, $\mathrm{LiFe_{2}F_{6}}$, and $\mathrm{La_{2}O_{3}Mn_{2}Se_{2}}$ (around their zone center) with respect to Eq.~(\ref{eq_alter_cof}). We focus on the band structures of these materials under a shear strain of $+2\%$. As shown in Fig.~\ref{fig_data_fitting_str}, our effective Hamiltonian correctly reproduce the band structures of these materials around their zone centers. The fitted parameters are summarized in Table~\ref{tab:paras}.

\section{Summary and outlook}
Based on symmetry arguments, we propose that $d$-wave altermagnets with $\lambda_{\alpha\beta}k_\alpha k_\beta \sigma_\chi$ momentum-dependent nonrelativistic ASSs should host strain-induced nonrelativistic ZSSs (described by $\tilde{\lambda}_{\alpha\beta}\eta_{\alpha\beta} \sigma_\chi$) as well. We demonstrate that 15 collinear SPGs host strain-induced nonrelativistic ZSSs, where these SPGs coincide with the 15 cases hosting $d$-wave nonrelativistic ASSs (see Ref.~\cite{vsmejkal2022beyond}). The two-band effective Hamiltonians involving effective mass terms, nonrelativistic ASS terms, and strain-induced nonrelativistic ZSS terms are derived for these SPGs (see Table~\ref{tab:AM}). By first-principles simulations, we show that $\eta_{xy}$ shear strain of $2\%$ creates sizable ZSSs up to 177, 100, and 102 meV~\footnote{By coincidence, the strain-induced nonrelativistic ZSSs in $\mathrm{LiFe_{2}F_{6}}$ and $\mathrm{La_{2}O_{3}Mn_{2}Se_{2}}$ are very close to each other.} in $\mathrm{CoF_{2}}$, $\mathrm{LiFe_{2}F_{6}}$ and $\mathrm{La_{2}O_{3}Mn_{2}Se_{2}}$ $d$-wave altermagnetic semiconductors, respectively. 
Compared with altermagnetic spin splittings, ZSSs offer another route towards the generation of spin current in antiferromagnets~\cite{ralph2008spin,maekawa2013spin, brataas2012current, maekawa2017spin}. 
In experiments, shear strains can be applied to a material by epitaxially growing the material on lattice-mismatched substrates or by inducing mechanical buckling to the material via flexible substrates~\cite{pandey2023straining, miao2021straintronics}. Furthermore, angle-resolved photoemission spectroscopy~\cite{zhang2022angle} can be employed to measure the intrinsic nonrelativistic ASSs and strained-induced nonrelativistic ZSSs in $d$-wave altermagnets. Several examples for the measurements of the ASSs in altermagnets can be found in Refs.~\cite{krempasky2024altermagnetic,lee2024broken, reimers2024direct,
hajlaoui2024temperature, zeng2024observation,yang2025three,ding2024large,fedchenko2024observation, song2025altermagnets}. As an outlook, our work implies that strain-induced nonrelativistic ZSSs in $d$-wave altermagnets can be utilized to design altermagnetic-based spintronic devices.

\section{Note added}
Recently, we are aware of a work on strain tunable magnetization in altermagnets~\cite{khodas2025tuning}. A comprehensive derivation of piezomagnetic energetic invariants in the nonrelativistic and relativistic regimes is provided by Ref.~\cite{khodas2025tuning}, where the nonrelativistic piezomagnetic terms are closely related to our discussed strain-induced nonrelativistic ZSSs. In particular, the $d$-wave piezomagnetic terms in Table I of Ref.~\cite{khodas2025tuning}  coincide with our Table~\ref{tab:AM}. Furthermore, Ref.~\cite{khodas2025tuning} discusses the shear strain induced band splittings at the zone center in  fluoride altermagnets (e.g., $^{\bar{1}}4/^{1}m^{\bar{1}}m^{1}m^{\infty m}1$ MnF$_2$). The results are consistent with our analysis (see Table~\ref{tab:AM}).\\

\section{Acknowledgement} 
This work is supported by the National Key Research and Development Program of China (Grants No.~2023YFA1406103 and No.~2024YFA1409904) and the National Natural Science Foundation of China (Grants No.~12374005 and No.~12274174). Y. Z. thanks the support from high-performance computing center of Jilin University. H. J. Z. thanks the support from ``Xiaomi YoungScholar'' Project.


\begin{thebibliography}{115}%
\makeatletter
\providecommand \@ifxundefined [1]{%
 \@ifx{#1\undefined}
}%
\providecommand \@ifnum [1]{%
 \ifnum #1\expandafter \@firstoftwo
 \else \expandafter \@secondoftwo
 \fi
}%
\providecommand \@ifx [1]{%
 \ifx #1\expandafter \@firstoftwo
 \else \expandafter \@secondoftwo
 \fi
}%
\providecommand \natexlab [1]{#1}%
\providecommand \enquote  [1]{``#1''}%
\providecommand \bibnamefont  [1]{#1}%
\providecommand \bibfnamefont [1]{#1}%
\providecommand \citenamefont [1]{#1}%
\providecommand \href@noop [0]{\@secondoftwo}%
\providecommand \href [0]{\begingroup \@sanitize@url \@href}%
\providecommand \@href[1]{\@@startlink{#1}\@@href}%
\providecommand \@@href[1]{\endgroup#1\@@endlink}%
\providecommand \@sanitize@url [0]{\catcode `\\12\catcode `\$12\catcode `\&12\catcode `\#12\catcode `\^12\catcode `\_12\catcode `\%12\relax}%
\providecommand \@@startlink[1]{}%
\providecommand \@@endlink[0]{}%
\providecommand \url  [0]{\begingroup\@sanitize@url \@url }%
\providecommand \@url [1]{\endgroup\@href {#1}{\urlprefix }}%
\providecommand \urlprefix  [0]{URL }%
\providecommand \Eprint [0]{\href }%
\providecommand \doibase [0]{https://doi.org/}%
\providecommand \selectlanguage [0]{\@gobble}%
\providecommand \bibinfo  [0]{\@secondoftwo}%
\providecommand \bibfield  [0]{\@secondoftwo}%
\providecommand \translation [1]{[#1]}%
\providecommand \BibitemOpen [0]{}%
\providecommand \bibitemStop [0]{}%
\providecommand \bibitemNoStop [0]{.\EOS\space}%
\providecommand \EOS [0]{\spacefactor3000\relax}%
\providecommand \BibitemShut  [1]{\csname bibitem#1\endcsname}%
\let\auto@bib@innerbib\@empty
\bibitem [{\citenamefont {Wolf}\ \emph {et~al.}(2001)\citenamefont {Wolf}, \citenamefont {Awschalom}, \citenamefont {Buhrman}, \citenamefont {Daughton}, \citenamefont {von Moln{\'a}r}, \citenamefont {Roukes}, \citenamefont {Chtchelkanova},\ and\ \citenamefont {Treger}}]{wolf2001spintronics}%
  \BibitemOpen
  \bibfield  {author} {\bibinfo {author} {\bibfnamefont {S.}~\bibnamefont {Wolf}}, \bibinfo {author} {\bibfnamefont {D.}~\bibnamefont {Awschalom}}, \bibinfo {author} {\bibfnamefont {R.}~\bibnamefont {Buhrman}}, \bibinfo {author} {\bibfnamefont {J.}~\bibnamefont {Daughton}}, \bibinfo {author} {\bibfnamefont {v.~S.}\ \bibnamefont {von Moln{\'a}r}}, \bibinfo {author} {\bibfnamefont {M.}~\bibnamefont {Roukes}}, \bibinfo {author} {\bibfnamefont {A.~Y.}\ \bibnamefont {Chtchelkanova}},\ and\ \bibinfo {author} {\bibfnamefont {D.}~\bibnamefont {Treger}},\ }\bibfield  {title} {\bibinfo {title} {Spintronics: a spin-based electronics vision for the future},\ }\href {https://www.science.org/doi/abs/10.1126/science.1065389} {\bibfield  {journal} {\bibinfo  {journal} {Science}\ }\textbf {\bibinfo {volume} {294}},\ \bibinfo {pages} {1488} (\bibinfo {year} {2001})}\BibitemShut {NoStop}%
\bibitem [{\citenamefont {{\v{Z}}uti{\'c}}\ \emph {et~al.}(2004)\citenamefont {{\v{Z}}uti{\'c}}, \citenamefont {Fabian},\ and\ \citenamefont {Sarma}}]{vzutic2004spintronics}%
  \BibitemOpen
  \bibfield  {author} {\bibinfo {author} {\bibfnamefont {I.}~\bibnamefont {{\v{Z}}uti{\'c}}}, \bibinfo {author} {\bibfnamefont {J.}~\bibnamefont {Fabian}},\ and\ \bibinfo {author} {\bibfnamefont {S.~D.}\ \bibnamefont {Sarma}},\ }\bibfield  {title} {\bibinfo {title} {Spintronics: Fundamentals and applications},\ }\href {https://journals.aps.org/rmp/abstract/10.1103/RevModPhys.76.323} {\bibfield  {journal} {\bibinfo  {journal} {Rev. Mod. Phys.}\ }\textbf {\bibinfo {volume} {76}},\ \bibinfo {pages} {323} (\bibinfo {year} {2004})}\BibitemShut {NoStop}%
\bibitem [{\citenamefont {Sharma}(2005)}]{sharma2005create}%
  \BibitemOpen
  \bibfield  {author} {\bibinfo {author} {\bibfnamefont {P.}~\bibnamefont {Sharma}},\ }\bibfield  {title} {\bibinfo {title} {How to create a spin current},\ }\href {https://www.science.org/doi/full/10.1126/science.1099388} {\bibfield  {journal} {\bibinfo  {journal} {Science}\ }\textbf {\bibinfo {volume} {307}},\ \bibinfo {pages} {531} (\bibinfo {year} {2005})}\BibitemShut {NoStop}%
\bibitem [{\citenamefont {Maekawa}\ \emph {et~al.}(2017)\citenamefont {Maekawa}, \citenamefont {Valenzuela},\ and\ \citenamefont {Saitoh}}]{maekawa2017spin}%
  \BibitemOpen
  \bibfield  {author} {\bibinfo {author} {\bibfnamefont {S.}~\bibnamefont {Maekawa}}, \bibinfo {author} {\bibfnamefont {S.~O.}\ \bibnamefont {Valenzuela}},\ and\ \bibinfo {author} {\bibfnamefont {E.}~\bibnamefont {Saitoh}},\ }\href {https://books.google.co.jp/books?hl=zh-CN&lr=&id=iZ83DwAAQBAJ&oi=fnd&pg=PP1&dq=Book+Review:+Spin+Current&ots=8Um15ESrHb&sig=pUQ1PSK-koQehVuJblf7x_I4HDo&redir_esc=y#v=onepage&q&f=false} {\emph {\bibinfo {title} {Spin current}}}\ (\bibinfo  {publisher} {Oxford University Press},\ \bibinfo {year} {2017})\BibitemShut {NoStop}%
\bibitem [{\citenamefont {Liu}\ \emph {et~al.}(2023)\citenamefont {Liu}, \citenamefont {Zhao}, \citenamefont {Bellaiche},\ and\ \citenamefont {Ma}}]{liu2023zeeman}%
  \BibitemOpen
  \bibfield  {author} {\bibinfo {author} {\bibfnamefont {X.}~\bibnamefont {Liu}}, \bibinfo {author} {\bibfnamefont {H.~J.}\ \bibnamefont {Zhao}}, \bibinfo {author} {\bibfnamefont {L.}~\bibnamefont {Bellaiche}},\ and\ \bibinfo {author} {\bibfnamefont {Y.}~\bibnamefont {Ma}},\ }\bibfield  {title} {\bibinfo {title} {Zeeman spin splittings with persistent spin textures in antiferromagnetic semiconductors},\ }\href {https://journals.aps.org/prb/abstract/10.1103/PhysRevB.108.125108} {\bibfield  {journal} {\bibinfo  {journal} {Phys. Rev. B}\ }\textbf {\bibinfo {volume} {108}},\ \bibinfo {pages} {125108} (\bibinfo {year} {2023})}\BibitemShut {NoStop}%
\bibitem [{\citenamefont {Milivojevi{\'c}}\ \emph {et~al.}(2024)\citenamefont {Milivojevi{\'c}}, \citenamefont {Orozovi{\'c}}, \citenamefont {Picozzi}, \citenamefont {Gmitra},\ and\ \citenamefont {Stavri{\'c}}}]{milivojevic2024interplay}%
  \BibitemOpen
  \bibfield  {author} {\bibinfo {author} {\bibfnamefont {M.}~\bibnamefont {Milivojevi{\'c}}}, \bibinfo {author} {\bibfnamefont {M.}~\bibnamefont {Orozovi{\'c}}}, \bibinfo {author} {\bibfnamefont {S.}~\bibnamefont {Picozzi}}, \bibinfo {author} {\bibfnamefont {M.}~\bibnamefont {Gmitra}},\ and\ \bibinfo {author} {\bibfnamefont {S.}~\bibnamefont {Stavri{\'c}}},\ }\bibfield  {title} {\bibinfo {title} {Interplay of altermagnetism and weak ferromagnetism in two-dimensional $\mathrm{RuF}_{4}$},\ }\href {https://iopscience.iop.org/article/10.1088/2053-1583/ad4c73/meta} {\bibfield  {journal} {\bibinfo  {journal} {2D Mater.}\ }\textbf {\bibinfo {volume} {11}},\ \bibinfo {pages} {035025} (\bibinfo {year} {2024})}\BibitemShut {NoStop}%
\bibitem [{\citenamefont {Liu}\ \emph {et~al.}(2024{\natexlab{a}})\citenamefont {Liu}, \citenamefont {Li}, \citenamefont {Liu},\ and\ \citenamefont {Liu}}]{liu2024unconventional}%
  \BibitemOpen
  \bibfield  {author} {\bibinfo {author} {\bibfnamefont {Y.}~\bibnamefont {Liu}}, \bibinfo {author} {\bibfnamefont {J.}~\bibnamefont {Li}}, \bibinfo {author} {\bibfnamefont {P.}~\bibnamefont {Liu}},\ and\ \bibinfo {author} {\bibfnamefont {Q.}~\bibnamefont {Liu}},\ }\bibfield  {title} {\bibinfo {title} {Unconventional spin textures emerging from a universal symmetry theory of spin-momentum locking},\ }\href {https://www.nature.com/articles/s41535-024-00682-y} {\bibfield  {journal} {\bibinfo  {journal} {Npj Quantum Mater.}\ }\textbf {\bibinfo {volume} {9}},\ \bibinfo {pages} {69} (\bibinfo {year} {2024}{\natexlab{a}})}\BibitemShut {NoStop}%
\bibitem [{\citenamefont {Zhao}\ \emph {et~al.}(2022)\citenamefont {Zhao}, \citenamefont {Liu}, \citenamefont {Wang}, \citenamefont {Yang}, \citenamefont {Bellaiche},\ and\ \citenamefont {Ma}}]{zhao2022zeeman}%
  \BibitemOpen
  \bibfield  {author} {\bibinfo {author} {\bibfnamefont {H.~J.}\ \bibnamefont {Zhao}}, \bibinfo {author} {\bibfnamefont {X.}~\bibnamefont {Liu}}, \bibinfo {author} {\bibfnamefont {Y.}~\bibnamefont {Wang}}, \bibinfo {author} {\bibfnamefont {Y.}~\bibnamefont {Yang}}, \bibinfo {author} {\bibfnamefont {L.}~\bibnamefont {Bellaiche}},\ and\ \bibinfo {author} {\bibfnamefont {Y.}~\bibnamefont {Ma}},\ }\bibfield  {title} {\bibinfo {title} {Zeeman effect in centrosymmetric antiferromagnetic semiconductors controlled by an electric field},\ }\href {https://journals.aps.org/prl/abstract/10.1103/PhysRevLett.129.187602} {\bibfield  {journal} {\bibinfo  {journal} {Phys. Rev. Lett.}\ }\textbf {\bibinfo {volume} {129}},\ \bibinfo {pages} {187602} (\bibinfo {year} {2022})}\BibitemShut {NoStop}%
\bibitem [{\citenamefont {Tao}\ \emph {et~al.}(2024)\citenamefont {Tao}, \citenamefont {Zhang}, \citenamefont {Li}, \citenamefont {Zhao}, \citenamefont {Wang}, \citenamefont {Song}, \citenamefont {Tsymbal},\ and\ \citenamefont {Bellaiche}}]{tao2024layer}%
  \BibitemOpen
  \bibfield  {author} {\bibinfo {author} {\bibfnamefont {L.}~\bibnamefont {Tao}}, \bibinfo {author} {\bibfnamefont {Q.}~\bibnamefont {Zhang}}, \bibinfo {author} {\bibfnamefont {H.}~\bibnamefont {Li}}, \bibinfo {author} {\bibfnamefont {H.~J.}\ \bibnamefont {Zhao}}, \bibinfo {author} {\bibfnamefont {X.}~\bibnamefont {Wang}}, \bibinfo {author} {\bibfnamefont {B.}~\bibnamefont {Song}}, \bibinfo {author} {\bibfnamefont {E.~Y.}\ \bibnamefont {Tsymbal}},\ and\ \bibinfo {author} {\bibfnamefont {L.}~\bibnamefont {Bellaiche}},\ }\bibfield  {title} {\bibinfo {title} {Layer hall detection of the n{\'e}el vector in centrosymmetric magnetoelectric antiferromagnets},\ }\href {https://journals.aps.org/prl/abstract/10.1103/PhysRevLett.133.096803} {\bibfield  {journal} {\bibinfo  {journal} {Phys. Rev. Lett.}\ }\textbf {\bibinfo {volume} {133}},\ \bibinfo {pages} {096803} (\bibinfo {year} {2024})}\BibitemShut {NoStop}%
\bibitem [{\citenamefont {Lou}\ \emph {et~al.}(2020)\citenamefont {Lou}, \citenamefont {Gu}, \citenamefont {Ji}, \citenamefont {Feng}, \citenamefont {Xiang},\ and\ \citenamefont {Stroppa}}]{lou2020tunable}%
  \BibitemOpen
  \bibfield  {author} {\bibinfo {author} {\bibfnamefont {F.}~\bibnamefont {Lou}}, \bibinfo {author} {\bibfnamefont {T.}~\bibnamefont {Gu}}, \bibinfo {author} {\bibfnamefont {J.}~\bibnamefont {Ji}}, \bibinfo {author} {\bibfnamefont {J.}~\bibnamefont {Feng}}, \bibinfo {author} {\bibfnamefont {H.}~\bibnamefont {Xiang}},\ and\ \bibinfo {author} {\bibfnamefont {A.}~\bibnamefont {Stroppa}},\ }\bibfield  {title} {\bibinfo {title} {Tunable spin textures in polar antiferromagnetic hybrid organic--inorganic perovskites by electric and magnetic fields},\ }\href {https://www.nature.com/articles/s41524-020-00374-8} {\bibfield  {journal} {\bibinfo  {journal} {Npj Comput. Mater.}\ }\textbf {\bibinfo {volume} {6}},\ \bibinfo {pages} {114} (\bibinfo {year} {2020})}\BibitemShut {NoStop}%
\bibitem [{\citenamefont {Maekawa}\ \emph {et~al.}(2013)\citenamefont {Maekawa}, \citenamefont {Adachi}, \citenamefont {Uchida}, \citenamefont {Ieda},\ and\ \citenamefont {Saitoh}}]{maekawa2013spin}%
  \BibitemOpen
  \bibfield  {author} {\bibinfo {author} {\bibfnamefont {S.}~\bibnamefont {Maekawa}}, \bibinfo {author} {\bibfnamefont {H.}~\bibnamefont {Adachi}}, \bibinfo {author} {\bibfnamefont {K.}~\bibnamefont {Uchida}}, \bibinfo {author} {\bibfnamefont {J.}~\bibnamefont {Ieda}},\ and\ \bibinfo {author} {\bibfnamefont {E.}~\bibnamefont {Saitoh}},\ }\bibfield  {title} {\bibinfo {title} {Spin current: Experimental and theoretical aspects},\ }\href {https://journals.jps.jp/doi/abs/10.7566/JPSJ.82.102002} {\bibfield  {journal} {\bibinfo  {journal} {J. Phys. Soc. Jpn.}\ }\textbf {\bibinfo {volume} {82}},\ \bibinfo {pages} {102002} (\bibinfo {year} {2013})}\BibitemShut {NoStop}%
\bibitem [{\citenamefont {Sch{\"a}pers}(2021)}]{schapers2021semiconductor}%
  \BibitemOpen
  \bibfield  {author} {\bibinfo {author} {\bibfnamefont {T.}~\bibnamefont {Sch{\"a}pers}},\ }\href {https://xs.gupiaoq.com/scholar?q=Thomas+Sch%C3%A4pers+Semiconductor+Spintronics+De+Gruyter+Graduate} {\emph {\bibinfo {title} {Semiconductor spintronics}}}\ (\bibinfo  {publisher} {Walter de Gruyter GmbH \& Co KG},\ \bibinfo {year} {2021})\BibitemShut {NoStop}%
\bibitem [{\citenamefont {Hayami}\ \emph {et~al.}(2019)\citenamefont {Hayami}, \citenamefont {Yanagi},\ and\ \citenamefont {Kusunose}}]{hayami2019momentum}%
  \BibitemOpen
  \bibfield  {author} {\bibinfo {author} {\bibfnamefont {S.}~\bibnamefont {Hayami}}, \bibinfo {author} {\bibfnamefont {Y.}~\bibnamefont {Yanagi}},\ and\ \bibinfo {author} {\bibfnamefont {H.}~\bibnamefont {Kusunose}},\ }\bibfield  {title} {\bibinfo {title} {Momentum-dependent spin splitting by collinear antiferromagnetic ordering},\ }\href {https://journals.jps.jp/doi/10.7566/JPSJ.88.123702} {\bibfield  {journal} {\bibinfo  {journal} {J. Phys. Soc. Jpn.}\ }\textbf {\bibinfo {volume} {88}},\ \bibinfo {pages} {123702} (\bibinfo {year} {2019})}\BibitemShut {NoStop}%
\bibitem [{\citenamefont {Krempask{\`y}}\ \emph {et~al.}(2024)\citenamefont {Krempask{\`y}}, \citenamefont {{\v{S}}mejkal}, \citenamefont {D’souza}, \citenamefont {Hajlaoui}, \citenamefont {Springholz}, \citenamefont {Uhl{\'\i}{\v{r}}ov{\'a}}, \citenamefont {Alarab}, \citenamefont {Constantinou}, \citenamefont {Strocov}, \citenamefont {Usanov} \emph {et~al.}}]{krempasky2024altermagnetic}%
  \BibitemOpen
  \bibfield  {author} {\bibinfo {author} {\bibfnamefont {J.}~\bibnamefont {Krempask{\`y}}}, \bibinfo {author} {\bibfnamefont {L.}~\bibnamefont {{\v{S}}mejkal}}, \bibinfo {author} {\bibfnamefont {S.}~\bibnamefont {D’souza}}, \bibinfo {author} {\bibfnamefont {M.}~\bibnamefont {Hajlaoui}}, \bibinfo {author} {\bibfnamefont {G.}~\bibnamefont {Springholz}}, \bibinfo {author} {\bibfnamefont {K.}~\bibnamefont {Uhl{\'\i}{\v{r}}ov{\'a}}}, \bibinfo {author} {\bibfnamefont {F.}~\bibnamefont {Alarab}}, \bibinfo {author} {\bibfnamefont {P.}~\bibnamefont {Constantinou}}, \bibinfo {author} {\bibfnamefont {V.}~\bibnamefont {Strocov}}, \bibinfo {author} {\bibfnamefont {D.}~\bibnamefont {Usanov}}, \emph {et~al.},\ }\bibfield  {title} {\bibinfo {title} {Altermagnetic lifting of kramers spin degeneracy},\ }\href {https://www.nature.com/articles/s41586-023-06907-7} {\bibfield  {journal} {\bibinfo  {journal} {Nature}\ }\textbf {\bibinfo {volume} {626}},\ \bibinfo {pages} {517} (\bibinfo {year} {2024})}\BibitemShut {NoStop}%
\bibitem [{\citenamefont {Lee}\ \emph {et~al.}(2024)\citenamefont {Lee}, \citenamefont {Lee}, \citenamefont {Jung}, \citenamefont {Jung}, \citenamefont {Kim}, \citenamefont {Lee}, \citenamefont {Seok}, \citenamefont {Kim}, \citenamefont {Park}, \citenamefont {{\v{S}}mejkal} \emph {et~al.}}]{lee2024broken}%
  \BibitemOpen
  \bibfield  {author} {\bibinfo {author} {\bibfnamefont {S.}~\bibnamefont {Lee}}, \bibinfo {author} {\bibfnamefont {S.}~\bibnamefont {Lee}}, \bibinfo {author} {\bibfnamefont {S.}~\bibnamefont {Jung}}, \bibinfo {author} {\bibfnamefont {J.}~\bibnamefont {Jung}}, \bibinfo {author} {\bibfnamefont {D.}~\bibnamefont {Kim}}, \bibinfo {author} {\bibfnamefont {Y.}~\bibnamefont {Lee}}, \bibinfo {author} {\bibfnamefont {B.}~\bibnamefont {Seok}}, \bibinfo {author} {\bibfnamefont {J.}~\bibnamefont {Kim}}, \bibinfo {author} {\bibfnamefont {B.~G.}\ \bibnamefont {Park}}, \bibinfo {author} {\bibfnamefont {L.}~\bibnamefont {{\v{S}}mejkal}}, \emph {et~al.},\ }\bibfield  {title} {\bibinfo {title} {Broken kramers degeneracy in altermagnetic {MnTe}},\ }\href {https://journals.aps.org/prl/abstract/10.1103/PhysRevLett.132.036702} {\bibfield  {journal} {\bibinfo  {journal} {Phys. Rev. Lett.}\ }\textbf {\bibinfo {volume} {132}},\ \bibinfo {pages} {036702} (\bibinfo {year} {2024})}\BibitemShut {NoStop}%
\bibitem [{\citenamefont {Reimers}\ \emph {et~al.}(2024)\citenamefont {Reimers}, \citenamefont {Odenbreit}, \citenamefont {{\v{S}}mejkal}, \citenamefont {Strocov}, \citenamefont {Constantinou}, \citenamefont {Hellenes}, \citenamefont {Jaeschke~Ubiergo}, \citenamefont {Campos}, \citenamefont {Bharadwaj}, \citenamefont {Chakraborty} \emph {et~al.}}]{reimers2024direct}%
  \BibitemOpen
  \bibfield  {author} {\bibinfo {author} {\bibfnamefont {S.}~\bibnamefont {Reimers}}, \bibinfo {author} {\bibfnamefont {L.}~\bibnamefont {Odenbreit}}, \bibinfo {author} {\bibfnamefont {L.}~\bibnamefont {{\v{S}}mejkal}}, \bibinfo {author} {\bibfnamefont {V.~N.}\ \bibnamefont {Strocov}}, \bibinfo {author} {\bibfnamefont {P.}~\bibnamefont {Constantinou}}, \bibinfo {author} {\bibfnamefont {A.~B.}\ \bibnamefont {Hellenes}}, \bibinfo {author} {\bibfnamefont {R.}~\bibnamefont {Jaeschke~Ubiergo}}, \bibinfo {author} {\bibfnamefont {W.~H.}\ \bibnamefont {Campos}}, \bibinfo {author} {\bibfnamefont {V.~K.}\ \bibnamefont {Bharadwaj}}, \bibinfo {author} {\bibfnamefont {A.}~\bibnamefont {Chakraborty}}, \emph {et~al.},\ }\bibfield  {title} {\bibinfo {title} {Direct observation of altermagnetic band splitting in {CrSb} thin films},\ }\href {https://www.nature.com/articles/s41467-024-46476-5} {\bibfield  {journal} {\bibinfo  {journal} {Nat. Commun.}\ }\textbf {\bibinfo {volume} {15}},\ \bibinfo {pages} {2116} (\bibinfo {year}
  {2024})}\BibitemShut {NoStop}%
\bibitem [{\citenamefont {Hajlaoui}\ \emph {et~al.}(2024)\citenamefont {Hajlaoui}, \citenamefont {Wilfred~D'Souza}, \citenamefont {{\v{S}}mejkal}, \citenamefont {Kriegner}, \citenamefont {Krizman}, \citenamefont {Zakusylo}, \citenamefont {Olszowska}, \citenamefont {Caha}, \citenamefont {Michali{\v{c}}ka}, \citenamefont {S{\'a}nchez~Barriga} \emph {et~al.}}]{hajlaoui2024temperature}%
  \BibitemOpen
  \bibfield  {author} {\bibinfo {author} {\bibfnamefont {M.}~\bibnamefont {Hajlaoui}}, \bibinfo {author} {\bibfnamefont {S.}~\bibnamefont {Wilfred~D'Souza}}, \bibinfo {author} {\bibfnamefont {L.}~\bibnamefont {{\v{S}}mejkal}}, \bibinfo {author} {\bibfnamefont {D.}~\bibnamefont {Kriegner}}, \bibinfo {author} {\bibfnamefont {G.}~\bibnamefont {Krizman}}, \bibinfo {author} {\bibfnamefont {T.}~\bibnamefont {Zakusylo}}, \bibinfo {author} {\bibfnamefont {N.}~\bibnamefont {Olszowska}}, \bibinfo {author} {\bibfnamefont {O.}~\bibnamefont {Caha}}, \bibinfo {author} {\bibfnamefont {J.}~\bibnamefont {Michali{\v{c}}ka}}, \bibinfo {author} {\bibfnamefont {J.}~\bibnamefont {S{\'a}nchez~Barriga}}, \emph {et~al.},\ }\bibfield  {title} {\bibinfo {title} {Temperature dependence of relativistic valence band splitting induced by an altermagnetic phase transition},\ }\href {https://advanced.onlinelibrary.wiley.com/doi/full/10.1002/adma.202314076} {\bibfield  {journal} {\bibinfo  {journal} {Adv. Mater.}\ }\textbf {\bibinfo {volume}
  {36}},\ \bibinfo {pages} {2314076} (\bibinfo {year} {2024})}\BibitemShut {NoStop}%
\bibitem [{\citenamefont {Zeng}\ \emph {et~al.}(2024)\citenamefont {Zeng}, \citenamefont {Zhu}, \citenamefont {Zhu}, \citenamefont {Liu}, \citenamefont {Ma}, \citenamefont {Hao}, \citenamefont {Liu}, \citenamefont {Qu}, \citenamefont {Yang}, \citenamefont {Jiang} \emph {et~al.}}]{zeng2024observation}%
  \BibitemOpen
  \bibfield  {author} {\bibinfo {author} {\bibfnamefont {M.}~\bibnamefont {Zeng}}, \bibinfo {author} {\bibfnamefont {M.~Y.}\ \bibnamefont {Zhu}}, \bibinfo {author} {\bibfnamefont {Y.~P.}\ \bibnamefont {Zhu}}, \bibinfo {author} {\bibfnamefont {X.~R.}\ \bibnamefont {Liu}}, \bibinfo {author} {\bibfnamefont {X.~M.}\ \bibnamefont {Ma}}, \bibinfo {author} {\bibfnamefont {Y.~J.}\ \bibnamefont {Hao}}, \bibinfo {author} {\bibfnamefont {P.}~\bibnamefont {Liu}}, \bibinfo {author} {\bibfnamefont {G.}~\bibnamefont {Qu}}, \bibinfo {author} {\bibfnamefont {Y.}~\bibnamefont {Yang}}, \bibinfo {author} {\bibfnamefont {Z.}~\bibnamefont {Jiang}}, \emph {et~al.},\ }\bibfield  {title} {\bibinfo {title} {Observation of spin splitting in room-temperature metallic antiferromagnet {CrSb}},\ }\href {https://advanced.onlinelibrary.wiley.com/doi/full/10.1002/advs.202406529} {\bibfield  {journal} {\bibinfo  {journal} {Adv. Sci.}\ }\textbf {\bibinfo {volume} {11}},\ \bibinfo {pages} {2406529} (\bibinfo {year} {2024})}\BibitemShut {NoStop}%
\bibitem [{\citenamefont {Yang}\ \emph {et~al.}(2025)\citenamefont {Yang}, \citenamefont {Li}, \citenamefont {Yang}, \citenamefont {Li}, \citenamefont {Zheng}, \citenamefont {Zhu}, \citenamefont {Pan}, \citenamefont {Xu}, \citenamefont {Cao}, \citenamefont {Zhao} \emph {et~al.}}]{yang2025three}%
  \BibitemOpen
  \bibfield  {author} {\bibinfo {author} {\bibfnamefont {G.}~\bibnamefont {Yang}}, \bibinfo {author} {\bibfnamefont {Z.}~\bibnamefont {Li}}, \bibinfo {author} {\bibfnamefont {S.}~\bibnamefont {Yang}}, \bibinfo {author} {\bibfnamefont {J.}~\bibnamefont {Li}}, \bibinfo {author} {\bibfnamefont {H.}~\bibnamefont {Zheng}}, \bibinfo {author} {\bibfnamefont {W.}~\bibnamefont {Zhu}}, \bibinfo {author} {\bibfnamefont {Z.}~\bibnamefont {Pan}}, \bibinfo {author} {\bibfnamefont {Y.}~\bibnamefont {Xu}}, \bibinfo {author} {\bibfnamefont {S.}~\bibnamefont {Cao}}, \bibinfo {author} {\bibfnamefont {W.}~\bibnamefont {Zhao}}, \emph {et~al.},\ }\bibfield  {title} {\bibinfo {title} {Three-dimensional mapping of the altermagnetic spin splitting in {CrSb}},\ }\href {https://www.nature.com/articles/s41467-025-56647-7} {\bibfield  {journal} {\bibinfo  {journal} {Nat. Commun.}\ }\textbf {\bibinfo {volume} {16}},\ \bibinfo {pages} {1442} (\bibinfo {year} {2025})}\BibitemShut {NoStop}%
\bibitem [{\citenamefont {Ding}\ \emph {et~al.}(2024)\citenamefont {Ding}, \citenamefont {Jiang}, \citenamefont {Chen}, \citenamefont {Tao}, \citenamefont {Liu}, \citenamefont {Li}, \citenamefont {Liu}, \citenamefont {Sun}, \citenamefont {Cheng}, \citenamefont {Liu} \emph {et~al.}}]{ding2024large}%
  \BibitemOpen
  \bibfield  {author} {\bibinfo {author} {\bibfnamefont {J.}~\bibnamefont {Ding}}, \bibinfo {author} {\bibfnamefont {Z.}~\bibnamefont {Jiang}}, \bibinfo {author} {\bibfnamefont {X.}~\bibnamefont {Chen}}, \bibinfo {author} {\bibfnamefont {Z.}~\bibnamefont {Tao}}, \bibinfo {author} {\bibfnamefont {Z.}~\bibnamefont {Liu}}, \bibinfo {author} {\bibfnamefont {T.}~\bibnamefont {Li}}, \bibinfo {author} {\bibfnamefont {J.}~\bibnamefont {Liu}}, \bibinfo {author} {\bibfnamefont {J.}~\bibnamefont {Sun}}, \bibinfo {author} {\bibfnamefont {J.}~\bibnamefont {Cheng}}, \bibinfo {author} {\bibfnamefont {J.}~\bibnamefont {Liu}}, \emph {et~al.},\ }\bibfield  {title} {\bibinfo {title} {Large band splitting in g-wave altermagnet {CrSb}},\ }\href {https://journals.aps.org/prl/abstract/10.1103/PhysRevLett.133.206401} {\bibfield  {journal} {\bibinfo  {journal} {Phys. Rev. Lett.}\ }\textbf {\bibinfo {volume} {133}},\ \bibinfo {pages} {206401} (\bibinfo {year} {2024})}\BibitemShut {NoStop}%
\bibitem [{\citenamefont {Fedchenko}\ \emph {et~al.}(2024)\citenamefont {Fedchenko}, \citenamefont {Min{\'a}r}, \citenamefont {Akashdeep}, \citenamefont {D’Souza}, \citenamefont {Vasilyev}, \citenamefont {Tkach}, \citenamefont {Odenbreit}, \citenamefont {Nguyen}, \citenamefont {Kutnyakhov}, \citenamefont {Wind} \emph {et~al.}}]{fedchenko2024observation}%
  \BibitemOpen
  \bibfield  {author} {\bibinfo {author} {\bibfnamefont {O.}~\bibnamefont {Fedchenko}}, \bibinfo {author} {\bibfnamefont {J.}~\bibnamefont {Min{\'a}r}}, \bibinfo {author} {\bibfnamefont {A.}~\bibnamefont {Akashdeep}}, \bibinfo {author} {\bibfnamefont {S.~W.}\ \bibnamefont {D’Souza}}, \bibinfo {author} {\bibfnamefont {D.}~\bibnamefont {Vasilyev}}, \bibinfo {author} {\bibfnamefont {O.}~\bibnamefont {Tkach}}, \bibinfo {author} {\bibfnamefont {L.}~\bibnamefont {Odenbreit}}, \bibinfo {author} {\bibfnamefont {Q.}~\bibnamefont {Nguyen}}, \bibinfo {author} {\bibfnamefont {D.}~\bibnamefont {Kutnyakhov}}, \bibinfo {author} {\bibfnamefont {N.}~\bibnamefont {Wind}}, \emph {et~al.},\ }\bibfield  {title} {\bibinfo {title} {Observation of time-reversal symmetry breaking in the band structure of altermagnetic $\mathrm{RuO}_{2}$},\ }\href {https://www.science.org/doi/full/10.1126/sciadv.adj4883} {\bibfield  {journal} {\bibinfo  {journal} {Sci. Adv.}\ }\textbf {\bibinfo {volume} {10}},\ \bibinfo {pages} {eadj4883} (\bibinfo
  {year} {2024})}\BibitemShut {NoStop}%
\bibitem [{\citenamefont {Zhang}\ \emph {et~al.}(2025)\citenamefont {Zhang}, \citenamefont {Cheng}, \citenamefont {Yin}, \citenamefont {Liu}, \citenamefont {Deng}, \citenamefont {Qiao}, \citenamefont {Shi}, \citenamefont {Zhang}, \citenamefont {Lin}, \citenamefont {Liu} \emph {et~al.}}]{zhang2025crystal}%
  \BibitemOpen
  \bibfield  {author} {\bibinfo {author} {\bibfnamefont {F.}~\bibnamefont {Zhang}}, \bibinfo {author} {\bibfnamefont {X.}~\bibnamefont {Cheng}}, \bibinfo {author} {\bibfnamefont {Z.}~\bibnamefont {Yin}}, \bibinfo {author} {\bibfnamefont {C.}~\bibnamefont {Liu}}, \bibinfo {author} {\bibfnamefont {L.}~\bibnamefont {Deng}}, \bibinfo {author} {\bibfnamefont {Y.}~\bibnamefont {Qiao}}, \bibinfo {author} {\bibfnamefont {Z.}~\bibnamefont {Shi}}, \bibinfo {author} {\bibfnamefont {S.}~\bibnamefont {Zhang}}, \bibinfo {author} {\bibfnamefont {J.}~\bibnamefont {Lin}}, \bibinfo {author} {\bibfnamefont {Z.}~\bibnamefont {Liu}}, \emph {et~al.},\ }\bibfield  {title} {\bibinfo {title} {Crystal-symmetry-paired spin-valley locking in a layered room-temperature metallic altermagnet candidate},\ }\href {https://www.nature.com/articles/s41567-025-02864-2} {\bibfield  {journal} {\bibinfo  {journal} {Nat. Phys.}\ }\textbf {\bibinfo {volume} {21}},\ \bibinfo {pages} {760} (\bibinfo {year} {2025})}\BibitemShut {NoStop}%
\bibitem [{\citenamefont {{\v{S}}mejkal}\ \emph {et~al.}(2022{\natexlab{a}})\citenamefont {{\v{S}}mejkal}, \citenamefont {Sinova},\ and\ \citenamefont {Jungwirth}}]{vsmejkal2022emerging}%
  \BibitemOpen
  \bibfield  {author} {\bibinfo {author} {\bibfnamefont {L.}~\bibnamefont {{\v{S}}mejkal}}, \bibinfo {author} {\bibfnamefont {J.}~\bibnamefont {Sinova}},\ and\ \bibinfo {author} {\bibfnamefont {T.}~\bibnamefont {Jungwirth}},\ }\bibfield  {title} {\bibinfo {title} {Emerging research landscape of altermagnetism},\ }\href {https://journals.aps.org/prx/abstract/10.1103/PhysRevX.12.040501} {\bibfield  {journal} {\bibinfo  {journal} {Phys. Rev. X}\ }\textbf {\bibinfo {volume} {12}},\ \bibinfo {pages} {040501} (\bibinfo {year} {2022}{\natexlab{a}})}\BibitemShut {NoStop}%
\bibitem [{\citenamefont {Mazin}\ and\ \citenamefont {Editors}(2022)}]{mazin2022altermagnetism}%
  \BibitemOpen
  \bibfield  {author} {\bibinfo {author} {\bibfnamefont {I.}~\bibnamefont {Mazin}}\ and\ \bibinfo {author} {\bibfnamefont {P.}~\bibnamefont {Editors}},\ }\bibfield  {title} {\bibinfo {title} {Altermagnetism—a new punch line of fundamental magnetism},\ }\href {https://journals.aps.org/prx/edannounce/10.1103/PhysRevX.12.040002} {\bibfield  {journal} {\bibinfo  {journal} {Phys. Rev. X}\ }\textbf {\bibinfo {volume} {12}},\ \bibinfo {pages} {040002} (\bibinfo {year} {2022})}\BibitemShut {NoStop}%
\bibitem [{\citenamefont {Tamang}\ \emph {et~al.}(2024)\citenamefont {Tamang}, \citenamefont {Gurung}, \citenamefont {Rai}, \citenamefont {Brahimi},\ and\ \citenamefont {Lounis}}]{tamang2024newly}%
  \BibitemOpen
  \bibfield  {author} {\bibinfo {author} {\bibfnamefont {R.}~\bibnamefont {Tamang}}, \bibinfo {author} {\bibfnamefont {S.}~\bibnamefont {Gurung}}, \bibinfo {author} {\bibfnamefont {D.}~\bibnamefont {Rai}}, \bibinfo {author} {\bibfnamefont {S.}~\bibnamefont {Brahimi}},\ and\ \bibinfo {author} {\bibfnamefont {S.}~\bibnamefont {Lounis}},\ }\bibfield  {title} {\bibinfo {title} {Newly discovered magnetic phase: A brief review on altermagnets},\ }\href {https://arxiv.org/abs/2412.05377} {\bibfield  {journal} {\bibinfo  {journal} {arXiv:2412.05377}\ } (\bibinfo {year} {2024})}\BibitemShut {NoStop}%
\bibitem [{\citenamefont {{\v{S}}mejkal}\ \emph {et~al.}(2020)\citenamefont {{\v{S}}mejkal}, \citenamefont {Gonz{\'a}lez-Hern{\'a}ndez}, \citenamefont {Jungwirth},\ and\ \citenamefont {Sinova}}]{vsmejkal2020crystal}%
  \BibitemOpen
  \bibfield  {author} {\bibinfo {author} {\bibfnamefont {L.}~\bibnamefont {{\v{S}}mejkal}}, \bibinfo {author} {\bibfnamefont {R.}~\bibnamefont {Gonz{\'a}lez-Hern{\'a}ndez}}, \bibinfo {author} {\bibfnamefont {T.}~\bibnamefont {Jungwirth}},\ and\ \bibinfo {author} {\bibfnamefont {J.}~\bibnamefont {Sinova}},\ }\bibfield  {title} {\bibinfo {title} {Crystal time-reversal symmetry breaking and spontaneous hall effect in collinear antiferromagnets},\ }\href {https://www.science.org/doi/full/10.1126/sciadv.aaz8809} {\bibfield  {journal} {\bibinfo  {journal} {Sci. Adv.}\ }\textbf {\bibinfo {volume} {6}},\ \bibinfo {pages} {eaaz8809} (\bibinfo {year} {2020})}\BibitemShut {NoStop}%
\bibitem [{\citenamefont {Reichlova}\ \emph {et~al.}(2024)\citenamefont {Reichlova}, \citenamefont {Lopes~Seeger}, \citenamefont {Gonz{\'a}lez-Hern{\'a}ndez}, \citenamefont {Kounta}, \citenamefont {Schlitz}, \citenamefont {Kriegner}, \citenamefont {Ritzinger}, \citenamefont {Lammel}, \citenamefont {Leivisk{\"a}}, \citenamefont {Birk~Hellenes} \emph {et~al.}}]{reichlova2024observation}%
  \BibitemOpen
  \bibfield  {author} {\bibinfo {author} {\bibfnamefont {H.}~\bibnamefont {Reichlova}}, \bibinfo {author} {\bibfnamefont {R.}~\bibnamefont {Lopes~Seeger}}, \bibinfo {author} {\bibfnamefont {R.}~\bibnamefont {Gonz{\'a}lez-Hern{\'a}ndez}}, \bibinfo {author} {\bibfnamefont {I.}~\bibnamefont {Kounta}}, \bibinfo {author} {\bibfnamefont {R.}~\bibnamefont {Schlitz}}, \bibinfo {author} {\bibfnamefont {D.}~\bibnamefont {Kriegner}}, \bibinfo {author} {\bibfnamefont {P.}~\bibnamefont {Ritzinger}}, \bibinfo {author} {\bibfnamefont {M.}~\bibnamefont {Lammel}}, \bibinfo {author} {\bibfnamefont {M.}~\bibnamefont {Leivisk{\"a}}}, \bibinfo {author} {\bibfnamefont {A.}~\bibnamefont {Birk~Hellenes}}, \emph {et~al.},\ }\bibfield  {title} {\bibinfo {title} {Observation of a spontaneous anomalous hall response in the $\mathrm{Mn}_{5}\mathrm{Si}_{3}$ $d$-wave altermagnet candidate},\ }\href {https://www.nature.com/articles/s41467-024-48493-w} {\bibfield  {journal} {\bibinfo  {journal} {Nat. Commun.}\ }\textbf {\bibinfo {volume}
  {15}},\ \bibinfo {pages} {4961} (\bibinfo {year} {2024})}\BibitemShut {NoStop}%
\bibitem [{\citenamefont {Feng}\ \emph {et~al.}(2022)\citenamefont {Feng}, \citenamefont {Zhou}, \citenamefont {{\v{S}}mejkal}, \citenamefont {Wu}, \citenamefont {Zhu}, \citenamefont {Guo}, \citenamefont {Gonz{\'a}lez-Hern{\'a}ndez}, \citenamefont {Wang}, \citenamefont {Yan}, \citenamefont {Qin} \emph {et~al.}}]{feng2022anomalous}%
  \BibitemOpen
  \bibfield  {author} {\bibinfo {author} {\bibfnamefont {Z.}~\bibnamefont {Feng}}, \bibinfo {author} {\bibfnamefont {X.}~\bibnamefont {Zhou}}, \bibinfo {author} {\bibfnamefont {L.}~\bibnamefont {{\v{S}}mejkal}}, \bibinfo {author} {\bibfnamefont {L.}~\bibnamefont {Wu}}, \bibinfo {author} {\bibfnamefont {Z.}~\bibnamefont {Zhu}}, \bibinfo {author} {\bibfnamefont {H.}~\bibnamefont {Guo}}, \bibinfo {author} {\bibfnamefont {R.}~\bibnamefont {Gonz{\'a}lez-Hern{\'a}ndez}}, \bibinfo {author} {\bibfnamefont {X.}~\bibnamefont {Wang}}, \bibinfo {author} {\bibfnamefont {H.}~\bibnamefont {Yan}}, \bibinfo {author} {\bibfnamefont {P.}~\bibnamefont {Qin}}, \emph {et~al.},\ }\bibfield  {title} {\bibinfo {title} {An anomalous hall effect in altermagnetic ruthenium dioxide},\ }\href {https://www.nature.com/articles/s41928-022-00866-z} {\bibfield  {journal} {\bibinfo  {journal} {Nat. Electron.}\ }\textbf {\bibinfo {volume} {5}},\ \bibinfo {pages} {735} (\bibinfo {year} {2022})}\BibitemShut {NoStop}%
\bibitem [{\citenamefont {Gonzalez~Betancourt}\ \emph {et~al.}(2023)\citenamefont {Gonzalez~Betancourt}, \citenamefont {Zub{\'a}{\v{c}}}, \citenamefont {Gonzalez-Hernandez}, \citenamefont {Geishendorf}, \citenamefont {{\v{S}}ob{\'a}{\v{n}}}, \citenamefont {Springholz}, \citenamefont {Olejn{\'\i}k}, \citenamefont {{\v{S}}mejkal}, \citenamefont {Sinova}, \citenamefont {Jungwirth} \emph {et~al.}}]{gonzalez2023spontaneous}%
  \BibitemOpen
  \bibfield  {author} {\bibinfo {author} {\bibfnamefont {R.}~\bibnamefont {Gonzalez~Betancourt}}, \bibinfo {author} {\bibfnamefont {J.}~\bibnamefont {Zub{\'a}{\v{c}}}}, \bibinfo {author} {\bibfnamefont {R.}~\bibnamefont {Gonzalez-Hernandez}}, \bibinfo {author} {\bibfnamefont {K.}~\bibnamefont {Geishendorf}}, \bibinfo {author} {\bibfnamefont {Z.}~\bibnamefont {{\v{S}}ob{\'a}{\v{n}}}}, \bibinfo {author} {\bibfnamefont {G.}~\bibnamefont {Springholz}}, \bibinfo {author} {\bibfnamefont {K.}~\bibnamefont {Olejn{\'\i}k}}, \bibinfo {author} {\bibfnamefont {L.}~\bibnamefont {{\v{S}}mejkal}}, \bibinfo {author} {\bibfnamefont {J.}~\bibnamefont {Sinova}}, \bibinfo {author} {\bibfnamefont {T.}~\bibnamefont {Jungwirth}}, \emph {et~al.},\ }\bibfield  {title} {\bibinfo {title} {Spontaneous anomalous hall effect arising from an unconventional compensated magnetic phase in a semiconductor},\ }\href {https://journals.aps.org/prl/abstract/10.1103/PhysRevLett.130.036702} {\bibfield  {journal} {\bibinfo  {journal} {Phys. Rev.
  Lett.}\ }\textbf {\bibinfo {volume} {130}},\ \bibinfo {pages} {036702} (\bibinfo {year} {2023})}\BibitemShut {NoStop}%
\bibitem [{\citenamefont {Han}\ \emph {et~al.}(2025)\citenamefont {Han}, \citenamefont {Fu}, \citenamefont {He}, \citenamefont {Dai}, \citenamefont {Zhu}, \citenamefont {Yang}, \citenamefont {Chen}, \citenamefont {Zhang}, \citenamefont {Zhu}, \citenamefont {Bai} \emph {et~al.}}]{han2025nonvolatile}%
  \BibitemOpen
  \bibfield  {author} {\bibinfo {author} {\bibfnamefont {L.}~\bibnamefont {Han}}, \bibinfo {author} {\bibfnamefont {X.}~\bibnamefont {Fu}}, \bibinfo {author} {\bibfnamefont {W.}~\bibnamefont {He}}, \bibinfo {author} {\bibfnamefont {J.}~\bibnamefont {Dai}}, \bibinfo {author} {\bibfnamefont {Y.}~\bibnamefont {Zhu}}, \bibinfo {author} {\bibfnamefont {W.}~\bibnamefont {Yang}}, \bibinfo {author} {\bibfnamefont {Y.}~\bibnamefont {Chen}}, \bibinfo {author} {\bibfnamefont {J.}~\bibnamefont {Zhang}}, \bibinfo {author} {\bibfnamefont {W.}~\bibnamefont {Zhu}}, \bibinfo {author} {\bibfnamefont {H.}~\bibnamefont {Bai}}, \emph {et~al.},\ }\bibfield  {title} {\bibinfo {title} {Nonvolatile anomalous nernst effect in $\mathrm{Mn}_{5}\mathrm{Si}_{3}$ with a collinear n{\'e}el vector},\ }\href {https://journals.aps.org/prapplied/abstract/10.1103/PhysRevApplied.23.044066} {\bibfield  {journal} {\bibinfo  {journal} {Phys. Rev. Appl.}\ }\textbf {\bibinfo {volume} {23}},\ \bibinfo {pages} {044066} (\bibinfo {year}
  {2025})}\BibitemShut {NoStop}%
\bibitem [{\citenamefont {Zhou}\ \emph {et~al.}(2024)\citenamefont {Zhou}, \citenamefont {Feng}, \citenamefont {Zhang}, \citenamefont {{\v{S}}mejkal}, \citenamefont {Sinova}, \citenamefont {Mokrousov},\ and\ \citenamefont {Yao}}]{zhou2024crystal}%
  \BibitemOpen
  \bibfield  {author} {\bibinfo {author} {\bibfnamefont {X.}~\bibnamefont {Zhou}}, \bibinfo {author} {\bibfnamefont {W.}~\bibnamefont {Feng}}, \bibinfo {author} {\bibfnamefont {R.~W.}\ \bibnamefont {Zhang}}, \bibinfo {author} {\bibfnamefont {L.}~\bibnamefont {{\v{S}}mejkal}}, \bibinfo {author} {\bibfnamefont {J.}~\bibnamefont {Sinova}}, \bibinfo {author} {\bibfnamefont {Y.}~\bibnamefont {Mokrousov}},\ and\ \bibinfo {author} {\bibfnamefont {Y.}~\bibnamefont {Yao}},\ }\bibfield  {title} {\bibinfo {title} {Crystal thermal transport in altermagnetic $\mathrm{RuO}_{2}$},\ }\href {https://journals.aps.org/prl/abstract/10.1103/PhysRevLett.132.056701} {\bibfield  {journal} {\bibinfo  {journal} {Phys. Rev. Lett.}\ }\textbf {\bibinfo {volume} {132}},\ \bibinfo {pages} {056701} (\bibinfo {year} {2024})}\BibitemShut {NoStop}%
\bibitem [{\citenamefont {Zhou}\ \emph {et~al.}(2025)\citenamefont {Zhou}, \citenamefont {Cheng}, \citenamefont {Hu}, \citenamefont {Chu}, \citenamefont {Bai}, \citenamefont {Han}, \citenamefont {Liu}, \citenamefont {Pan},\ and\ \citenamefont {Song}}]{zhou2025manipulation}%
  \BibitemOpen
  \bibfield  {author} {\bibinfo {author} {\bibfnamefont {Z.}~\bibnamefont {Zhou}}, \bibinfo {author} {\bibfnamefont {X.}~\bibnamefont {Cheng}}, \bibinfo {author} {\bibfnamefont {M.}~\bibnamefont {Hu}}, \bibinfo {author} {\bibfnamefont {R.}~\bibnamefont {Chu}}, \bibinfo {author} {\bibfnamefont {H.}~\bibnamefont {Bai}}, \bibinfo {author} {\bibfnamefont {L.}~\bibnamefont {Han}}, \bibinfo {author} {\bibfnamefont {J.}~\bibnamefont {Liu}}, \bibinfo {author} {\bibfnamefont {F.}~\bibnamefont {Pan}},\ and\ \bibinfo {author} {\bibfnamefont {C.}~\bibnamefont {Song}},\ }\bibfield  {title} {\bibinfo {title} {Manipulation of the altermagnetic order in {CrSb} via crystal symmetry},\ }\href {https://www.nature.com/articles/s41586-024-08436-3} {\bibfield  {journal} {\bibinfo  {journal} {Nature}\ }\textbf {\bibinfo {volume} {638}},\ \bibinfo {pages} {645} (\bibinfo {year} {2025})}\BibitemShut {NoStop}%
\bibitem [{\citenamefont {Han}\ \emph {et~al.}(2024)\citenamefont {Han}, \citenamefont {Fu}, \citenamefont {Peng}, \citenamefont {Cheng}, \citenamefont {Dai}, \citenamefont {Liu}, \citenamefont {Li}, \citenamefont {Zhang}, \citenamefont {Zhu}, \citenamefont {Bai} \emph {et~al.}}]{han2024electrical}%
  \BibitemOpen
  \bibfield  {author} {\bibinfo {author} {\bibfnamefont {L.}~\bibnamefont {Han}}, \bibinfo {author} {\bibfnamefont {X.}~\bibnamefont {Fu}}, \bibinfo {author} {\bibfnamefont {R.}~\bibnamefont {Peng}}, \bibinfo {author} {\bibfnamefont {X.}~\bibnamefont {Cheng}}, \bibinfo {author} {\bibfnamefont {J.}~\bibnamefont {Dai}}, \bibinfo {author} {\bibfnamefont {L.}~\bibnamefont {Liu}}, \bibinfo {author} {\bibfnamefont {Y.}~\bibnamefont {Li}}, \bibinfo {author} {\bibfnamefont {Y.}~\bibnamefont {Zhang}}, \bibinfo {author} {\bibfnamefont {W.}~\bibnamefont {Zhu}}, \bibinfo {author} {\bibfnamefont {H.}~\bibnamefont {Bai}}, \emph {et~al.},\ }\bibfield  {title} {\bibinfo {title} {Electrical 180 switching of n{\'e}el vector in spin-splitting antiferromagnet},\ }\href {https://www.science.org/doi/10.1126/sciadv.adn0479} {\bibfield  {journal} {\bibinfo  {journal} {Sci. Adv.}\ }\textbf {\bibinfo {volume} {10}},\ \bibinfo {pages} {eadn0479} (\bibinfo {year} {2024})}\BibitemShut {NoStop}%
\bibitem [{\citenamefont {Song}\ \emph {et~al.}(2025)\citenamefont {Song}, \citenamefont {Bai}, \citenamefont {Zhou}, \citenamefont {Han}, \citenamefont {Reichlova}, \citenamefont {Dil}, \citenamefont {Liu}, \citenamefont {Chen},\ and\ \citenamefont {Pan}}]{song2025altermagnets}%
  \BibitemOpen
  \bibfield  {author} {\bibinfo {author} {\bibfnamefont {C.}~\bibnamefont {Song}}, \bibinfo {author} {\bibfnamefont {H.}~\bibnamefont {Bai}}, \bibinfo {author} {\bibfnamefont {Z.}~\bibnamefont {Zhou}}, \bibinfo {author} {\bibfnamefont {L.}~\bibnamefont {Han}}, \bibinfo {author} {\bibfnamefont {H.}~\bibnamefont {Reichlova}}, \bibinfo {author} {\bibfnamefont {J.~H.}\ \bibnamefont {Dil}}, \bibinfo {author} {\bibfnamefont {J.}~\bibnamefont {Liu}}, \bibinfo {author} {\bibfnamefont {X.}~\bibnamefont {Chen}},\ and\ \bibinfo {author} {\bibfnamefont {F.}~\bibnamefont {Pan}},\ }\bibfield  {title} {\bibinfo {title} {Altermagnets as a new class of functional materials},\ }\href {https://www.nature.com/articles/s41578-025-00779-1} {\bibfield  {journal} {\bibinfo  {journal} {Nat. Rev. Mater.}\ }\textbf {\bibinfo {volume} {10}},\ \bibinfo {pages} {473} (\bibinfo {year} {2025})}\BibitemShut {NoStop}%
\bibitem [{\citenamefont {Takahashi}\ \emph {et~al.}(2025)\citenamefont {Takahashi}, \citenamefont {Steward}, \citenamefont {Ogata}, \citenamefont {Fernandes},\ and\ \citenamefont {Schmalian}}]{takahashi2025elasto}%
  \BibitemOpen
  \bibfield  {author} {\bibinfo {author} {\bibfnamefont {K.}~\bibnamefont {Takahashi}}, \bibinfo {author} {\bibfnamefont {C.~R.}\ \bibnamefont {Steward}}, \bibinfo {author} {\bibfnamefont {M.}~\bibnamefont {Ogata}}, \bibinfo {author} {\bibfnamefont {R.~M.}\ \bibnamefont {Fernandes}},\ and\ \bibinfo {author} {\bibfnamefont {J.}~\bibnamefont {Schmalian}},\ }\bibfield  {title} {\bibinfo {title} {Elasto-hall conductivity and the anomalous hall effect in altermagnets},\ }\href {https://journals.aps.org/prb/abstract/10.1103/PhysRevB.111.184408} {\bibfield  {journal} {\bibinfo  {journal} {Phys. Rev. B}\ }\textbf {\bibinfo {volume} {111}},\ \bibinfo {pages} {184408} (\bibinfo {year} {2025})}\BibitemShut {NoStop}%
\bibitem [{\citenamefont {Shao}\ \emph {et~al.}(2021)\citenamefont {Shao}, \citenamefont {Zhang}, \citenamefont {Li}, \citenamefont {Eom},\ and\ \citenamefont {Tsymbal}}]{shao2021spin}%
  \BibitemOpen
  \bibfield  {author} {\bibinfo {author} {\bibfnamefont {D.~F.}\ \bibnamefont {Shao}}, \bibinfo {author} {\bibfnamefont {S.~H.}\ \bibnamefont {Zhang}}, \bibinfo {author} {\bibfnamefont {M.}~\bibnamefont {Li}}, \bibinfo {author} {\bibfnamefont {C.~B.}\ \bibnamefont {Eom}},\ and\ \bibinfo {author} {\bibfnamefont {E.~Y.}\ \bibnamefont {Tsymbal}},\ }\bibfield  {title} {\bibinfo {title} {Spin-neutral currents for spintronics},\ }\href {https://www.nature.com/articles/s41467-021-26915-3} {\bibfield  {journal} {\bibinfo  {journal} {Nat. Commun.}\ }\textbf {\bibinfo {volume} {12}},\ \bibinfo {pages} {7061} (\bibinfo {year} {2021})}\BibitemShut {NoStop}%
\bibitem [{\citenamefont {Shao}\ and\ \citenamefont {Tsymbal}(2024)}]{shao2024antiferromagnetic}%
  \BibitemOpen
  \bibfield  {author} {\bibinfo {author} {\bibfnamefont {D.~F.}\ \bibnamefont {Shao}}\ and\ \bibinfo {author} {\bibfnamefont {E.~Y.}\ \bibnamefont {Tsymbal}},\ }\bibfield  {title} {\bibinfo {title} {Antiferromagnetic tunnel junctions for spintronics},\ }\href {https://www.nature.com/articles/s44306-024-00014-7} {\bibfield  {journal} {\bibinfo  {journal} {Npj Spintron.}\ }\textbf {\bibinfo {volume} {2}},\ \bibinfo {pages} {13} (\bibinfo {year} {2024})}\BibitemShut {NoStop}%
\bibitem [{\citenamefont {Jiang}\ \emph {et~al.}(2023)\citenamefont {Jiang}, \citenamefont {Wang}, \citenamefont {Samanta}, \citenamefont {Zhang}, \citenamefont {Xiao}, \citenamefont {Lu}, \citenamefont {Sun}, \citenamefont {Tsymbal},\ and\ \citenamefont {Shao}}]{jiang2023prediction}%
  \BibitemOpen
  \bibfield  {author} {\bibinfo {author} {\bibfnamefont {Y.~Y.}\ \bibnamefont {Jiang}}, \bibinfo {author} {\bibfnamefont {Z.~A.}\ \bibnamefont {Wang}}, \bibinfo {author} {\bibfnamefont {K.}~\bibnamefont {Samanta}}, \bibinfo {author} {\bibfnamefont {S.~H.}\ \bibnamefont {Zhang}}, \bibinfo {author} {\bibfnamefont {R.~C.}\ \bibnamefont {Xiao}}, \bibinfo {author} {\bibfnamefont {W.}~\bibnamefont {Lu}}, \bibinfo {author} {\bibfnamefont {Y.}~\bibnamefont {Sun}}, \bibinfo {author} {\bibfnamefont {E.~Y.}\ \bibnamefont {Tsymbal}},\ and\ \bibinfo {author} {\bibfnamefont {D.~F.}\ \bibnamefont {Shao}},\ }\bibfield  {title} {\bibinfo {title} {Prediction of giant tunneling magnetoresistance in $\mathrm{RuO}_{2}/\mathrm{TiO}_{2}/\mathrm{RuO}_{2}$ (110) antiferromagnetic tunnel junctions},\ }\href {https://journals.aps.org/prb/abstract/10.1103/PhysRevB.108.174439} {\bibfield  {journal} {\bibinfo  {journal} {Phys. Rev. B}\ }\textbf {\bibinfo {volume} {108}},\ \bibinfo {pages} {174439} (\bibinfo {year} {2023})}\BibitemShut
  {NoStop}%
\bibitem [{\citenamefont {Samanta}\ \emph {et~al.}(2025)\citenamefont {Samanta}, \citenamefont {Shao},\ and\ \citenamefont {Tsymbal}}]{samanta2025spin}%
  \BibitemOpen
  \bibfield  {author} {\bibinfo {author} {\bibfnamefont {K.}~\bibnamefont {Samanta}}, \bibinfo {author} {\bibfnamefont {D.~F.}\ \bibnamefont {Shao}},\ and\ \bibinfo {author} {\bibfnamefont {E.~Y.}\ \bibnamefont {Tsymbal}},\ }\bibfield  {title} {\bibinfo {title} {Spin filtering with insulating altermagnets},\ }\href {https://pubs.acs.org/doi/full/10.1021/acs.nanolett.4c05672} {\bibfield  {journal} {\bibinfo  {journal} {Nano Lett.}\ }\textbf {\bibinfo {volume} {25}},\ \bibinfo {pages} {3150} (\bibinfo {year} {2025})}\BibitemShut {NoStop}%
\bibitem [{\citenamefont {{\v{S}}mejkal}\ \emph {et~al.}(2022{\natexlab{b}})\citenamefont {{\v{S}}mejkal}, \citenamefont {Hellenes}, \citenamefont {Gonz{\'a}lez-Hern{\'a}ndez}, \citenamefont {Sinova},\ and\ \citenamefont {Jungwirth}}]{vsmejkal2022giant}%
  \BibitemOpen
  \bibfield  {author} {\bibinfo {author} {\bibfnamefont {L.}~\bibnamefont {{\v{S}}mejkal}}, \bibinfo {author} {\bibfnamefont {A.~B.}\ \bibnamefont {Hellenes}}, \bibinfo {author} {\bibfnamefont {R.}~\bibnamefont {Gonz{\'a}lez-Hern{\'a}ndez}}, \bibinfo {author} {\bibfnamefont {J.}~\bibnamefont {Sinova}},\ and\ \bibinfo {author} {\bibfnamefont {T.}~\bibnamefont {Jungwirth}},\ }\bibfield  {title} {\bibinfo {title} {Giant and tunneling magnetoresistance in unconventional collinear antiferromagnets with nonrelativistic spin-momentum coupling},\ }\href {https://journals.aps.org/prx/abstract/10.1103/PhysRevX.12.011028} {\bibfield  {journal} {\bibinfo  {journal} {Phys. Rev. X}\ }\textbf {\bibinfo {volume} {12}},\ \bibinfo {pages} {011028} (\bibinfo {year} {2022}{\natexlab{b}})}\BibitemShut {NoStop}%
\bibitem [{\citenamefont {Liu}\ \emph {et~al.}(2024{\natexlab{b}})\citenamefont {Liu}, \citenamefont {Zhang}, \citenamefont {Yuan}, \citenamefont {Liu}, \citenamefont {Zhu}, \citenamefont {Lu},\ and\ \citenamefont {Xiong}}]{liu2024giant}%
  \BibitemOpen
  \bibfield  {author} {\bibinfo {author} {\bibfnamefont {F.}~\bibnamefont {Liu}}, \bibinfo {author} {\bibfnamefont {Z.}~\bibnamefont {Zhang}}, \bibinfo {author} {\bibfnamefont {X.}~\bibnamefont {Yuan}}, \bibinfo {author} {\bibfnamefont {Y.}~\bibnamefont {Liu}}, \bibinfo {author} {\bibfnamefont {S.}~\bibnamefont {Zhu}}, \bibinfo {author} {\bibfnamefont {Z.}~\bibnamefont {Lu}},\ and\ \bibinfo {author} {\bibfnamefont {R.}~\bibnamefont {Xiong}},\ }\bibfield  {title} {\bibinfo {title} {Giant tunneling magnetoresistance in insulated altermagnet/ferromagnet junctions induced by spin-dependent tunneling effect},\ }\href {https://journals.aps.org/prb/abstract/10.1103/PhysRevB.110.134437} {\bibfield  {journal} {\bibinfo  {journal} {Phys. Rev. B}\ }\textbf {\bibinfo {volume} {110}},\ \bibinfo {pages} {134437} (\bibinfo {year} {2024}{\natexlab{b}})}\BibitemShut {NoStop}%
\bibitem [{\citenamefont {Baltz}\ \emph {et~al.}(2018)\citenamefont {Baltz}, \citenamefont {Manchon}, \citenamefont {Tsoi}, \citenamefont {Moriyama}, \citenamefont {Ono},\ and\ \citenamefont {Tserkovnyak}}]{baltz2018antiferromagnetic}%
  \BibitemOpen
  \bibfield  {author} {\bibinfo {author} {\bibfnamefont {V.}~\bibnamefont {Baltz}}, \bibinfo {author} {\bibfnamefont {A.}~\bibnamefont {Manchon}}, \bibinfo {author} {\bibfnamefont {M.}~\bibnamefont {Tsoi}}, \bibinfo {author} {\bibfnamefont {T.}~\bibnamefont {Moriyama}}, \bibinfo {author} {\bibfnamefont {T.}~\bibnamefont {Ono}},\ and\ \bibinfo {author} {\bibfnamefont {Y.}~\bibnamefont {Tserkovnyak}},\ }\bibfield  {title} {\bibinfo {title} {Antiferromagnetic spintronics},\ }\href {https://journals.aps.org/rmp/abstract/10.1103/RevModPhys.90.015005} {\bibfield  {journal} {\bibinfo  {journal} {Rev. Mod. Phys.}\ }\textbf {\bibinfo {volume} {90}},\ \bibinfo {pages} {015005} (\bibinfo {year} {2018})}\BibitemShut {NoStop}%
\bibitem [{\citenamefont {Jungwirth}\ \emph {et~al.}(2016)\citenamefont {Jungwirth}, \citenamefont {Marti}, \citenamefont {Wadley},\ and\ \citenamefont {Wunderlich}}]{jungwirth2016antiferromagnetic}%
  \BibitemOpen
  \bibfield  {author} {\bibinfo {author} {\bibfnamefont {T.}~\bibnamefont {Jungwirth}}, \bibinfo {author} {\bibfnamefont {X.}~\bibnamefont {Marti}}, \bibinfo {author} {\bibfnamefont {P.}~\bibnamefont {Wadley}},\ and\ \bibinfo {author} {\bibfnamefont {J.}~\bibnamefont {Wunderlich}},\ }\bibfield  {title} {\bibinfo {title} {Antiferromagnetic spintronics},\ }\href {https://www.nature.com/articles/nnano.2016.18} {\bibfield  {journal} {\bibinfo  {journal} {Nat. Nanotechnol.}\ }\textbf {\bibinfo {volume} {11}},\ \bibinfo {pages} {231} (\bibinfo {year} {2016})}\BibitemShut {NoStop}%
\bibitem [{\citenamefont {Jungwirth}\ \emph {et~al.}(2018)\citenamefont {Jungwirth}, \citenamefont {Sinova}, \citenamefont {Manchon}, \citenamefont {Marti}, \citenamefont {Wunderlich},\ and\ \citenamefont {Felser}}]{jungwirth2018multiple}%
  \BibitemOpen
  \bibfield  {author} {\bibinfo {author} {\bibfnamefont {T.}~\bibnamefont {Jungwirth}}, \bibinfo {author} {\bibfnamefont {J.}~\bibnamefont {Sinova}}, \bibinfo {author} {\bibfnamefont {A.}~\bibnamefont {Manchon}}, \bibinfo {author} {\bibfnamefont {X.}~\bibnamefont {Marti}}, \bibinfo {author} {\bibfnamefont {J.}~\bibnamefont {Wunderlich}},\ and\ \bibinfo {author} {\bibfnamefont {C.}~\bibnamefont {Felser}},\ }\bibfield  {title} {\bibinfo {title} {The multiple directions of antiferromagnetic spintronics},\ }\href {https://www.nature.com/articles/s41567-018-0063-6} {\bibfield  {journal} {\bibinfo  {journal} {Nat. Phys.}\ }\textbf {\bibinfo {volume} {14}},\ \bibinfo {pages} {200} (\bibinfo {year} {2018})}\BibitemShut {NoStop}%
\bibitem [{\citenamefont {{\v{S}}mejkal}\ \emph {et~al.}(2022{\natexlab{c}})\citenamefont {{\v{S}}mejkal}, \citenamefont {Sinova},\ and\ \citenamefont {Jungwirth}}]{vsmejkal2022beyond}%
  \BibitemOpen
  \bibfield  {author} {\bibinfo {author} {\bibfnamefont {L.}~\bibnamefont {{\v{S}}mejkal}}, \bibinfo {author} {\bibfnamefont {J.}~\bibnamefont {Sinova}},\ and\ \bibinfo {author} {\bibfnamefont {T.}~\bibnamefont {Jungwirth}},\ }\bibfield  {title} {\bibinfo {title} {Beyond conventional ferromagnetism and antiferromagnetism: A phase with nonrelativistic spin and crystal rotation symmetry},\ }\href {https://journals.aps.org/prx/abstract/10.1103/PhysRevX.12.031042} {\bibfield  {journal} {\bibinfo  {journal} {Phys. Rev. X}\ }\textbf {\bibinfo {volume} {12}},\ \bibinfo {pages} {031042} (\bibinfo {year} {2022}{\natexlab{c}})}\BibitemShut {NoStop}%
\bibitem [{\citenamefont {Gonz{\'a}lez~Hern{\'a}ndez}\ \emph {et~al.}(2021)\citenamefont {Gonz{\'a}lez~Hern{\'a}ndez}, \citenamefont {{\v{S}}mejkal}, \citenamefont {V{\`y}born{\`y}}, \citenamefont {Yahagi}, \citenamefont {Sinova}, \citenamefont {Jungwirth},\ and\ \citenamefont {{\v{Z}}elezn{\`y}}}]{gonzalez2021efficient}%
  \BibitemOpen
  \bibfield  {author} {\bibinfo {author} {\bibfnamefont {R.}~\bibnamefont {Gonz{\'a}lez~Hern{\'a}ndez}}, \bibinfo {author} {\bibfnamefont {L.}~\bibnamefont {{\v{S}}mejkal}}, \bibinfo {author} {\bibfnamefont {K.}~\bibnamefont {V{\`y}born{\`y}}}, \bibinfo {author} {\bibfnamefont {Y.}~\bibnamefont {Yahagi}}, \bibinfo {author} {\bibfnamefont {J.}~\bibnamefont {Sinova}}, \bibinfo {author} {\bibfnamefont {T.}~\bibnamefont {Jungwirth}},\ and\ \bibinfo {author} {\bibfnamefont {J.}~\bibnamefont {{\v{Z}}elezn{\`y}}},\ }\bibfield  {title} {\bibinfo {title} {Efficient electrical spin splitter based on nonrelativistic collinear antiferromagnetism},\ }\href {https://journals.aps.org/prl/abstract/10.1103/PhysRevLett.126.127701} {\bibfield  {journal} {\bibinfo  {journal} {Phys. Rev. Lett.}\ }\textbf {\bibinfo {volume} {126}},\ \bibinfo {pages} {127701} (\bibinfo {year} {2021})}\BibitemShut {NoStop}%
\bibitem [{\citenamefont {Bose}\ \emph {et~al.}(2022)\citenamefont {Bose}, \citenamefont {Schreiber}, \citenamefont {Jain}, \citenamefont {Shao}, \citenamefont {Nair}, \citenamefont {Sun}, \citenamefont {Zhang}, \citenamefont {Muller}, \citenamefont {Tsymbal}, \citenamefont {Schlom} \emph {et~al.}}]{bose2022tilted}%
  \BibitemOpen
  \bibfield  {author} {\bibinfo {author} {\bibfnamefont {A.}~\bibnamefont {Bose}}, \bibinfo {author} {\bibfnamefont {N.~J.}\ \bibnamefont {Schreiber}}, \bibinfo {author} {\bibfnamefont {R.}~\bibnamefont {Jain}}, \bibinfo {author} {\bibfnamefont {D.~F.}\ \bibnamefont {Shao}}, \bibinfo {author} {\bibfnamefont {H.~P.}\ \bibnamefont {Nair}}, \bibinfo {author} {\bibfnamefont {J.}~\bibnamefont {Sun}}, \bibinfo {author} {\bibfnamefont {X.~S.}\ \bibnamefont {Zhang}}, \bibinfo {author} {\bibfnamefont {D.~A.}\ \bibnamefont {Muller}}, \bibinfo {author} {\bibfnamefont {E.~Y.}\ \bibnamefont {Tsymbal}}, \bibinfo {author} {\bibfnamefont {D.~G.}\ \bibnamefont {Schlom}}, \emph {et~al.},\ }\bibfield  {title} {\bibinfo {title} {Tilted spin current generated by the collinear antiferromagnet ruthenium dioxide},\ }\href {https://www.nature.com/articles/s41928-022-00744-8} {\bibfield  {journal} {\bibinfo  {journal} {Nat. Electron.}\ }\textbf {\bibinfo {volume} {5}},\ \bibinfo {pages} {267} (\bibinfo {year} {2022})}\BibitemShut
  {NoStop}%
\bibitem [{\citenamefont {Bai}\ \emph {et~al.}(2022)\citenamefont {Bai}, \citenamefont {Han}, \citenamefont {Feng}, \citenamefont {Zhou}, \citenamefont {Su}, \citenamefont {Wang}, \citenamefont {Liao}, \citenamefont {Zhu}, \citenamefont {Chen}, \citenamefont {Pan} \emph {et~al.}}]{bai2022observation}%
  \BibitemOpen
  \bibfield  {author} {\bibinfo {author} {\bibfnamefont {H.}~\bibnamefont {Bai}}, \bibinfo {author} {\bibfnamefont {L.}~\bibnamefont {Han}}, \bibinfo {author} {\bibfnamefont {X.}~\bibnamefont {Feng}}, \bibinfo {author} {\bibfnamefont {Y.}~\bibnamefont {Zhou}}, \bibinfo {author} {\bibfnamefont {R.}~\bibnamefont {Su}}, \bibinfo {author} {\bibfnamefont {Q.}~\bibnamefont {Wang}}, \bibinfo {author} {\bibfnamefont {L.}~\bibnamefont {Liao}}, \bibinfo {author} {\bibfnamefont {W.}~\bibnamefont {Zhu}}, \bibinfo {author} {\bibfnamefont {X.}~\bibnamefont {Chen}}, \bibinfo {author} {\bibfnamefont {F.}~\bibnamefont {Pan}}, \emph {et~al.},\ }\bibfield  {title} {\bibinfo {title} {Observation of spin splitting torque in a collinear antiferromagnet $\mathrm{RuO}_{2}$},\ }\href {https://journals.aps.org/prl/abstract/10.1103/PhysRevLett.128.197202} {\bibfield  {journal} {\bibinfo  {journal} {Phys. Rev. Lett.}\ }\textbf {\bibinfo {volume} {128}},\ \bibinfo {pages} {197202} (\bibinfo {year} {2022})}\BibitemShut {NoStop}%
\bibitem [{\citenamefont {Karube}\ \emph {et~al.}(2022)\citenamefont {Karube}, \citenamefont {Tanaka}, \citenamefont {Sugawara}, \citenamefont {Kadoguchi}, \citenamefont {Kohda},\ and\ \citenamefont {Nitta}}]{karube2022observation}%
  \BibitemOpen
  \bibfield  {author} {\bibinfo {author} {\bibfnamefont {S.}~\bibnamefont {Karube}}, \bibinfo {author} {\bibfnamefont {T.}~\bibnamefont {Tanaka}}, \bibinfo {author} {\bibfnamefont {D.}~\bibnamefont {Sugawara}}, \bibinfo {author} {\bibfnamefont {N.}~\bibnamefont {Kadoguchi}}, \bibinfo {author} {\bibfnamefont {M.}~\bibnamefont {Kohda}},\ and\ \bibinfo {author} {\bibfnamefont {J.}~\bibnamefont {Nitta}},\ }\bibfield  {title} {\bibinfo {title} {Observation of spin-splitter torque in collinear antiferromagnetic $\mathrm{RuO}_{2}$},\ }\href {https://journals.aps.org/prl/abstract/10.1103/PhysRevLett.129.137201} {\bibfield  {journal} {\bibinfo  {journal} {Phys. Rev. Lett.}\ }\textbf {\bibinfo {volume} {129}},\ \bibinfo {pages} {137201} (\bibinfo {year} {2022})}\BibitemShut {NoStop}%
\bibitem [{\citenamefont {Naka}\ \emph {et~al.}(2019)\citenamefont {Naka}, \citenamefont {Hayami}, \citenamefont {Kusunose}, \citenamefont {Yanagi}, \citenamefont {Motome},\ and\ \citenamefont {Seo}}]{naka2019spin}%
  \BibitemOpen
  \bibfield  {author} {\bibinfo {author} {\bibfnamefont {M.}~\bibnamefont {Naka}}, \bibinfo {author} {\bibfnamefont {S.}~\bibnamefont {Hayami}}, \bibinfo {author} {\bibfnamefont {H.}~\bibnamefont {Kusunose}}, \bibinfo {author} {\bibfnamefont {Y.}~\bibnamefont {Yanagi}}, \bibinfo {author} {\bibfnamefont {Y.}~\bibnamefont {Motome}},\ and\ \bibinfo {author} {\bibfnamefont {H.}~\bibnamefont {Seo}},\ }\bibfield  {title} {\bibinfo {title} {Spin current generation in organic antiferromagnets},\ }\href {https://www.nature.com/articles/s41467-019-12229-y} {\bibfield  {journal} {\bibinfo  {journal} {Nat. Commun.}\ }\textbf {\bibinfo {volume} {10}},\ \bibinfo {pages} {4305} (\bibinfo {year} {2019})}\BibitemShut {NoStop}%
\bibitem [{\citenamefont {Naka}\ \emph {et~al.}(2021)\citenamefont {Naka}, \citenamefont {Motome},\ and\ \citenamefont {Seo}}]{naka2021perovskite}%
  \BibitemOpen
  \bibfield  {author} {\bibinfo {author} {\bibfnamefont {M.}~\bibnamefont {Naka}}, \bibinfo {author} {\bibfnamefont {Y.}~\bibnamefont {Motome}},\ and\ \bibinfo {author} {\bibfnamefont {H.}~\bibnamefont {Seo}},\ }\bibfield  {title} {\bibinfo {title} {Perovskite as a spin current generator},\ }\href {https://journals.aps.org/prb/abstract/10.1103/PhysRevB.103.125114} {\bibfield  {journal} {\bibinfo  {journal} {Phys. Rev. B}\ }\textbf {\bibinfo {volume} {103}},\ \bibinfo {pages} {125114} (\bibinfo {year} {2021})}\BibitemShut {NoStop}%
\bibitem [{\citenamefont {Ezawa}(2025)}]{ezawa2025third}%
  \BibitemOpen
  \bibfield  {author} {\bibinfo {author} {\bibfnamefont {M.}~\bibnamefont {Ezawa}},\ }\bibfield  {title} {\bibinfo {title} {Third-order and fifth-order nonlinear spin-current generation in $g$-wave and $i$-wave altermagnets and perfectly nonreciprocal spin current in $f$-wave magnets},\ }\href {https://journals.aps.org/prb/abstract/10.1103/PhysRevB.111.125420} {\bibfield  {journal} {\bibinfo  {journal} {Phys. Rev. B}\ }\textbf {\bibinfo {volume} {111}},\ \bibinfo {pages} {125420} (\bibinfo {year} {2025})}\BibitemShut {NoStop}%
\bibitem [{\citenamefont {Belashchenko}(2025)}]{belashchenko2025giant}%
  \BibitemOpen
  \bibfield  {author} {\bibinfo {author} {\bibfnamefont {K.}~\bibnamefont {Belashchenko}},\ }\bibfield  {title} {\bibinfo {title} {Giant strain-induced spin splitting effect in {MnTe}, a g-wave altermagnetic semiconductor},\ }\href {https://journals.aps.org/prl/abstract/10.1103/PhysRevLett.134.086701} {\bibfield  {journal} {\bibinfo  {journal} {Phys. Rev. Lett.}\ }\textbf {\bibinfo {volume} {134}},\ \bibinfo {pages} {086701} (\bibinfo {year} {2025})}\BibitemShut {NoStop}%
\bibitem [{\citenamefont {Roig}\ \emph {et~al.}(2024)\citenamefont {Roig}, \citenamefont {Kreisel}, \citenamefont {Yu}, \citenamefont {Andersen},\ and\ \citenamefont {Agterberg}}]{roig2024minimal}%
  \BibitemOpen
  \bibfield  {author} {\bibinfo {author} {\bibfnamefont {M.}~\bibnamefont {Roig}}, \bibinfo {author} {\bibfnamefont {A.}~\bibnamefont {Kreisel}}, \bibinfo {author} {\bibfnamefont {Y.}~\bibnamefont {Yu}}, \bibinfo {author} {\bibfnamefont {B.~M.}\ \bibnamefont {Andersen}},\ and\ \bibinfo {author} {\bibfnamefont {D.~F.}\ \bibnamefont {Agterberg}},\ }\bibfield  {title} {\bibinfo {title} {Minimal models for altermagnetism},\ }\href {https://journals.aps.org/prb/abstract/10.1103/PhysRevB.110.144412} {\bibfield  {journal} {\bibinfo  {journal} {Phys. Rev. B}\ }\textbf {\bibinfo {volume} {110}},\ \bibinfo {pages} {144412} (\bibinfo {year} {2024})}\BibitemShut {NoStop}%
\bibitem [{\citenamefont {Voon}\ and\ \citenamefont {Willatzen}(2009)}]{voon2009kp}%
  \BibitemOpen
  \bibfield  {author} {\bibinfo {author} {\bibfnamefont {L.~C. L.~Y.}\ \bibnamefont {Voon}}\ and\ \bibinfo {author} {\bibfnamefont {M.}~\bibnamefont {Willatzen}},\ }\href {https://link.springer.com/book/10.1007/978-3-540-92872-0} {\emph {\bibinfo {title} {The kp method: electronic properties of semiconductors}}}\ (\bibinfo  {publisher} {Springer Science \& Business Media},\ \bibinfo {year} {2009})\BibitemShut {NoStop}%
\bibitem [{\citenamefont {Ma}\ \emph {et~al.}(2021)\citenamefont {Ma}, \citenamefont {Hu}, \citenamefont {Li}, \citenamefont {Liu}, \citenamefont {Yao}, \citenamefont {Jia},\ and\ \citenamefont {Liu}}]{ma2021multifunctional}%
  \BibitemOpen
  \bibfield  {author} {\bibinfo {author} {\bibfnamefont {H.~Y.}\ \bibnamefont {Ma}}, \bibinfo {author} {\bibfnamefont {M.}~\bibnamefont {Hu}}, \bibinfo {author} {\bibfnamefont {N.}~\bibnamefont {Li}}, \bibinfo {author} {\bibfnamefont {J.}~\bibnamefont {Liu}}, \bibinfo {author} {\bibfnamefont {W.}~\bibnamefont {Yao}}, \bibinfo {author} {\bibfnamefont {J.-F.}\ \bibnamefont {Jia}},\ and\ \bibinfo {author} {\bibfnamefont {J.}~\bibnamefont {Liu}},\ }\bibfield  {title} {\bibinfo {title} {Multifunctional antiferromagnetic materials with giant piezomagnetism and noncollinear spin current},\ }\href {https://www.nature.com/articles/s41467-021-23127-7} {\bibfield  {journal} {\bibinfo  {journal} {Nat. Commun.}\ }\textbf {\bibinfo {volume} {12}},\ \bibinfo {pages} {2846} (\bibinfo {year} {2021})}\BibitemShut {NoStop}%
\bibitem [{\citenamefont {Zhu}\ \emph {et~al.}(2023)\citenamefont {Zhu}, \citenamefont {Chen}, \citenamefont {Li}, \citenamefont {Qiao}, \citenamefont {Ma}, \citenamefont {Liu}, \citenamefont {Hu}, \citenamefont {Gao},\ and\ \citenamefont {Ren}}]{zhu2023multipiezo}%
  \BibitemOpen
  \bibfield  {author} {\bibinfo {author} {\bibfnamefont {Y.}~\bibnamefont {Zhu}}, \bibinfo {author} {\bibfnamefont {T.}~\bibnamefont {Chen}}, \bibinfo {author} {\bibfnamefont {Y.}~\bibnamefont {Li}}, \bibinfo {author} {\bibfnamefont {L.}~\bibnamefont {Qiao}}, \bibinfo {author} {\bibfnamefont {X.}~\bibnamefont {Ma}}, \bibinfo {author} {\bibfnamefont {C.}~\bibnamefont {Liu}}, \bibinfo {author} {\bibfnamefont {T.}~\bibnamefont {Hu}}, \bibinfo {author} {\bibfnamefont {H.}~\bibnamefont {Gao}},\ and\ \bibinfo {author} {\bibfnamefont {W.}~\bibnamefont {Ren}},\ }\bibfield  {title} {\bibinfo {title} {Multipiezo effect in altermagnetic $\mathrm{V_2SeTeO}$ monolayer},\ }\href {https://pubs.acs.org/doi/full/10.1021/acs.nanolett.3c04330} {\bibfield  {journal} {\bibinfo  {journal} {Nano Lett.}\ }\textbf {\bibinfo {volume} {24}},\ \bibinfo {pages} {472} (\bibinfo {year} {2023})}\BibitemShut {NoStop}%
\bibitem [{\citenamefont {Wu}\ \emph {et~al.}(2024)\citenamefont {Wu}, \citenamefont {Deng}, \citenamefont {Yin}, \citenamefont {Tong}, \citenamefont {Tian},\ and\ \citenamefont {Zhang}}]{wu2024valley}%
  \BibitemOpen
  \bibfield  {author} {\bibinfo {author} {\bibfnamefont {Y.}~\bibnamefont {Wu}}, \bibinfo {author} {\bibfnamefont {L.}~\bibnamefont {Deng}}, \bibinfo {author} {\bibfnamefont {X.}~\bibnamefont {Yin}}, \bibinfo {author} {\bibfnamefont {J.}~\bibnamefont {Tong}}, \bibinfo {author} {\bibfnamefont {F.}~\bibnamefont {Tian}},\ and\ \bibinfo {author} {\bibfnamefont {X.}~\bibnamefont {Zhang}},\ }\bibfield  {title} {\bibinfo {title} {Valley-related multipiezo effect and noncollinear spin current in an altermagnet $\mathrm{Fe_2Se_2O}$ monolayer},\ }\href {https://pubs.acs.org/doi/full/10.1021/acs.nanolett.4c02554} {\bibfield  {journal} {\bibinfo  {journal} {Nano Lett.}\ }\textbf {\bibinfo {volume} {24}},\ \bibinfo {pages} {10534} (\bibinfo {year} {2024})}\BibitemShut {NoStop}%
\bibitem [{\citenamefont {Ma}\ \emph {et~al.}(2025)\citenamefont {Ma}, \citenamefont {Wang}, \citenamefont {Yang},\ and\ \citenamefont {Liu}}]{yu2024typical}%
  \BibitemOpen
  \bibfield  {author} {\bibinfo {author} {\bibfnamefont {Q.}~\bibnamefont {Ma}}, \bibinfo {author} {\bibfnamefont {B.}~\bibnamefont {Wang}}, \bibinfo {author} {\bibfnamefont {G.}~\bibnamefont {Yang}},\ and\ \bibinfo {author} {\bibfnamefont {Y.}~\bibnamefont {Liu}},\ }\bibfield  {title} {\bibinfo {title} {Multifunctional altermagnet with large out-of-plane piezoelectric response in janus $\mathrm{V_2AsBrO}$ monolayer},\ }\href {https://arxiv.org/abs/2502.10055} {\bibfield  {journal} {\bibinfo  {journal} {arXiv:2502.10055}\ } (\bibinfo {year} {2025})}\BibitemShut {NoStop}%
\bibitem [{\citenamefont {Guo}\ \emph {et~al.}(2023{\natexlab{a}})\citenamefont {Guo}, \citenamefont {Guo}, \citenamefont {Cheng}, \citenamefont {Wang},\ and\ \citenamefont {Ang}}]{guo2023piezoelectric}%
  \BibitemOpen
  \bibfield  {author} {\bibinfo {author} {\bibfnamefont {S.~D.}\ \bibnamefont {Guo}}, \bibinfo {author} {\bibfnamefont {X.~S.}\ \bibnamefont {Guo}}, \bibinfo {author} {\bibfnamefont {K.}~\bibnamefont {Cheng}}, \bibinfo {author} {\bibfnamefont {K.}~\bibnamefont {Wang}},\ and\ \bibinfo {author} {\bibfnamefont {Y.~S.}\ \bibnamefont {Ang}},\ }\bibfield  {title} {\bibinfo {title} {Piezoelectric altermagnetism and spin-valley polarization in janus monolayer $\mathrm{Cr_2SO}$},\ }\href {https://pubs.aip.org/aip/apl/article/123/8/082401/2907421} {\bibfield  {journal} {\bibinfo  {journal} {Appl. Phys. Lett.}\ }\textbf {\bibinfo {volume} {123}},\ \bibinfo {pages} {082401} (\bibinfo {year} {2023}{\natexlab{a}})}\BibitemShut {NoStop}%
\bibitem [{\citenamefont {Xie}\ \emph {et~al.}(2025)\citenamefont {Xie}, \citenamefont {Xu}, \citenamefont {Yue}, \citenamefont {Xia},\ and\ \citenamefont {Wang}}]{xie2025piezovalley}%
  \BibitemOpen
  \bibfield  {author} {\bibinfo {author} {\bibfnamefont {W.}~\bibnamefont {Xie}}, \bibinfo {author} {\bibfnamefont {X.}~\bibnamefont {Xu}}, \bibinfo {author} {\bibfnamefont {Y.}~\bibnamefont {Yue}}, \bibinfo {author} {\bibfnamefont {H.}~\bibnamefont {Xia}},\ and\ \bibinfo {author} {\bibfnamefont {H.}~\bibnamefont {Wang}},\ }\bibfield  {title} {\bibinfo {title} {Piezovalley effect and magnetovalley coupling in altermagnetic semiconductors studied by first-principles calculations},\ }\href {https://journals.aps.org/prb/abstract/10.1103/PhysRevB.111.134429} {\bibfield  {journal} {\bibinfo  {journal} {Phys. Rev. B}\ }\textbf {\bibinfo {volume} {111}},\ \bibinfo {pages} {134429} (\bibinfo {year} {2025})}\BibitemShut {NoStop}%
\bibitem [{\citenamefont {Li}\ \emph {et~al.}(2024)\citenamefont {Li}, \citenamefont {Fan}, \citenamefont {Wang}, \citenamefont {Zhang},\ and\ \citenamefont {Li}}]{li2024strain}%
  \BibitemOpen
  \bibfield  {author} {\bibinfo {author} {\bibfnamefont {J.~Y.}\ \bibnamefont {Li}}, \bibinfo {author} {\bibfnamefont {A.~D.}\ \bibnamefont {Fan}}, \bibinfo {author} {\bibfnamefont {Y.~K.}\ \bibnamefont {Wang}}, \bibinfo {author} {\bibfnamefont {Y.}~\bibnamefont {Zhang}},\ and\ \bibinfo {author} {\bibfnamefont {S.}~\bibnamefont {Li}},\ }\bibfield  {title} {\bibinfo {title} {Strain-induced valley polarization, topological states, and piezomagnetism in two-dimensional altermagnetic $\mathrm{V_2Te_2O}$, $\mathrm{V_2STeO}$, $\mathrm{V_2SSeO}$, and $\mathrm{V_2S_2O}$},\ }\href {https://pubs.aip.org/aip/apl/article/125/22/222404/3322459} {\bibfield  {journal} {\bibinfo  {journal} {Appl. Phys. Lett.}\ }\textbf {\bibinfo {volume} {125}},\ \bibinfo {pages} {222404} (\bibinfo {year} {2024})}\BibitemShut {NoStop}%
\bibitem [{\citenamefont {Chen}\ \emph {et~al.}(2024{\natexlab{a}})\citenamefont {Chen}, \citenamefont {He}, \citenamefont {Xiong}, \citenamefont {Quan}, \citenamefont {Hou}, \citenamefont {Ji}, \citenamefont {Yang},\ and\ \citenamefont {Li}}]{chen2024strain}%
  \BibitemOpen
  \bibfield  {author} {\bibinfo {author} {\bibfnamefont {C.}~\bibnamefont {Chen}}, \bibinfo {author} {\bibfnamefont {X.}~\bibnamefont {He}}, \bibinfo {author} {\bibfnamefont {Q.}~\bibnamefont {Xiong}}, \bibinfo {author} {\bibfnamefont {C.}~\bibnamefont {Quan}}, \bibinfo {author} {\bibfnamefont {H.}~\bibnamefont {Hou}}, \bibinfo {author} {\bibfnamefont {S.}~\bibnamefont {Ji}}, \bibinfo {author} {\bibfnamefont {J.}~\bibnamefont {Yang}},\ and\ \bibinfo {author} {\bibfnamefont {X.}~\bibnamefont {Li}},\ }\bibfield  {title} {\bibinfo {title} {Strain-modulated valley polarization and piezomagnetic effects in altermagnetic $\mathrm{Cr_2S_2}$},\ }\href {https://arxiv.org/abs/2410.17686} {\bibfield  {journal} {\bibinfo  {journal} {arXiv:2410.17686}\ } (\bibinfo {year} {2024}{\natexlab{a}})}\BibitemShut {NoStop}%
\bibitem [{\citenamefont {Xun}\ \emph {et~al.}(2025)\citenamefont {Xun}, \citenamefont {Liu}, \citenamefont {Zhang}, \citenamefont {Wu},\ and\ \citenamefont {Li}}]{xun2025stacking}%
  \BibitemOpen
  \bibfield  {author} {\bibinfo {author} {\bibfnamefont {W.}~\bibnamefont {Xun}}, \bibinfo {author} {\bibfnamefont {X.}~\bibnamefont {Liu}}, \bibinfo {author} {\bibfnamefont {Y.}~\bibnamefont {Zhang}}, \bibinfo {author} {\bibfnamefont {Y.~Z.}\ \bibnamefont {Wu}},\ and\ \bibinfo {author} {\bibfnamefont {P.}~\bibnamefont {Li}},\ }\bibfield  {title} {\bibinfo {title} {Stacking-, strain-engineering induced altermagnetism, multipiezo effect, and topological state in two-dimensional materials},\ }\href {https://pubs.aip.org/aip/apl/article/126/16/161903/3344954} {\bibfield  {journal} {\bibinfo  {journal} {Appl. Phys. Lett.}\ }\textbf {\bibinfo {volume} {126}} (\bibinfo {year} {2025})}\BibitemShut {NoStop}%
\bibitem [{\citenamefont {Hu}\ \emph {et~al.}(2025{\natexlab{a}})\citenamefont {Hu}, \citenamefont {Zhao}, \citenamefont {Xia}, \citenamefont {Sun}, \citenamefont {Wu}, \citenamefont {Wu},\ and\ \citenamefont {Li}}]{hu2025valley}%
  \BibitemOpen
  \bibfield  {author} {\bibinfo {author} {\bibfnamefont {X.}~\bibnamefont {Hu}}, \bibinfo {author} {\bibfnamefont {W.}~\bibnamefont {Zhao}}, \bibinfo {author} {\bibfnamefont {W.}~\bibnamefont {Xia}}, \bibinfo {author} {\bibfnamefont {H.}~\bibnamefont {Sun}}, \bibinfo {author} {\bibfnamefont {C.}~\bibnamefont {Wu}}, \bibinfo {author} {\bibfnamefont {Y.-Z.}\ \bibnamefont {Wu}},\ and\ \bibinfo {author} {\bibfnamefont {P.}~\bibnamefont {Li}},\ }\bibfield  {title} {\bibinfo {title} {Valley polarization and anomalous valley hall effect in altermagnet $\mathrm{Ti_2Se_2S}$ with multipiezo properties},\ }\href {https://arxiv.org/abs/2504.18308} {\bibfield  {journal} {\bibinfo  {journal} {arXiv:2504.18308}\ } (\bibinfo {year} {2025}{\natexlab{a}})}\BibitemShut {NoStop}%
\bibitem [{\citenamefont {Guo}\ \emph {et~al.}(2023{\natexlab{b}})\citenamefont {Guo}, \citenamefont {Liu},\ and\ \citenamefont {Lu}}]{guo2023quantum}%
  \BibitemOpen
  \bibfield  {author} {\bibinfo {author} {\bibfnamefont {P.~J.}\ \bibnamefont {Guo}}, \bibinfo {author} {\bibfnamefont {Z.~X.}\ \bibnamefont {Liu}},\ and\ \bibinfo {author} {\bibfnamefont {Z.~Y.}\ \bibnamefont {Lu}},\ }\bibfield  {title} {\bibinfo {title} {Quantum anomalous hall effect in collinear antiferromagnetism},\ }\href {https://www.nature.com/articles/s41524-023-01025-4} {\bibfield  {journal} {\bibinfo  {journal} {Npj Comput. Mater.}\ }\textbf {\bibinfo {volume} {9}},\ \bibinfo {pages} {70} (\bibinfo {year} {2023}{\natexlab{b}})}\BibitemShut {NoStop}%
\bibitem [{\citenamefont {Tian}\ \emph {et~al.}(2025)\citenamefont {Tian}, \citenamefont {Li}, \citenamefont {Liu}, \citenamefont {Li}, \citenamefont {Liu}, \citenamefont {Li}, \citenamefont {Li}, \citenamefont {Liu},\ and\ \citenamefont {Shi}}]{tian2025spin}%
  \BibitemOpen
  \bibfield  {author} {\bibinfo {author} {\bibfnamefont {J.}~\bibnamefont {Tian}}, \bibinfo {author} {\bibfnamefont {J.}~\bibnamefont {Li}}, \bibinfo {author} {\bibfnamefont {H.}~\bibnamefont {Liu}}, \bibinfo {author} {\bibfnamefont {Y.}~\bibnamefont {Li}}, \bibinfo {author} {\bibfnamefont {Z.}~\bibnamefont {Liu}}, \bibinfo {author} {\bibfnamefont {L.}~\bibnamefont {Li}}, \bibinfo {author} {\bibfnamefont {J.}~\bibnamefont {Li}}, \bibinfo {author} {\bibfnamefont {G.}~\bibnamefont {Liu}},\ and\ \bibinfo {author} {\bibfnamefont {J.}~\bibnamefont {Shi}},\ }\bibfield  {title} {\bibinfo {title} {Spin-layer coupling in an altermagnetic multilayer: A design principle for spintronics},\ }\href {https://journals.aps.org/prb/abstract/10.1103/PhysRevB.111.035437} {\bibfield  {journal} {\bibinfo  {journal} {Phys. Rev. B}\ }\textbf {\bibinfo {volume} {111}},\ \bibinfo {pages} {035437} (\bibinfo {year} {2025})}\BibitemShut {NoStop}%
\bibitem [{\citenamefont {Fernandes}\ \emph {et~al.}(2024)\citenamefont {Fernandes}, \citenamefont {De~Carvalho}, \citenamefont {Birol},\ and\ \citenamefont {Pereira}}]{fernandes2024topological}%
  \BibitemOpen
  \bibfield  {author} {\bibinfo {author} {\bibfnamefont {R.~M.}\ \bibnamefont {Fernandes}}, \bibinfo {author} {\bibfnamefont {V.~S.}\ \bibnamefont {De~Carvalho}}, \bibinfo {author} {\bibfnamefont {T.}~\bibnamefont {Birol}},\ and\ \bibinfo {author} {\bibfnamefont {R.~G.}\ \bibnamefont {Pereira}},\ }\bibfield  {title} {\bibinfo {title} {Topological transition from nodal to nodeless zeeman splitting in altermagnets},\ }\href {https://journals.aps.org/prb/abstract/10.1103/PhysRevB.109.024404} {\bibfield  {journal} {\bibinfo  {journal} {Phys. Rev. B}\ }\textbf {\bibinfo {volume} {109}},\ \bibinfo {pages} {024404} (\bibinfo {year} {2024})}\BibitemShut {NoStop}%
\bibitem [{\citenamefont {Hu}\ \emph {et~al.}(2025{\natexlab{b}})\citenamefont {Hu}, \citenamefont {Cheng}, \citenamefont {Huang},\ and\ \citenamefont {Liu}}]{hu2025catalog}%
  \BibitemOpen
  \bibfield  {author} {\bibinfo {author} {\bibfnamefont {M.}~\bibnamefont {Hu}}, \bibinfo {author} {\bibfnamefont {X.}~\bibnamefont {Cheng}}, \bibinfo {author} {\bibfnamefont {Z.}~\bibnamefont {Huang}},\ and\ \bibinfo {author} {\bibfnamefont {J.}~\bibnamefont {Liu}},\ }\bibfield  {title} {\bibinfo {title} {Catalog of c-paired spin-momentum locking in antiferromagnetic systems},\ }\href {https://journals.aps.org/prx/abstract/10.1103/PhysRevX.15.021083} {\bibfield  {journal} {\bibinfo  {journal} {Phys. Rev. X}\ }\textbf {\bibinfo {volume} {15}},\ \bibinfo {pages} {021083} (\bibinfo {year} {2025}{\natexlab{b}})}\BibitemShut {NoStop}%
\bibitem [{\citenamefont {Chang}\ \emph {et~al.}(2025)\citenamefont {Chang}, \citenamefont {Wu}, \citenamefont {Deng}, \citenamefont {Yin},\ and\ \citenamefont {Zhang}}]{chang2025valley}%
  \BibitemOpen
  \bibfield  {author} {\bibinfo {author} {\bibfnamefont {Y.}~\bibnamefont {Chang}}, \bibinfo {author} {\bibfnamefont {Y.}~\bibnamefont {Wu}}, \bibinfo {author} {\bibfnamefont {L.}~\bibnamefont {Deng}}, \bibinfo {author} {\bibfnamefont {X.}~\bibnamefont {Yin}},\ and\ \bibinfo {author} {\bibfnamefont {X.}~\bibnamefont {Zhang}},\ }\bibfield  {title} {\bibinfo {title} {Valley-related multipiezo effect in altermagnet monolayer $\mathrm{V_2STeO}$},\ }\href {https://www.mdpi.com/1996-1944/18/3/527} {\bibfield  {journal} {\bibinfo  {journal} {Materials}\ }\textbf {\bibinfo {volume} {18}},\ \bibinfo {pages} {527} (\bibinfo {year} {2025})}\BibitemShut {NoStop}%
\bibitem [{\citenamefont {Kresse}\ and\ \citenamefont {Furthm{\"u}ller}(1996)}]{kresse1996efficient}%
  \BibitemOpen
  \bibfield  {author} {\bibinfo {author} {\bibfnamefont {G.}~\bibnamefont {Kresse}}\ and\ \bibinfo {author} {\bibfnamefont {J.}~\bibnamefont {Furthm{\"u}ller}},\ }\bibfield  {title} {\bibinfo {title} {Efficient iterative schemes for ab initio total-energy calculations using a plane-wave basis set},\ }\href {https://journals.aps.org/prb/abstract/10.1103/PhysRevB.54.11169} {\bibfield  {journal} {\bibinfo  {journal} {Phys. Rev. B}\ }\textbf {\bibinfo {volume} {54}},\ \bibinfo {pages} {11169} (\bibinfo {year} {1996})}\BibitemShut {NoStop}%
\bibitem [{\citenamefont {Kresse}\ and\ \citenamefont {Joubert}(1999)}]{kresse1999ultrasoft}%
  \BibitemOpen
  \bibfield  {author} {\bibinfo {author} {\bibfnamefont {G.}~\bibnamefont {Kresse}}\ and\ \bibinfo {author} {\bibfnamefont {D.}~\bibnamefont {Joubert}},\ }\bibfield  {title} {\bibinfo {title} {From ultrasoft pseudopotentials to the projector augmented-wave method},\ }\href {https://journals.aps.org/prb/abstract/10.1103/PhysRevB.59.1758} {\bibfield  {journal} {\bibinfo  {journal} {Phys. Rev. B}\ }\textbf {\bibinfo {volume} {59}},\ \bibinfo {pages} {1758} (\bibinfo {year} {1999})}\BibitemShut {NoStop}%
\bibitem [{\citenamefont {Perdew}\ \emph {et~al.}(1996)\citenamefont {Perdew}, \citenamefont {Burke},\ and\ \citenamefont {Ernzerhof}}]{perdew1996generalized}%
  \BibitemOpen
  \bibfield  {author} {\bibinfo {author} {\bibfnamefont {J.~P.}\ \bibnamefont {Perdew}}, \bibinfo {author} {\bibfnamefont {K.}~\bibnamefont {Burke}},\ and\ \bibinfo {author} {\bibfnamefont {M.}~\bibnamefont {Ernzerhof}},\ }\bibfield  {title} {\bibinfo {title} {Generalized gradient approximation made simple},\ }\href {https://journals.aps.org/prl/abstract/10.1103/PhysRevLett.77.3865} {\bibfield  {journal} {\bibinfo  {journal} {Phys. Rev. Lett.}\ }\textbf {\bibinfo {volume} {77}},\ \bibinfo {pages} {3865} (\bibinfo {year} {1996})}\BibitemShut {NoStop}%
\bibitem [{\citenamefont {Bl{\"o}chl}(1994)}]{blochl1994projector}%
  \BibitemOpen
  \bibfield  {author} {\bibinfo {author} {\bibfnamefont {P.~E.}\ \bibnamefont {Bl{\"o}chl}},\ }\bibfield  {title} {\bibinfo {title} {Projector augmented-wave method},\ }\href {https://journals.aps.org/prb/abstract/10.1103/PhysRevB.50.17953} {\bibfield  {journal} {\bibinfo  {journal} {Phys. Rev. B}\ }\textbf {\bibinfo {volume} {50}},\ \bibinfo {pages} {17953} (\bibinfo {year} {1994})}\BibitemShut {NoStop}%
\bibitem [{\citenamefont {Dudarev}\ \emph {et~al.}(1998)\citenamefont {Dudarev}, \citenamefont {Botton}, \citenamefont {Savrasov}, \citenamefont {Humphreys},\ and\ \citenamefont {Sutton}}]{dudarev1998electron}%
  \BibitemOpen
  \bibfield  {author} {\bibinfo {author} {\bibfnamefont {S.~L.}\ \bibnamefont {Dudarev}}, \bibinfo {author} {\bibfnamefont {G.~A.}\ \bibnamefont {Botton}}, \bibinfo {author} {\bibfnamefont {S.~Y.}\ \bibnamefont {Savrasov}}, \bibinfo {author} {\bibfnamefont {C.}~\bibnamefont {Humphreys}},\ and\ \bibinfo {author} {\bibfnamefont {A.~P.}\ \bibnamefont {Sutton}},\ }\bibfield  {title} {\bibinfo {title} {Electron-energy-loss spectra and the structural stability of nickel oxide: An lsda+ u study},\ }\href {https://journals.aps.org/prb/abstract/10.1103/PhysRevB.57.1505} {\bibfield  {journal} {\bibinfo  {journal} {Phys. Rev. B}\ }\textbf {\bibinfo {volume} {57}},\ \bibinfo {pages} {1505} (\bibinfo {year} {1998})}\BibitemShut {NoStop}%
\bibitem [{\citenamefont {Wei}\ \emph {et~al.}(2025)\citenamefont {Wei}, \citenamefont {Li}, \citenamefont {Hatt}, \citenamefont {Huai}, \citenamefont {Liu}, \citenamefont {Singh}, \citenamefont {Kim}, \citenamefont {Fernandes}, \citenamefont {Cardon}, \citenamefont {Zhao} \emph {et~al.}}]{wei20252}%
  \BibitemOpen
  \bibfield  {author} {\bibinfo {author} {\bibfnamefont {C.~C.}\ \bibnamefont {Wei}}, \bibinfo {author} {\bibfnamefont {X.}~\bibnamefont {Li}}, \bibinfo {author} {\bibfnamefont {S.}~\bibnamefont {Hatt}}, \bibinfo {author} {\bibfnamefont {X.}~\bibnamefont {Huai}}, \bibinfo {author} {\bibfnamefont {J.}~\bibnamefont {Liu}}, \bibinfo {author} {\bibfnamefont {B.}~\bibnamefont {Singh}}, \bibinfo {author} {\bibfnamefont {K.~M.}\ \bibnamefont {Kim}}, \bibinfo {author} {\bibfnamefont {R.~M.}\ \bibnamefont {Fernandes}}, \bibinfo {author} {\bibfnamefont {P.}~\bibnamefont {Cardon}}, \bibinfo {author} {\bibfnamefont {L.}~\bibnamefont {Zhao}}, \emph {et~al.},\ }\bibfield  {title} {\bibinfo {title} {$\mathrm{La}_{2}\mathrm{O}_{3}\mathrm{Mn}_{2}\mathrm{Se}_{2}$: A correlated insulating layered d-wave altermagnet},\ }\href {https://journals.aps.org/prmaterials/abstract/10.1103/PhysRevMaterials.9.024402} {\bibfield  {journal} {\bibinfo  {journal} {Phys. Rev. Mater.}\ }\textbf {\bibinfo {volume} {9}},\ \bibinfo {pages} {024402}
  (\bibinfo {year} {2025})}\BibitemShut {NoStop}%
\bibitem [{\citenamefont {Lin}\ \emph {et~al.}(2017)\citenamefont {Lin}, \citenamefont {Xu}, \citenamefont {Zhang}, \citenamefont {Zhang}, \citenamefont {Liang},\ and\ \citenamefont {Dong}}]{lin2017ferroelectric}%
  \BibitemOpen
  \bibfield  {author} {\bibinfo {author} {\bibfnamefont {L.~F.}\ \bibnamefont {Lin}}, \bibinfo {author} {\bibfnamefont {Q.~R.}\ \bibnamefont {Xu}}, \bibinfo {author} {\bibfnamefont {Y.}~\bibnamefont {Zhang}}, \bibinfo {author} {\bibfnamefont {J.~J.}\ \bibnamefont {Zhang}}, \bibinfo {author} {\bibfnamefont {Y.~P.}\ \bibnamefont {Liang}},\ and\ \bibinfo {author} {\bibfnamefont {S.}~\bibnamefont {Dong}},\ }\bibfield  {title} {\bibinfo {title} {Ferroelectric ferrimagnetic $\mathrm{Li}\mathrm{Fe}_{2}\mathrm{F}_{6}$: Charge-ordering-mediated magnetoelectricity},\ }\href {https://journals.aps.org/prmaterials/abstract/10.1103/PhysRevMaterials.1.071401} {\bibfield  {journal} {\bibinfo  {journal} {Phys. Rev. Mater.}\ }\textbf {\bibinfo {volume} {1}},\ \bibinfo {pages} {071401} (\bibinfo {year} {2017})}\BibitemShut {NoStop}%
\bibitem [{\citenamefont {Wang}\ \emph {et~al.}(2021)\citenamefont {Wang}, \citenamefont {Xu}, \citenamefont {Liu}, \citenamefont {Tang},\ and\ \citenamefont {Geng}}]{wang2021vaspkit}%
  \BibitemOpen
  \bibfield  {author} {\bibinfo {author} {\bibfnamefont {V.}~\bibnamefont {Wang}}, \bibinfo {author} {\bibfnamefont {N.}~\bibnamefont {Xu}}, \bibinfo {author} {\bibfnamefont {J.~C.}\ \bibnamefont {Liu}}, \bibinfo {author} {\bibfnamefont {G.}~\bibnamefont {Tang}},\ and\ \bibinfo {author} {\bibfnamefont {W.~T.}\ \bibnamefont {Geng}},\ }\bibfield  {title} {\bibinfo {title} {Vaspkit: A user-friendly interface facilitating high-throughput computing and analysis using vasp code},\ }\href {https://www.sciencedirect.com/science/article/pii/S0010465521001454?via%3Dihub} {\bibfield  {journal} {\bibinfo  {journal} {Comput. Phys. Commun.}\ }\textbf {\bibinfo {volume} {267}},\ \bibinfo {pages} {108033} (\bibinfo {year} {2021})}\BibitemShut {NoStop}%
\bibitem [{\citenamefont {VASPKIT}()}]{vaspkit}%
  \BibitemOpen
  \bibfield  {author} {\bibinfo {author} {\bibnamefont {VASPKIT}},\ }\href@noop {} {}\bibinfo {note} {{\color{blue} \href{https://vaspkit.com}{https://vaspkit.com}}}\BibitemShut {NoStop}%
\bibitem [{\citenamefont {Momma}\ and\ \citenamefont {Izumi}(2011)}]{momma2011vesta}%
  \BibitemOpen
  \bibfield  {author} {\bibinfo {author} {\bibfnamefont {K.}~\bibnamefont {Momma}}\ and\ \bibinfo {author} {\bibfnamefont {F.}~\bibnamefont {Izumi}},\ }\bibfield  {title} {\bibinfo {title} {Vesta 3 for three-dimensional visualization of crystal, volumetric and morphology data},\ }\href {https://journals.iucr.org/paper?S0021889811038970} {\bibfield  {journal} {\bibinfo  {journal} {J Appl. Crystallogr.}\ }\textbf {\bibinfo {volume} {44}},\ \bibinfo {pages} {1272} (\bibinfo {year} {2011})}\BibitemShut {NoStop}%
\bibitem [{\citenamefont {See}()}]{mathematica}%
  \BibitemOpen
  \bibfield  {author} {\bibinfo {author} {\bibnamefont {See}},\ }\href@noop {} {}\bibinfo {note} {{\color{blue} \href{https://www.wolfram.com/mathematica}{https://www.wolfram.com/mathematica}}}\BibitemShut {NoStop}%
\bibitem [{\citenamefont {Aroyo}\ \emph {et~al.}(2011)\citenamefont {Aroyo}, \citenamefont {Perez~Mato}, \citenamefont {Orobengoa}, \citenamefont {Tasci}, \citenamefont {de~la Flor},\ and\ \citenamefont {Kirov}}]{aroyo2011crystallography}%
  \BibitemOpen
  \bibfield  {author} {\bibinfo {author} {\bibfnamefont {M.~I.}\ \bibnamefont {Aroyo}}, \bibinfo {author} {\bibfnamefont {J.~M.}\ \bibnamefont {Perez~Mato}}, \bibinfo {author} {\bibfnamefont {D.}~\bibnamefont {Orobengoa}}, \bibinfo {author} {\bibfnamefont {E.}~\bibnamefont {Tasci}}, \bibinfo {author} {\bibfnamefont {G.}~\bibnamefont {de~la Flor}},\ and\ \bibinfo {author} {\bibfnamefont {A.}~\bibnamefont {Kirov}},\ }\bibfield  {title} {\bibinfo {title} {Crystallography online: Bilbao crystallographic server},\ }\href {http://www.bcc.bas.bg/BCC_Volumes/Volume_43_Number_2_2011/Volume_43_Number_2_2011_PDF/2011_43_2_1.pdf} {\bibfield  {journal} {\bibinfo  {journal} {Bulg. Chem. Commun.}\ }\textbf {\bibinfo {volume} {43}},\ \bibinfo {pages} {183} (\bibinfo {year} {2011})}\BibitemShut {NoStop}%
\bibitem [{\citenamefont {Aroyo}\ \emph {et~al.}(2006{\natexlab{a}})\citenamefont {Aroyo}, \citenamefont {Perez~Mato}, \citenamefont {Capillas}, \citenamefont {Kroumova}, \citenamefont {Ivantchev}, \citenamefont {Madariaga}, \citenamefont {Kirov},\ and\ \citenamefont {Wondratschek}}]{aroyo2006bilbao1}%
  \BibitemOpen
  \bibfield  {author} {\bibinfo {author} {\bibfnamefont {M.~I.}\ \bibnamefont {Aroyo}}, \bibinfo {author} {\bibfnamefont {J.~M.}\ \bibnamefont {Perez~Mato}}, \bibinfo {author} {\bibfnamefont {C.}~\bibnamefont {Capillas}}, \bibinfo {author} {\bibfnamefont {E.}~\bibnamefont {Kroumova}}, \bibinfo {author} {\bibfnamefont {S.}~\bibnamefont {Ivantchev}}, \bibinfo {author} {\bibfnamefont {G.}~\bibnamefont {Madariaga}}, \bibinfo {author} {\bibfnamefont {A.}~\bibnamefont {Kirov}},\ and\ \bibinfo {author} {\bibfnamefont {H.}~\bibnamefont {Wondratschek}},\ }\bibfield  {title} {\bibinfo {title} {Bilbao crystallographic server: {I}. databases and crystallographic computing programs},\ }\href {https://www.degruyter.com/document/doi/10.1524/zkri.2006.221.1.15/html} {\bibfield  {journal} {\bibinfo  {journal} {Z. Kristallogr. Cryst. Mater.}\ }\textbf {\bibinfo {volume} {221}},\ \bibinfo {pages} {15} (\bibinfo {year} {2006}{\natexlab{a}})}\BibitemShut {NoStop}%
\bibitem [{\citenamefont {Aroyo}\ \emph {et~al.}(2006{\natexlab{b}})\citenamefont {Aroyo}, \citenamefont {Kirov}, \citenamefont {Capillas}, \citenamefont {Perez-Mato},\ and\ \citenamefont {Wondratschek}}]{aroyo2006bilbao2}%
  \BibitemOpen
  \bibfield  {author} {\bibinfo {author} {\bibfnamefont {M.~I.}\ \bibnamefont {Aroyo}}, \bibinfo {author} {\bibfnamefont {A.}~\bibnamefont {Kirov}}, \bibinfo {author} {\bibfnamefont {C.}~\bibnamefont {Capillas}}, \bibinfo {author} {\bibfnamefont {J.}~\bibnamefont {Perez-Mato}},\ and\ \bibinfo {author} {\bibfnamefont {H.}~\bibnamefont {Wondratschek}},\ }\bibfield  {title} {\bibinfo {title} {Bilbao crystallographic server. {II}. representations of crystallographic point groups and space groups},\ }\href {https://journals.iucr.org/paper?S0108767305040286} {\bibfield  {journal} {\bibinfo  {journal} {Acta Crystallogr. Sect. A: Found. Crystallogr.}\ }\textbf {\bibinfo {volume} {62}},\ \bibinfo {pages} {115} (\bibinfo {year} {2006}{\natexlab{b}})}\BibitemShut {NoStop}%
\bibitem [{\citenamefont {Gallego}\ \emph {et~al.}(2016{\natexlab{a}})\citenamefont {Gallego}, \citenamefont {Perez~Mato}, \citenamefont {Elcoro}, \citenamefont {Tasci}, \citenamefont {Hanson}, \citenamefont {Momma}, \citenamefont {Aroyo},\ and\ \citenamefont {Madariaga}}]{gallego2016magndata1}%
  \BibitemOpen
  \bibfield  {author} {\bibinfo {author} {\bibfnamefont {S.~V.}\ \bibnamefont {Gallego}}, \bibinfo {author} {\bibfnamefont {J.~M.}\ \bibnamefont {Perez~Mato}}, \bibinfo {author} {\bibfnamefont {L.}~\bibnamefont {Elcoro}}, \bibinfo {author} {\bibfnamefont {E.~S.}\ \bibnamefont {Tasci}}, \bibinfo {author} {\bibfnamefont {R.~M.}\ \bibnamefont {Hanson}}, \bibinfo {author} {\bibfnamefont {K.}~\bibnamefont {Momma}}, \bibinfo {author} {\bibfnamefont {M.~I.}\ \bibnamefont {Aroyo}},\ and\ \bibinfo {author} {\bibfnamefont {G.}~\bibnamefont {Madariaga}},\ }\bibfield  {title} {\bibinfo {title} {Magndata: towards a database of magnetic structures. {I}. the commensurate case},\ }\href {https://journals.iucr.org/j/issues/2016/05/00/ks5532/ks5532.pdf} {\bibfield  {journal} {\bibinfo  {journal} {J. Appl. Crystallogr.}\ }\textbf {\bibinfo {volume} {49}},\ \bibinfo {pages} {1750} (\bibinfo {year} {2016}{\natexlab{a}})}\BibitemShut {NoStop}%
\bibitem [{\citenamefont {Gallego}\ \emph {et~al.}(2016{\natexlab{b}})\citenamefont {Gallego}, \citenamefont {Perez~Mato}, \citenamefont {Elcoro}, \citenamefont {Tasci}, \citenamefont {Hanson}, \citenamefont {Aroyo},\ and\ \citenamefont {Madariaga}}]{gallego2016magndata2}%
  \BibitemOpen
  \bibfield  {author} {\bibinfo {author} {\bibfnamefont {S.~V.}\ \bibnamefont {Gallego}}, \bibinfo {author} {\bibfnamefont {J.~M.}\ \bibnamefont {Perez~Mato}}, \bibinfo {author} {\bibfnamefont {L.}~\bibnamefont {Elcoro}}, \bibinfo {author} {\bibfnamefont {E.~S.}\ \bibnamefont {Tasci}}, \bibinfo {author} {\bibfnamefont {R.~M.}\ \bibnamefont {Hanson}}, \bibinfo {author} {\bibfnamefont {M.~I.}\ \bibnamefont {Aroyo}},\ and\ \bibinfo {author} {\bibfnamefont {G.}~\bibnamefont {Madariaga}},\ }\bibfield  {title} {\bibinfo {title} {Magndata: towards a database of magnetic structures. ii. the incommensurate case},\ }\href {https://journals.iucr.org/paper?S1600576716015491} {\bibfield  {journal} {\bibinfo  {journal} {J. Appl. Crystallogr.}\ }\textbf {\bibinfo {volume} {49}},\ \bibinfo {pages} {1941} (\bibinfo {year} {2016}{\natexlab{b}})}\BibitemShut {NoStop}%
\bibitem [{mag()}]{magndata}%
  \BibitemOpen
  \href@noop {} {\bibinfo {title} {Magndata}},\ \bibinfo {note} {{\color{blue} \href{http://webbdcrista1.ehu.es/magndata}{http://webbdcrista1.ehu.es/magndata}}}\BibitemShut {NoStop}%
\bibitem [{\citenamefont {Herath}\ \emph {et~al.}(2020)\citenamefont {Herath}, \citenamefont {Tavadze}, \citenamefont {He}, \citenamefont {Bousquet}, \citenamefont {Singh}, \citenamefont {Mu{\~n}oz},\ and\ \citenamefont {Romero}}]{herath2020pyprocar}%
  \BibitemOpen
  \bibfield  {author} {\bibinfo {author} {\bibfnamefont {U.}~\bibnamefont {Herath}}, \bibinfo {author} {\bibfnamefont {P.}~\bibnamefont {Tavadze}}, \bibinfo {author} {\bibfnamefont {X.}~\bibnamefont {He}}, \bibinfo {author} {\bibfnamefont {E.}~\bibnamefont {Bousquet}}, \bibinfo {author} {\bibfnamefont {S.}~\bibnamefont {Singh}}, \bibinfo {author} {\bibfnamefont {F.}~\bibnamefont {Mu{\~n}oz}},\ and\ \bibinfo {author} {\bibfnamefont {A.~H.}\ \bibnamefont {Romero}},\ }\bibfield  {title} {\bibinfo {title} {Pyprocar: A python library for electronic structure pre/post-processing},\ }\href {https://www.sciencedirect.com/science/article/pii/S0010465519303935} {\bibfield  {journal} {\bibinfo  {journal} {Comput. Phys. Commun.}\ }\textbf {\bibinfo {volume} {251}},\ \bibinfo {pages} {107080} (\bibinfo {year} {2020})}\BibitemShut {NoStop}%
\bibitem [{\citenamefont {Pyprocar}()}]{pyprocar}%
  \BibitemOpen
  \bibfield  {author} {\bibinfo {author} {\bibnamefont {Pyprocar}},\ }\href@noop {} {}\bibinfo {note} {{\color{blue} \href{https://romerogroup.github.io/pyprocar/index.html}{https://romerogroup.github.io/pyprocar/ind- ex.html}}}\BibitemShut {NoStop}%
\bibitem [{\citenamefont {Jain}\ \emph {et~al.}(2013)\citenamefont {Jain}, \citenamefont {Ong}, \citenamefont {Hautier}, \citenamefont {Chen}, \citenamefont {Richards}, \citenamefont {Dacek}, \citenamefont {Cholia}, \citenamefont {Gunter}, \citenamefont {Skinner}, \citenamefont {Ceder} \emph {et~al.}}]{jain2013commentary}%
  \BibitemOpen
  \bibfield  {author} {\bibinfo {author} {\bibfnamefont {A.}~\bibnamefont {Jain}}, \bibinfo {author} {\bibfnamefont {S.~P.}\ \bibnamefont {Ong}}, \bibinfo {author} {\bibfnamefont {G.}~\bibnamefont {Hautier}}, \bibinfo {author} {\bibfnamefont {W.}~\bibnamefont {Chen}}, \bibinfo {author} {\bibfnamefont {W.~D.}\ \bibnamefont {Richards}}, \bibinfo {author} {\bibfnamefont {S.}~\bibnamefont {Dacek}}, \bibinfo {author} {\bibfnamefont {S.}~\bibnamefont {Cholia}}, \bibinfo {author} {\bibfnamefont {D.}~\bibnamefont {Gunter}}, \bibinfo {author} {\bibfnamefont {D.}~\bibnamefont {Skinner}}, \bibinfo {author} {\bibfnamefont {G.}~\bibnamefont {Ceder}}, \emph {et~al.},\ }\bibfield  {title} {\bibinfo {title} {Commentary: The materials project: A materials genome approach to accelerating materials innovation},\ }\href {https://pubs.aip.org/aip/apm/article/1/1/011002/119685/Commentary-The-Materials-Project-A-materials} {\bibfield  {journal} {\bibinfo  {journal} {APL Mater.}\ }\textbf {\bibinfo {volume} {1}},\ \bibinfo {pages}
  {011002} (\bibinfo {year} {2013})}\BibitemShut {NoStop}%
\bibitem [{\citenamefont {Chen}\ \emph {et~al.}(2024{\natexlab{b}})\citenamefont {Chen}, \citenamefont {Ren}, \citenamefont {Zhu}, \citenamefont {Yu}, \citenamefont {Zhang}, \citenamefont {Liu}, \citenamefont {Li}, \citenamefont {Liu}, \citenamefont {Li},\ and\ \citenamefont {Liu}}]{chen2024enumeration}%
  \BibitemOpen
  \bibfield  {author} {\bibinfo {author} {\bibfnamefont {X.}~\bibnamefont {Chen}}, \bibinfo {author} {\bibfnamefont {J.}~\bibnamefont {Ren}}, \bibinfo {author} {\bibfnamefont {Y.}~\bibnamefont {Zhu}}, \bibinfo {author} {\bibfnamefont {Y.}~\bibnamefont {Yu}}, \bibinfo {author} {\bibfnamefont {A.}~\bibnamefont {Zhang}}, \bibinfo {author} {\bibfnamefont {P.}~\bibnamefont {Liu}}, \bibinfo {author} {\bibfnamefont {J.}~\bibnamefont {Li}}, \bibinfo {author} {\bibfnamefont {Y.}~\bibnamefont {Liu}}, \bibinfo {author} {\bibfnamefont {C.}~\bibnamefont {Li}},\ and\ \bibinfo {author} {\bibfnamefont {Q.}~\bibnamefont {Liu}},\ }\bibfield  {title} {\bibinfo {title} {Enumeration and representation theory of spin space groups},\ }\href {https://journals.aps.org/prx/abstract/10.1103/PhysRevX.14.031038} {\bibfield  {journal} {\bibinfo  {journal} {Phys. Rev. X.}\ }\textbf {\bibinfo {volume} {14}},\ \bibinfo {pages} {031038} (\bibinfo {year} {2024}{\natexlab{b}})}\BibitemShut {NoStop}%
\bibitem [{\citenamefont {Hunter}(2007)}]{hunter2007matplotlib}%
  \BibitemOpen
  \bibfield  {author} {\bibinfo {author} {\bibfnamefont {J.~D.}\ \bibnamefont {Hunter}},\ }\bibfield  {title} {\bibinfo {title} {Matplotlib: A 2d graphics environment},\ }\href {https://www.computer.org/csdl/magazine/cs/2007/03/c3090/13rRUwbJD0A} {\bibfield  {journal} {\bibinfo  {journal} {Comput. Sci. Eng.}\ }\textbf {\bibinfo {volume} {9}},\ \bibinfo {pages} {90} (\bibinfo {year} {2007})}\BibitemShut {NoStop}%
\bibitem [{\citenamefont {Litvin}\ and\ \citenamefont {Opechowski}(1974)}]{litvin1974spin}%
  \BibitemOpen
  \bibfield  {author} {\bibinfo {author} {\bibfnamefont {D.~B.}\ \bibnamefont {Litvin}}\ and\ \bibinfo {author} {\bibfnamefont {W.}~\bibnamefont {Opechowski}},\ }\bibfield  {title} {\bibinfo {title} {Spin groups},\ }\href {https://www.sciencedirect.com/science/article/abs/pii/0031891474901578} {\bibfield  {journal} {\bibinfo  {journal} {Physica}\ }\textbf {\bibinfo {volume} {76}},\ \bibinfo {pages} {538} (\bibinfo {year} {1974})}\BibitemShut {NoStop}%
\bibitem [{\citenamefont {Litvin}(1977)}]{litvin1977spin}%
  \BibitemOpen
  \bibfield  {author} {\bibinfo {author} {\bibfnamefont {D.~B.}\ \bibnamefont {Litvin}},\ }\bibfield  {title} {\bibinfo {title} {Spin point groups},\ }\href {https://onlinelibrary.wiley.com/doi/10.1107/S0567739477000709} {\bibfield  {journal} {\bibinfo  {journal} {Acta Crystallogr. Sect. A: Found. Crystallogr.}\ }\textbf {\bibinfo {volume} {33}},\ \bibinfo {pages} {279} (\bibinfo {year} {1977})}\BibitemShut {NoStop}%
\bibitem [{\citenamefont {Liu}\ \emph {et~al.}(2022)\citenamefont {Liu}, \citenamefont {Li}, \citenamefont {Han}, \citenamefont {Wan},\ and\ \citenamefont {Liu}}]{liu2022spin}%
  \BibitemOpen
  \bibfield  {author} {\bibinfo {author} {\bibfnamefont {P.}~\bibnamefont {Liu}}, \bibinfo {author} {\bibfnamefont {J.}~\bibnamefont {Li}}, \bibinfo {author} {\bibfnamefont {J.}~\bibnamefont {Han}}, \bibinfo {author} {\bibfnamefont {X.}~\bibnamefont {Wan}},\ and\ \bibinfo {author} {\bibfnamefont {Q.}~\bibnamefont {Liu}},\ }\bibfield  {title} {\bibinfo {title} {Spin-group symmetry in magnetic materials with negligible spin-orbit coupling},\ }\href {https://journals.aps.org/prx/abstract/10.1103/PhysRevX.12.021016} {\bibfield  {journal} {\bibinfo  {journal} {Phys. Rev. X}\ }\textbf {\bibinfo {volume} {12}},\ \bibinfo {pages} {021016} (\bibinfo {year} {2022})}\BibitemShut {NoStop}%
\bibitem [{\citenamefont {Schiff}\ \emph {et~al.}(2025)\citenamefont {Schiff}, \citenamefont {Corticelli}, \citenamefont {Guerreiro}, \citenamefont {Romh{\'a}nyi},\ and\ \citenamefont {McClarty}}]{schiff2025crystallographic}%
  \BibitemOpen
  \bibfield  {author} {\bibinfo {author} {\bibfnamefont {H.}~\bibnamefont {Schiff}}, \bibinfo {author} {\bibfnamefont {A.}~\bibnamefont {Corticelli}}, \bibinfo {author} {\bibfnamefont {A.}~\bibnamefont {Guerreiro}}, \bibinfo {author} {\bibfnamefont {J.}~\bibnamefont {Romh{\'a}nyi}},\ and\ \bibinfo {author} {\bibfnamefont {P.~A.}\ \bibnamefont {McClarty}},\ }\bibfield  {title} {\bibinfo {title} {The crystallographic spin point groups and their representations},\ }\href {https://scipost.org/10.21468/SciPostPhys.18.3.109} {\bibfield  {journal} {\bibinfo  {journal} {SciPost Phys.}\ }\textbf {\bibinfo {volume} {18}},\ \bibinfo {pages} {109} (\bibinfo {year} {2025})}\BibitemShut {NoStop}%
\bibitem [{\citenamefont {S{\o}dequist}\ and\ \citenamefont {Olsen}(2024)}]{sodequist2024two}%
  \BibitemOpen
  \bibfield  {author} {\bibinfo {author} {\bibfnamefont {J.}~\bibnamefont {S{\o}dequist}}\ and\ \bibinfo {author} {\bibfnamefont {T.}~\bibnamefont {Olsen}},\ }\bibfield  {title} {\bibinfo {title} {Two-dimensional altermagnets from high throughput computational screening: Symmetry requirements, chiral magnons, and spin-orbit effects},\ }\href {https://pubs.aip.org/aip/apl/article/124/18/182409/3288014} {\bibfield  {journal} {\bibinfo  {journal} {Appl. Phys. Lett.}\ }\textbf {\bibinfo {volume} {124}} (\bibinfo {year} {2024})}\BibitemShut {NoStop}%
\bibitem [{\citenamefont {Guo}\ \emph {et~al.}(2023{\natexlab{c}})\citenamefont {Guo}, \citenamefont {Liu}, \citenamefont {Janson}, \citenamefont {Fulga}, \citenamefont {van~den Brink},\ and\ \citenamefont {Facio}}]{guo2023spin}%
  \BibitemOpen
  \bibfield  {author} {\bibinfo {author} {\bibfnamefont {Y.}~\bibnamefont {Guo}}, \bibinfo {author} {\bibfnamefont {H.}~\bibnamefont {Liu}}, \bibinfo {author} {\bibfnamefont {O.}~\bibnamefont {Janson}}, \bibinfo {author} {\bibfnamefont {I.~C.}\ \bibnamefont {Fulga}}, \bibinfo {author} {\bibfnamefont {J.}~\bibnamefont {van~den Brink}},\ and\ \bibinfo {author} {\bibfnamefont {J.~I.}\ \bibnamefont {Facio}},\ }\bibfield  {title} {\bibinfo {title} {Spin-split collinear antiferromagnets: A large-scale ab-initio study},\ }\href {https://www.sciencedirect.com/science/article/abs/pii/S2542529323000275} {\bibfield  {journal} {\bibinfo  {journal} {Mater. Today Phys.}\ }\textbf {\bibinfo {volume} {32}},\ \bibinfo {pages} {100991} (\bibinfo {year} {2023}{\natexlab{c}})}\BibitemShut {NoStop}%
\bibitem [{\citenamefont {Gu}\ \emph {et~al.}(2025)\citenamefont {Gu}, \citenamefont {Liu}, \citenamefont {Zhu}, \citenamefont {Yananose}, \citenamefont {Chen}, \citenamefont {Hu}, \citenamefont {Stroppa},\ and\ \citenamefont {Liu}}]{gu2025ferroelectric}%
  \BibitemOpen
  \bibfield  {author} {\bibinfo {author} {\bibfnamefont {M.}~\bibnamefont {Gu}}, \bibinfo {author} {\bibfnamefont {Y.}~\bibnamefont {Liu}}, \bibinfo {author} {\bibfnamefont {H.}~\bibnamefont {Zhu}}, \bibinfo {author} {\bibfnamefont {K.}~\bibnamefont {Yananose}}, \bibinfo {author} {\bibfnamefont {X.}~\bibnamefont {Chen}}, \bibinfo {author} {\bibfnamefont {Y.}~\bibnamefont {Hu}}, \bibinfo {author} {\bibfnamefont {A.}~\bibnamefont {Stroppa}},\ and\ \bibinfo {author} {\bibfnamefont {Q.}~\bibnamefont {Liu}},\ }\bibfield  {title} {\bibinfo {title} {Ferroelectric switchable altermagnetism},\ }\href {https://journals.aps.org/prl/abstract/10.1103/PhysRevLett.134.106802} {\bibfield  {journal} {\bibinfo  {journal} {Phys. Rev. Lett.}\ }\textbf {\bibinfo {volume} {134}},\ \bibinfo {pages} {106802} (\bibinfo {year} {2025})}\BibitemShut {NoStop}%
\bibitem [{\citenamefont {Bai}\ \emph {et~al.}(2024)\citenamefont {Bai}, \citenamefont {Feng}, \citenamefont {Liu}, \citenamefont {{\v{S}}mejkal}, \citenamefont {Mokrousov},\ and\ \citenamefont {Yao}}]{bai2024altermagnetism}%
  \BibitemOpen
  \bibfield  {author} {\bibinfo {author} {\bibfnamefont {L.}~\bibnamefont {Bai}}, \bibinfo {author} {\bibfnamefont {W.}~\bibnamefont {Feng}}, \bibinfo {author} {\bibfnamefont {S.}~\bibnamefont {Liu}}, \bibinfo {author} {\bibfnamefont {L.}~\bibnamefont {{\v{S}}mejkal}}, \bibinfo {author} {\bibfnamefont {Y.}~\bibnamefont {Mokrousov}},\ and\ \bibinfo {author} {\bibfnamefont {Y.}~\bibnamefont {Yao}},\ }\bibfield  {title} {\bibinfo {title} {Altermagnetism: Exploring new frontiers in magnetism and spintronics},\ }\href {https://advanced.onlinelibrary.wiley.com/doi/full/10.1002/adfm.202409327} {\bibfield  {journal} {\bibinfo  {journal} {Adv. Funct. Mater.}\ }\textbf {\bibinfo {volume} {34}},\ \bibinfo {pages} {2409327} (\bibinfo {year} {2024})}\BibitemShut {NoStop}%
\bibitem [{\citenamefont {Gao}\ \emph {et~al.}(2025)\citenamefont {Gao}, \citenamefont {Qu}, \citenamefont {Zeng}, \citenamefont {Liu}, \citenamefont {Wen}, \citenamefont {Sun}, \citenamefont {Guo},\ and\ \citenamefont {Lu}}]{gao2025ai}%
  \BibitemOpen
  \bibfield  {author} {\bibinfo {author} {\bibfnamefont {Z.~F.}\ \bibnamefont {Gao}}, \bibinfo {author} {\bibfnamefont {S.}~\bibnamefont {Qu}}, \bibinfo {author} {\bibfnamefont {B.}~\bibnamefont {Zeng}}, \bibinfo {author} {\bibfnamefont {Y.}~\bibnamefont {Liu}}, \bibinfo {author} {\bibfnamefont {J.~R.}\ \bibnamefont {Wen}}, \bibinfo {author} {\bibfnamefont {H.}~\bibnamefont {Sun}}, \bibinfo {author} {\bibfnamefont {P.~J.}\ \bibnamefont {Guo}},\ and\ \bibinfo {author} {\bibfnamefont {Z.~Y.}\ \bibnamefont {Lu}},\ }\bibfield  {title} {\bibinfo {title} {{AI}-accelerated discovery of altermagnetic materials},\ }\href {https://academic.oup.com/nsr/article/12/4/nwaf066/8030545?login=true} {\bibfield  {journal} {\bibinfo  {journal} {Natl. Sci. Rev.}\ }\textbf {\bibinfo {volume} {12}},\ \bibinfo {pages} {nwaf066} (\bibinfo {year} {2025})}\BibitemShut {NoStop}%
\bibitem [{\citenamefont {Guo}\ \emph {et~al.}(2023{\natexlab{d}})\citenamefont {Guo}, \citenamefont {Gu}, \citenamefont {Gao},\ and\ \citenamefont {Lu}}]{guo2023altermagnetic}%
  \BibitemOpen
  \bibfield  {author} {\bibinfo {author} {\bibfnamefont {P.~J.}\ \bibnamefont {Guo}}, \bibinfo {author} {\bibfnamefont {Y.}~\bibnamefont {Gu}}, \bibinfo {author} {\bibfnamefont {Z.~F.}\ \bibnamefont {Gao}},\ and\ \bibinfo {author} {\bibfnamefont {Z.~Y.}\ \bibnamefont {Lu}},\ }\bibfield  {title} {\bibinfo {title} {Altermagnetic ferroelectric $\mathrm{Li}\mathrm{Fe}_{2}\mathrm{F}_{6}$ and spin-triplet excitonic insulator phase},\ }\href {https://arxiv.org/abs/2312.13911} {\bibfield  {journal} {\bibinfo  {journal} {arXiv:2312.13911}\ } (\bibinfo {year} {2023}{\natexlab{d}})}\BibitemShut {NoStop}%
\bibitem [{\citenamefont {Zhang}\ \emph {et~al.}(2024)\citenamefont {Zhang}, \citenamefont {Cui}, \citenamefont {Li}, \citenamefont {Duan}, \citenamefont {Li}, \citenamefont {Yu},\ and\ \citenamefont {Yao}}]{zhang2024predictable}%
  \BibitemOpen
  \bibfield  {author} {\bibinfo {author} {\bibfnamefont {R.~W.}\ \bibnamefont {Zhang}}, \bibinfo {author} {\bibfnamefont {C.}~\bibnamefont {Cui}}, \bibinfo {author} {\bibfnamefont {R.}~\bibnamefont {Li}}, \bibinfo {author} {\bibfnamefont {J.}~\bibnamefont {Duan}}, \bibinfo {author} {\bibfnamefont {L.}~\bibnamefont {Li}}, \bibinfo {author} {\bibfnamefont {Z.~M.}\ \bibnamefont {Yu}},\ and\ \bibinfo {author} {\bibfnamefont {Y.}~\bibnamefont {Yao}},\ }\bibfield  {title} {\bibinfo {title} {Predictable gate-field control of spin in altermagnets with spin-layer coupling},\ }\href {https://journals.aps.org/prl/abstract/10.1103/PhysRevLett.133.056401} {\bibfield  {journal} {\bibinfo  {journal} {Phys. Rev. Lett.}\ }\textbf {\bibinfo {volume} {133}},\ \bibinfo {pages} {056401} (\bibinfo {year} {2024})}\BibitemShut {NoStop}%
\bibitem [{\citenamefont {Dresselhaus}\ \emph {et~al.}(2007)\citenamefont {Dresselhaus}, \citenamefont {Dresselhaus},\ and\ \citenamefont {Jorio}}]{dresselhaus2007group}%
  \BibitemOpen
  \bibfield  {author} {\bibinfo {author} {\bibfnamefont {M.~S.}\ \bibnamefont {Dresselhaus}}, \bibinfo {author} {\bibfnamefont {G.}~\bibnamefont {Dresselhaus}},\ and\ \bibinfo {author} {\bibfnamefont {A.}~\bibnamefont {Jorio}},\ }\href {https://link.springer.com/book/10.1007/978-3-540-32899-5} {\emph {\bibinfo {title} {Group theory: application to the physics of condensed matter}}}\ (\bibinfo  {publisher} {Springer Science \& Business Media},\ \bibinfo {year} {2007})\BibitemShut {NoStop}%
\bibitem [{\citenamefont {Geilhufe}\ and\ \citenamefont {Hergert}(2018)}]{geilhufe2018gtpack}%
  \BibitemOpen
  \bibfield  {author} {\bibinfo {author} {\bibfnamefont {R.~M.}\ \bibnamefont {Geilhufe}}\ and\ \bibinfo {author} {\bibfnamefont {W.}~\bibnamefont {Hergert}},\ }\bibfield  {title} {\bibinfo {title} {Gtpack: A mathematica group theory package for application in solid-state physics and photonics},\ }\href {https://www.frontiersin.org/journals/physics/articles/10.3389/fphy.2018.00086/full} {\bibfield  {journal} {\bibinfo  {journal} {Front. Phys.}\ }\textbf {\bibinfo {volume} {6}},\ \bibinfo {pages} {86} (\bibinfo {year} {2018})}\BibitemShut {NoStop}%
\bibitem [{\citenamefont {Strempfer}\ \emph {et~al.}(2004)\citenamefont {Strempfer}, \citenamefont {R{\"u}tt}, \citenamefont {Bayrakci}, \citenamefont {Br{\"u}ckel},\ and\ \citenamefont {Jauch}}]{strempfer2004magnetic}%
  \BibitemOpen
  \bibfield  {author} {\bibinfo {author} {\bibfnamefont {J.}~\bibnamefont {Strempfer}}, \bibinfo {author} {\bibfnamefont {U.}~\bibnamefont {R{\"u}tt}}, \bibinfo {author} {\bibfnamefont {S.}~\bibnamefont {Bayrakci}}, \bibinfo {author} {\bibfnamefont {T.}~\bibnamefont {Br{\"u}ckel}},\ and\ \bibinfo {author} {\bibfnamefont {W.}~\bibnamefont {Jauch}},\ }\bibfield  {title} {\bibinfo {title} {Magnetic properties of transition metal fluorides $\mathrm{M}\mathrm{F}_{2}$ ($\mathrm{M}$= $\mathrm{Mn}$, $\mathrm{Fe}$, $\mathrm{Co}$, $\mathrm{Ni}$) via high-energy photon diffraction},\ }\href {https://journals.aps.org/prb/abstract/10.1103/PhysRevB.69.014417} {\bibfield  {journal} {\bibinfo  {journal} {Phys. Rev. B}\ }\textbf {\bibinfo {volume} {69}},\ \bibinfo {pages} {014417} (\bibinfo {year} {2004})}\BibitemShut {NoStop}%
\bibitem [{\citenamefont {Shachar}\ \emph {et~al.}(1972)\citenamefont {Shachar}, \citenamefont {Makovsky},\ and\ \citenamefont {Shaked}}]{shachar1972neutron}%
  \BibitemOpen
  \bibfield  {author} {\bibinfo {author} {\bibfnamefont {G.}~\bibnamefont {Shachar}}, \bibinfo {author} {\bibfnamefont {J.}~\bibnamefont {Makovsky}},\ and\ \bibinfo {author} {\bibfnamefont {H.}~\bibnamefont {Shaked}},\ }\bibfield  {title} {\bibinfo {title} {Neutron-diffraction study of the magnetic structure of the trirutile $\mathrm{Li}\mathrm{Fe}_{2}\mathrm{F}_{6}$},\ }\href {https://journals.aps.org/prb/abstract/10.1103/PhysRevB.6.1968} {\bibfield  {journal} {\bibinfo  {journal} {Phys. Rev. B}\ }\textbf {\bibinfo {volume} {6}},\ \bibinfo {pages} {1968} (\bibinfo {year} {1972})}\BibitemShut {NoStop}%
\bibitem [{\citenamefont {Fourquet}\ \emph {et~al.}(1988)\citenamefont {Fourquet}, \citenamefont {Le~Samedi},\ and\ \citenamefont {Calage}}]{fourquet1988trirutile}%
  \BibitemOpen
  \bibfield  {author} {\bibinfo {author} {\bibfnamefont {J.}~\bibnamefont {Fourquet}}, \bibinfo {author} {\bibfnamefont {E.}~\bibnamefont {Le~Samedi}},\ and\ \bibinfo {author} {\bibfnamefont {Y.}~\bibnamefont {Calage}},\ }\bibfield  {title} {\bibinfo {title} {Le trirutile ordonn{\'e} $\mathrm{Li}\mathrm{Fe}_{2}\mathrm{F}_{6}$: Croissance cristalline et {\'e}tude structurale},\ }\href {https://www.sciencedirect.com/science/article/abs/pii/002245968890093X} {\bibfield  {journal} {\bibinfo  {journal} {J. Solid State Chem.}\ }\textbf {\bibinfo {volume} {77}},\ \bibinfo {pages} {84} (\bibinfo {year} {1988})}\BibitemShut {NoStop}%
\bibitem [{Note1()}]{Note1}%
  \BibitemOpen
  \bibinfo {note} {By coincidence, the strain-induced nonrelativistic ZSSs in $\protect \mathrm {LiFe_{2}F_{6}}$ and $\protect \mathrm {La_{2}O_{3}Mn_{2}Se_{2}}$ are very close to each other.}\BibitemShut {Stop}%
\bibitem [{\citenamefont {Ralph}\ and\ \citenamefont {Stiles}(2008)}]{ralph2008spin}%
  \BibitemOpen
  \bibfield  {author} {\bibinfo {author} {\bibfnamefont {D.~C.}\ \bibnamefont {Ralph}}\ and\ \bibinfo {author} {\bibfnamefont {M.~D.}\ \bibnamefont {Stiles}},\ }\bibfield  {title} {\bibinfo {title} {Spin transfer torques},\ }\href {https://www.sciencedirect.com/science/article/abs/pii/S0304885307010116} {\bibfield  {journal} {\bibinfo  {journal} {J. Magn. Magn. Mater.}\ }\textbf {\bibinfo {volume} {320}},\ \bibinfo {pages} {1190} (\bibinfo {year} {2008})}\BibitemShut {NoStop}%
\bibitem [{\citenamefont {Brataas}\ \emph {et~al.}(2012)\citenamefont {Brataas}, \citenamefont {Kent},\ and\ \citenamefont {Ohno}}]{brataas2012current}%
  \BibitemOpen
  \bibfield  {author} {\bibinfo {author} {\bibfnamefont {A.}~\bibnamefont {Brataas}}, \bibinfo {author} {\bibfnamefont {A.~D.}\ \bibnamefont {Kent}},\ and\ \bibinfo {author} {\bibfnamefont {H.}~\bibnamefont {Ohno}},\ }\bibfield  {title} {\bibinfo {title} {Current-induced torques in magnetic materials},\ }\href {https://www.nature.com/articles/nmat3311} {\bibfield  {journal} {\bibinfo  {journal} {Nat. Mater.}\ }\textbf {\bibinfo {volume} {11}},\ \bibinfo {pages} {372} (\bibinfo {year} {2012})}\BibitemShut {NoStop}%
\bibitem [{\citenamefont {Pandey}\ \emph {et~al.}(2023)\citenamefont {Pandey}, \citenamefont {Pandey}, \citenamefont {Ahuja},\ and\ \citenamefont {Kumar}}]{pandey2023straining}%
  \BibitemOpen
  \bibfield  {author} {\bibinfo {author} {\bibfnamefont {M.}~\bibnamefont {Pandey}}, \bibinfo {author} {\bibfnamefont {C.}~\bibnamefont {Pandey}}, \bibinfo {author} {\bibfnamefont {R.}~\bibnamefont {Ahuja}},\ and\ \bibinfo {author} {\bibfnamefont {R.}~\bibnamefont {Kumar}},\ }\bibfield  {title} {\bibinfo {title} {Straining techniques for strain engineering of 2d materials towards flexible straintronic applications},\ }\href {https://www.sciencedirect.com/science/article/pii/S2211285523001143} {\bibfield  {journal} {\bibinfo  {journal} {Nano Energ.}\ }\textbf {\bibinfo {volume} {109}},\ \bibinfo {pages} {108278} (\bibinfo {year} {2023})}\BibitemShut {NoStop}%
\bibitem [{\citenamefont {Miao}\ \emph {et~al.}(2021)\citenamefont {Miao}, \citenamefont {Liang},\ and\ \citenamefont {Cheng}}]{miao2021straintronics}%
  \BibitemOpen
  \bibfield  {author} {\bibinfo {author} {\bibfnamefont {F.}~\bibnamefont {Miao}}, \bibinfo {author} {\bibfnamefont {S.~J.}\ \bibnamefont {Liang}},\ and\ \bibinfo {author} {\bibfnamefont {B.}~\bibnamefont {Cheng}},\ }\bibfield  {title} {\bibinfo {title} {Straintronics with van der waals materials},\ }\href {https://www.nature.com/articles/s41535-021-00360-3?utm_source=xmol&utm_medium=affiliate&utm_content=meta&utm_campaign=DDCN_1_GL01_metadata} {\bibfield  {journal} {\bibinfo  {journal} {Npj Quantum Mater.}\ }\textbf {\bibinfo {volume} {6}},\ \bibinfo {pages} {59} (\bibinfo {year} {2021})}\BibitemShut {NoStop}%
\bibitem [{\citenamefont {Zhang}\ \emph {et~al.}(2022)\citenamefont {Zhang}, \citenamefont {Pincelli}, \citenamefont {Jozwiak}, \citenamefont {Kondo}, \citenamefont {Ernstorfer}, \citenamefont {Sato},\ and\ \citenamefont {Zhou}}]{zhang2022angle}%
  \BibitemOpen
  \bibfield  {author} {\bibinfo {author} {\bibfnamefont {H.}~\bibnamefont {Zhang}}, \bibinfo {author} {\bibfnamefont {T.}~\bibnamefont {Pincelli}}, \bibinfo {author} {\bibfnamefont {C.}~\bibnamefont {Jozwiak}}, \bibinfo {author} {\bibfnamefont {T.}~\bibnamefont {Kondo}}, \bibinfo {author} {\bibfnamefont {R.}~\bibnamefont {Ernstorfer}}, \bibinfo {author} {\bibfnamefont {T.}~\bibnamefont {Sato}},\ and\ \bibinfo {author} {\bibfnamefont {S.}~\bibnamefont {Zhou}},\ }\bibfield  {title} {\bibinfo {title} {Angle-resolved photoemission spectroscopy},\ }\href {https://www.nature.com/articles/s43586-022-00133-7} {\bibfield  {journal} {\bibinfo  {journal} {Nat. Rev. Methods Primers}\ }\textbf {\bibinfo {volume} {2}},\ \bibinfo {pages} {54} (\bibinfo {year} {2022})}\BibitemShut {NoStop}%
\bibitem [{\citenamefont {Khodas}\ \emph {et~al.}(2025)\citenamefont {Khodas}, \citenamefont {Mu}, \citenamefont {Mazin},\ and\ \citenamefont {Belashchenko}}]{khodas2025tuning}%
  \BibitemOpen
  \bibfield  {author} {\bibinfo {author} {\bibfnamefont {M.}~\bibnamefont {Khodas}}, \bibinfo {author} {\bibfnamefont {S.}~\bibnamefont {Mu}}, \bibinfo {author} {\bibfnamefont {I.}~\bibnamefont {Mazin}},\ and\ \bibinfo {author} {\bibfnamefont {K.}~\bibnamefont {Belashchenko}},\ }\bibfield  {title} {\bibinfo {title} {Tuning of altermagnetism by strain},\ }\href {https://arxiv.org/abs/2506.06257} {\bibfield  {journal} {\bibinfo  {journal} {arXiv:2506.06257}\ } (\bibinfo {year} {2025})}\BibitemShut {NoStop}%
\end{thebibliography}
\providecommand{\noopsort}[1]{}\providecommand{\singleletter}[1]{#1}%

\end{document}